\documentclass[10pt]{article}

\usepackage{graphicx}    
\usepackage{amsfonts}
\usepackage{amssymb}
\usepackage{graphics}
\usepackage{mathrsfs}
\usepackage{hyperref}
\usepackage{color}
\newtheorem{remark}{Remark}[section]

\newtheorem{theorem}{Theorem}[section]

\newtheorem{corollary}{Corollary}[section]
\newtheorem{lemma}{Lemma}[section]

\newtheorem{definition}{Definition}[section]

\long\def\symbolfootnote[#1]#2{\begingroup%
\def\thefootnote{\fnsymbol{footnote}}\footnote[#1]{#2}\endgroup} 

\begin{document}

%
\begin{center}
\Large \textbf{Morphogenesis of Chaos}
\end{center}
{\par\noindent}
\vspace{-0.3cm}
\begin{center}
\normalsize \textbf{M.U. Akhmet$ \symbolfootnote[1]{Corresponding Author Tel.: +90 312 210 5355,  Fax: +90 312 210 2972, E-mail: marat@metu.edu.tr}^{,a}$, M.O. Fen$^a$} \\
\vspace{0.2cm}
\textit{\textbf{\footnotesize$^a$Department of Mathematics, Middle East Technical University, 06800, Ankara, Turkey}}
\vspace{0.1cm}
\end{center}
{\par\noindent}

~~~~~~~~~~~~~~~~~~~~~~~~~~~~~~~~~~~~~~~~~~~~~~~~~~~~~~~~~~~~~~~~~~~~~~~~~~~~~~~~~~~~~~~~~~~~~~~~~~~~~~~~~ {\footnotesize \textit{\textbf{Ordo Ab Chao}}} 
{\par\noindent}
{\par\noindent}

\begin{center}
\line(1,0){469}
\end{center}

\begin{flushleft}
\textbf{Abstract}
\end{flushleft}

\noindent\ignorespaces
Morphogenesis, as it is understood in a wide sense by  Ren$\acute{e}$ Thom \cite{Thom83}, is considered for various types of chaos. That is, those, obtained by period-doubling cascade, Devaney's and Li-Yorke chaos. Moreover, in discussion form we consider inheritance of intermittency, the double-scroll Chua's attractor and quasiperiodical motions as a possible skeleton of a chaotic attractor. To  make  our  introduction of the paper more clear, we have to  say  that one may consider  other various accompanying concepts of chaos such that a structure of the chaotic attractor, its fractal dimension, form of the bifurcation diagram, the spectra of Lyapunov exponents, etc.  We make comparison of the main concept of our paper with Turing's morphogenesis and John von Neumann automata, considering that this may be not only formal one, but will give ideas for the chaos development in the morphogenesis of Turing and for self-replicating machines. 

To provide rigorous study of the subject, we introduce new definitions such as chaotic sets of functions, the generator and replicator of chaos, and precise description of ingredients  for Devaney and Li-Yorke chaos in continuous dynamics. Appropriate simulations which illustrate the morphogenesis phenomenon are provided.
{\par\noindent}
 
\noindent\ignorespaces \textit{Keywords:} Morphogenesis, Self-replication, Hyperbolic set of functions, Chaotic set of functions, Period-doubling cascade, Devaney's chaos, Li-Yorke chaos, Intermittency, Chaotic attractor, Chaos control, The double-scroll Chua's attractor, Quasiperiodicity   

\begin{center}
\line(1,0){469}
\end{center}

\textbf{There are indicating properties of different types of chaos (sometimes they are called ingredients). Morphogenesis, which is understood as replication of a prior chaos to the large community of systems, saving the indicating properties is proposed and mathematically approved. }


\section{Introduction}

\begin{figure}[ht]
\centering
\includegraphics[height=1.0cm]{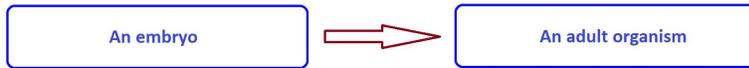} 
\caption{\footnotesize{Morphogenesis of an organism}}
\label{organism}
\end{figure}

The study of morphogenesis is one of the oldest of all the sciences, dating back to ancient Greece where Aristotle described the broad features of morphogenesis in birds, fish, and cephalopods. Even at this early stage of scientific thinking, he understood that an animal's egg had the ``potential" for its final form, but that it did not contain a miniature version of the form itself \cite{Davies}. The structure of morphogenesis in an organism is presented in Figure $\ref{organism}.$

In our paper, we try to follow the mechanism of morphogenesis, accepted as in classical and original meaning in biology \cite{Lafond}-\cite{Bozhkov}, not only in the sense of just replication of chaos of given type, but also saving the main assumption of Aristotle considering complexification of the form during its extension. That is, we try to establish the similarity between development from embryos to adult organisms considering development of chaos from simple observable systems to more complex ones. In the same time, we try to use the term morphogenesis not considering  Aristotle's idea strictly, but issuing from the sense of words \textit{morph} meaning ``form" and \textit{genesis} meaning ``creation" \cite{Davies}. In other words, similar to the ideas of Ren$\acute{e}$ Thom \cite{Thom83}, we employ the word \textit{morphogenesis} as its etymology indicates, to denote \textit{processes creating forms}. One of the purposes of our paper is to consider just the self-replication of chaotic systems without an emergence point of chaos, that is, we consider just the beginning of a technical task of formation of chaos of a given type through reproduction not necessarily considering complexification. This is different from Aristotle's idea in the sense that morphogenesis is effectively used in fields such as urban studies \cite{Courtat}, architecture \cite{Roudavski}, mechanics \cite{Taber}, computer science \cite{Bourgine}, linguistics \cite{Hagege} and sociology \cite{Archer,Buckley}.

\begin{figure}[ht]
\centering 
\includegraphics[height=1.0cm]{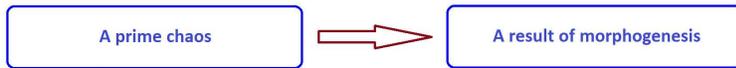} 
\caption{\footnotesize{Morphogenesis of chaos}}
\label{chaos_morphogenesis}
\end{figure}

The way of morphogenesis of chaos by the usage of a prime chaos as an \textit{embryo} is presented in Figure $\ref{chaos_morphogenesis}$. We understand the prime chaos as that one which rigorously examined for the presence of chaos of a certain type. For instance a chaos of the logistic map obtained through period-doubling cascade can be considered as a prime chaos, and a system which provides the chaos will be called the generator (system). Similarly, the system which shares the type of chaos with the generator being influenced will be called the replicator (system). In the paper, we consider the replicator as \textit{non-chaotic} and the sense of generator and replicator systems will be explained in the next section. 

Morphogenesis was deeply involved in mathematical discussions through Turing's investigations \cite{Turing} as well as the relation of the concept of structural stability \cite{Thom72}.  The paper \cite{Turing} was one of the first papers that considers mathematically the self-replicating forms using a set of reaction-diffusion equations \cite{Schiff}. 

Morphogenesis of chaos as understood in our paper, is a form-generating mechanism emerging from a dynamical process which is based on self-replication of chaos. Here, we accept the form (morph) not only as a type of chaos, but also accompanying concepts as the structure of the chaotic attractor, its fractal dimension, form of the bifurcation diagram, the spectra of Lyapunov exponents, inheritance of intermittency, etc. Not all of the properties have been discussed in our paper, but we suppose that they will be under investigation in our next papers.

The concept of self-replicating machines, in the abstract sense, starts with the ideas of von Neumann \cite{Neumann} and these ideas are supposed to be the origins of cellular automata theory \cite{Schiff}. According to von Neumann, it is feasible in principle to create a self-replicating machine, which he refers as an ``automaton", by starting with a machine $A,$ which has the ability to construct any other machine once it is furnished with a set of instructions, and then attaching to $A$ another component $B$ that can make a copy of any instruction supplied to it. Together with a third component labeled $C,$ it is possible to create a machine, denoted by $R,$ with components $A,B$ and $C$ such that $C$ is responsible to initiate $A$ to construct a machine as described by instructions, then make $B$ to create a copy of the instructions, and supply the copy of the instructions to the entire apparatus. The component $C$ is referred as ``control mechanism". It is the resulting machine $R',$ obtained by furnishing the machine $R$ by instructions $I_R,$ that is capable of replicating itself. Multiple usage of the set of instructions $I_R$ is crucial in the mechanism of self-replication. First, the instructions must be fulfilled by the machine $A,$ then they must be copied by $B,$ and finally the copy must be attached to machine $R$ to form the system $R'$ once again \cite{Schiff,Neumann}.

Our theory of morphogenesis of chaos relates the ideas of von Neumann about self-replicating machines in the following sense.
Initially, we take into account a system of differential equations (the generator) which plays the role of machine $A$ as in the ideas of von Neumann, and we use this system to influence in a unidirectional way, another system (the replicator) in the role of machine $B,$ in such a manner that the replicator mimics the same ingredients of chaos furnished to the generator. In the present paper, we use such ingredients in the form of period-doubling cascade, Devaney's and Li-Yorke chaos. In conclusion, the generator system  with the replicator counterpart together, that is the result-system, admits ingredients of the generator. In other words, a known type of chaos is self-replicated.

To understand the concept of our paper better, let us consider morphogenesis of fractal structures \cite{Mandelbrot77,Mandelbrot82}. It is important to say that Mandelbrot fractal structures exhibit the appearance of fractal hierarchy looking \textbf{in}, that is, \textbf{a part is similar to the whole}. Examples for this are the Julia sets \cite{Milnor06,Devaney_edt} and the Sierpinski carpet \cite{Peitgen04}. In our morphogenesis both directions, \textbf{in} and \textbf{out}, are present. Indeed, the fractal structure of the prior chaos has hierarchy looking \textbf{in}, and the structure for the result-system is obtained considering hierarchy looking \textbf{out}, that is, when \textbf{the whole is similar to the part}.

\begin{figure}[ht] 
\centering
\includegraphics[height=1.6cm]{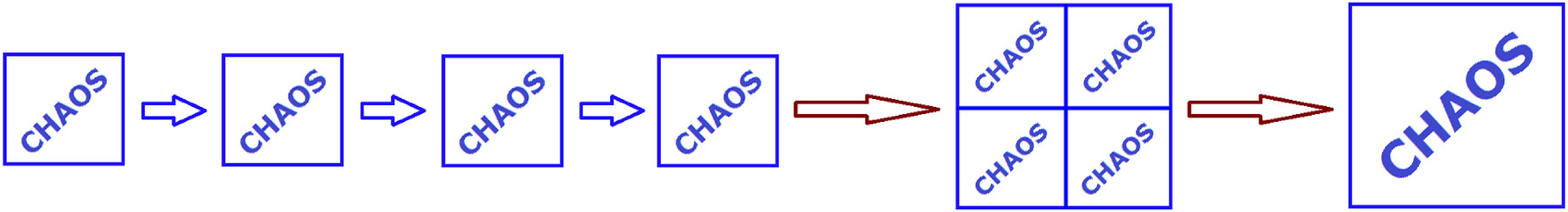} 
\caption{\footnotesize{Morphogenesis of chaos through consecutive replications}}
\label{box1}
\end{figure}

\begin{figure}[ht]
\centering
\includegraphics[height=3.4cm]{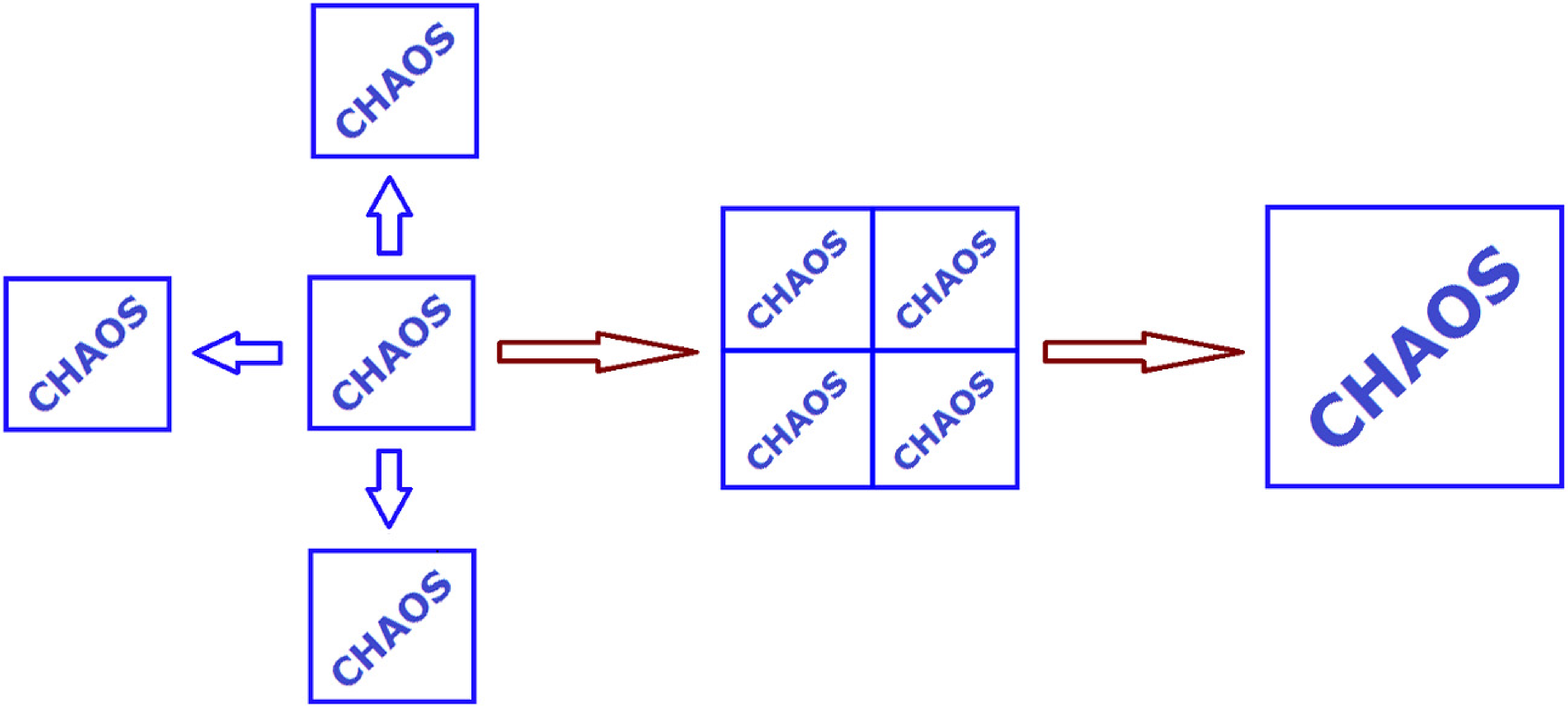} 
\caption{\footnotesize{Morphogenesis of chaos from a prior chaos as a core}}
\label{box2}
\end{figure}

Two different mechanisms of morphogenesis are considered in our paper. In the first one, the generator is a first link of a chain where all other systems are consecutive links. This mechanism is illustrated in  Figure \ref{box1}. In another way, the generator connected with all other systems directly such that it is rather core than a beginning element and this method is described in Figure \ref{box2}. In our paper we do not discuss theoretically constraints on the dimension of the result-system, but under certain conditions it seems the dimension is not restricted for both mechanisms. However, this is definitely true for the core mechanism even with infinite dimension.   We call the two ways as \textit{the chain} and \textit{the core} mechanisms, respectively.   We will discuss and simulate the chain mechanism in our paper, mainly, since the core mechanism can be discussed very similarly. One can invent other mechanisms of the morphogenesis, for example, by considering ``composition" of the two mechanisms proposed presently.

Successive applications of the morphogenesis procedure resemble the episodes of embryogenesis \cite{Lafond}, in the biological sense. That is, in each ``morphogenetic step", the regeneration of the same type of chaos in an arbitrary higher dimensional system is obtained. Moreover, elements inside the chaotic attractor of the generator, presented in the functional sense in our paper, play a role similar to the form-generating chemical substances, called morphogens, which are responsible for the morphogenetic processes \cite{Turing}. In other words, these elements are responsible for the morphogenesis of chaos.

In our paper, we are describing a process  involving the replication of chaos which does not occur in the course of time, but instead an \textbf{instantaneous} one. In other words, the prior chaos is mimicked in all existing replicators such that the generating mechanism works through arranging connections between systems not with the lapse of time.

Symbolic dynamics, whose earliest examples were constructed by Hadamard \cite{Hadamard1898} and Morse \cite{Morse38}, is one of the oldest techniques for the study of chaos. Symbolic dynamical systems are systems whose phase space consists of one-sided or two-sided infinite sequences of symbols chosen from a finite alphabet. Such dynamics arises in a variety of situations such as in horseshoe maps and the logistic map. The set of allowed sequences is invariant under the shift map, which is the most important ingredient in symbolic dynamics \cite{Wiggins}-\cite{Grebogi97}. Moreover, it is known that the symbolic dynamics admits the chaos in the sense of both Devaney and Li-Yorke \cite{Dev90,Rob95,Akh5,Akh8,Akh2}. The Smale Horseshoe map is first studied by Smale \cite{Smale67} and it is an example of a diffeomorphism which is structurally stable and possesses a chaotic invariant set \cite{Wiggins2, Dev90, Kennedy01b}. The horseshoe arises whenever one has transverse homoclinic orbits, as in the case of the Duffing equation \cite{Holmes}. People used the symbolic dynamics to discover chaos, but we suppose that it can serve as an ``embryo"  for the morphogenesis of chaos. 

The phenomenon of the form recognition for chaotic processes began in pioneering papers already. It was Lorenz \cite{Lorenz63} who discovered that the dynamics of an infinite dimensional system being reduced to three dimensional equation can be next analyzed in its chaotic appearances by application of the simple unimodal one dimensional map. Smale \cite{Smale67} explained that the geometry of the horseshoe map is underneath of the Van der Pol equation's complex dynamics investigated by Cartwright and Littlewood \cite{Cartwright1} and later by Levinson \cite{Levinson}. Nowadays, the Smale horseshoes with its chaotic dynamics, is one of the basic instruments when one tries to recognize a chaos in a process. Guckenheimer and Williams gave a geometric description of the flow of Lorenz attractor to show the structural stability of codimension 2 \cite{Guc79}. In addition to this, it was found out that the topology of the Lorenz attractor is considerably more complicated than the topology of the horseshoe \cite{Holmes}. Moreover, Levi used a geometric approach for a simplified version of the Van der pol equation to show existence of horseshoes embedded within the Van der Pol map and how the horseshoes fit in the phase plane \cite{Levi}. In other words, all the mentioned results say about chaos recognition, by reducing complex behavior to the structure with recognizable chaos. In \cite{Akh5,Akh2,Akh1,Akh3,Akh4,Akh6,Akh7}, we provide a different and constructive way when a recognized chaos can be extended saving the form of chaos to a multidimensional system. In the present study we generalize the idea to the morphogenesis of chaos. 

Now, to illustrate the concept discussed in our paper, let us visualize the morphogenesis of chaos  through simulating the Poincar$\acute{e}$ sections of both generator and replicator systems. For our purposes we shall take into account the Duffing's chaotic oscillator \cite{Th02}, represented by the differential equation
\begin{eqnarray} 
\begin{array}{l}
x''+0.05x'+x^3=7.5cos(t).\label{ueda1}
\end{array}
\end{eqnarray}
Defining the variables $x_1=x$ and $x_2=x',$ equation $(\ref{ueda1})$ can be reduced to the system
\begin{eqnarray} 
\begin{array}{l}
x'_1=x_2 \\  \label{ueda2}
x'_2=-0.05x_2-x_1^3+7.5cos(t).
\end{array}
\end{eqnarray}
Next, we attach three replicator systems, by the chain mechanism shown in Figure \ref{box1}, to the generator $(\ref{ueda2})$  and take into account the $8-$dimensional system
\begin{eqnarray} 
\begin{array}{l}
x'_1=x_2 \\  \label{ueda3}
x'_2=-0.05x_2-x_1^3+7.5cos(t) \\
x'_3=x_4+x_1 \\
x'_4=-3x_3-2x_4-0.008x_3^3+x_2 \\
x'_5=x_6+x_3 \\
x'_6=-3x_5-2.1x_6-0.007x_5^3+x_4 \\
x'_7=x_8+x_5 \\
x'_8=-3.1x_7-2.2x_8-0.006x_7^3+x_6. 
\end{array}
\end{eqnarray}

\begin{figure}[ht] 
\centering
\includegraphics[width=12.0cm]{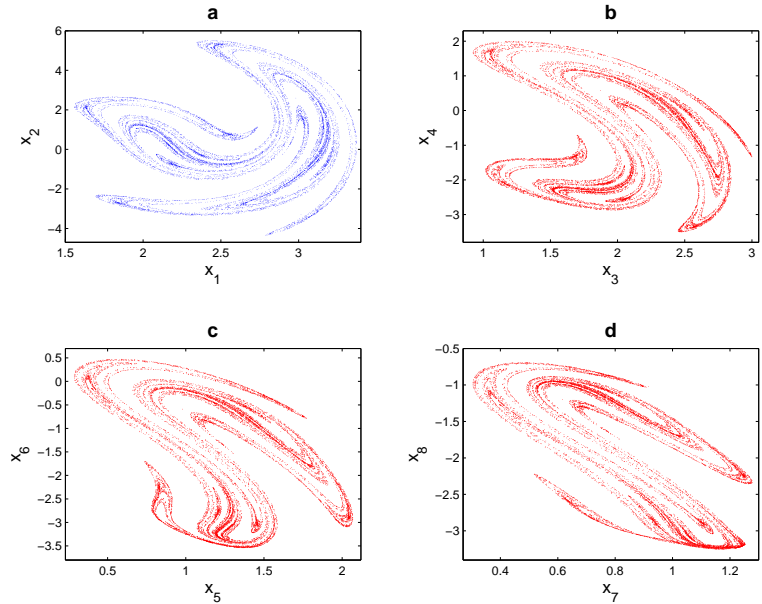}
\caption{\footnotesize{The picture in $(a)$ not only represents the projection of the whole attractor on the $x_1-x_2$ plane, but also the strange attractor of the generator. In a similar way, the pictures introduced in $(b),(c)$ and $(d)$ represent the chaotic attractors of the first, second and the third replicators, respectively. The resemblance of the presented shapes of the attractors between the generator and the replicator systems reveal the morphogenesis of chaos.}}
\label{poincare_section1}
\end{figure}

We suppose that system $(\ref{ueda3})$ admits a chaotic attractor in the $8-$dimensional phase space. By marking the trajectory of this system with the initial data $x_1(0)=2, x_2(0)=3, x_3(0)=x_5(0)=x_7(0)=-1, x_4(0)=x_6(0)=x_8(0)=1$ stroboscopically at times that are integer multiples of $2\pi,$ we obtain the Poincar$\acute{e}$ section inside the $8-$dimensional space and in Figure \ref{poincare_section1}, where the morphogenesis is apparent, we illustrate the $2-$dimensional projections of the whole Poincar$\acute{e}$ section.  Figure \ref{poincare_section1}(a) represents the projection of the Poincar$\acute{e}$ section on the $x_1-x_2$ plane, and we note that this projection is in fact the strange attractor of the original generator system $(\ref{ueda2}).$ On the other hand, the projections on the planes $x_3-x_4,$ $x_5-x_6,$ and $x_7-x_8$ presented in Figure $\ref{poincare_section1} (b),$ Figure $\ref{poincare_section1} (c)$ and Figure $\ref{poincare_section1} (d),$ respectively, are the attractors of the replicator systems. 

One can see that the attractors indicated in Figure $\ref{poincare_section1}, (b)-(d)$ repeated the structure of the attractor shown in Figure $\ref{poincare_section1}(a),$ and the similarity between these pictures is a manifestation of the morphogenesis of chaos.

\begin{figure}[ht] 
\centering
\includegraphics[width=14.0cm]{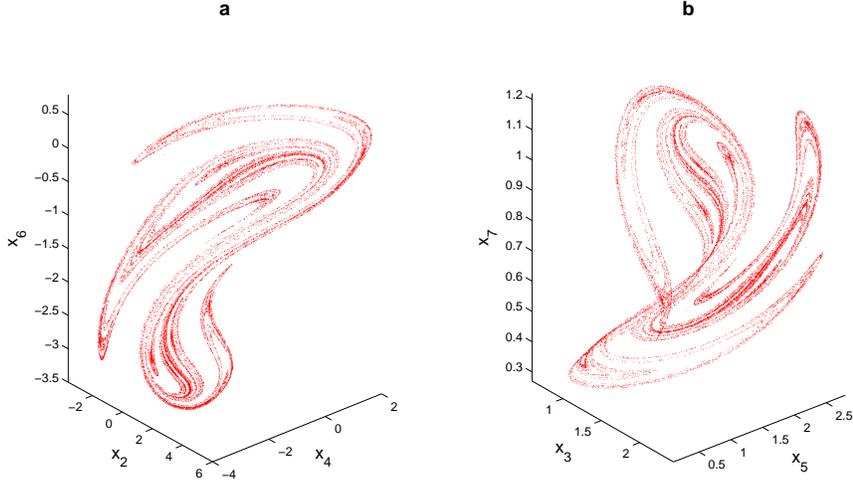}
\caption{ \footnotesize{In $(a)$ and $(b)$ projections of the result chaotic attractor on $x_2-x_4-x_6$ and $x_3-x_5-x_7$ spaces are respectively presented. One can see in $(a)$ and $(b)$ the additional \textit{foldings} which are not possible to observe in the  $2-$dimensional  picture of  the prior  classical  chaos  shown in Figure $\ref{poincare_section1}, (a).$ In the same time, the shape of the original attractor is seen in the resulting chaos. The illustrations in $(a)$ and $(b)$ repeat the structure of the attractor of the generator and the similarity between these pictures is a manifestation of the morphogenesis of chaos.}}
\label{poincare_section2}
\end{figure}

Next, let us illustrate in Figure $\ref{poincare_section2},$ which also informs us about morphogenesis, the $3-$dimensional projections of the whole Poincar$\acute{e}$ section on the $x_2-x_4-x_6$ and $x_3-x_5-x_7$ spaces. One can see in Figure $\ref{poincare_section2} (a)$ and Figure $\ref{poincare_section2} (b)$ the additional \textit{foldings} which is not possible to observe in the classical strange attractor shown in Figure $\ref{poincare_section1}, (a).$

Despite we are restricted to make illustrations at most in $3-$dimensional spaces, taking inspiration from Figure \ref{poincare_section1} and Figure \ref{poincare_section2}, one can imagine that the structure of the original Poincar$\acute{e}$ section in the $8-$dimensional space will be similar through its fractal structure, but more beautiful and impressive than its projections. From this point of view, we are not surprised since these facts have been proved theoretically in the paper.

To explain the morphogenesis procedure of our paper, let us give the following information. It is known that if one considers the evolution equation $u'=L[u]+F(t,u),$ where $L[u]$ is a linear operator with spectra placed in the left half of the complex plane, then a function $F(t)$ being considered as input with a certain property (boundedness, periodicity, almost periodicity) produces through the equation the output, a solution with a similar property, boundedness/periodicity/almost periodicity. In particular, in our paper, we solved a similar problem when the linear system has eigenvalues with negative real parts and input is considered as a chaotic set of functions with a known type. Our results are different in the sense that the input and the output are not single functions, but \textit{a collection of functions}. In other words, we prove that both the input and the output are \textit{chaos} of the same type for the discussed equation. The way of our investigation is arranged in the well accepted traditional mathematical fashion, but with a new and a more complex way of arranging of the connections between the input and the output. The same is true for the control results discussed in the paper.

One of the usage area of master-slave systems is the study of synchronization of chaotic systems \cite{Gon04,Pecora90}. In 1990, Pecora and Carroll \cite{Pecora90} realized that two identical chaotic systems can be synchronized under appropriate unidirectional coupling schemes. Let us consider the system
\begin{eqnarray}  
\begin{array}{l} \label{master_synch}
x'=G(x),
\end{array}
\end{eqnarray} 
as the master, where $x \in \mathbb R^d,$ such that the steady evolution of the system occurs in a chaotic attractor. Now, let us take into consideration the slave system whose dynamics is governed by the equation
\begin{eqnarray}
\begin{array}{l}
y'=H(x,y), \label{slave_synch}
\end{array}
\end{eqnarray} 
when the unidirectional drive is established, with $H(x,y)$ verifying the condition
\begin{eqnarray}
\begin{array}{l}
H(x,y)=G(x), \label{condition_identical}
\end{array}
\end{eqnarray}
for $y=x,$ and takes the form 
\begin{eqnarray} 
\begin{array}{l} 
y'=G(y),
\end{array}
\end{eqnarray}
which is a copy of system $(\ref{master_synch}),$ in the absence of driving. In such couplings, the signals of the master system acts on the slave system, but the converse is not true. Moreover, this action becomes null when the two systems follow identical trajectories \cite{Gon04}. The continuous control scheme \cite{Kapitaniak94c,Ding94} and the method of replacement of variables \cite{Pecora91,Cuomo93} can be used to obtain couplings in the form of the system $(\ref{master_synch})+(\ref{slave_synch}).$
Synchronization of a slave system to a master system, under the condition $(\ref{condition_identical}),$ is known as identical synchronization and it occurs when there are sets of initial data $\mathscr{X} \subset \mathbb R^d$ and $\mathscr{Y} \subset \mathbb R^d$ for the master and slave systems, respectively, such that the equation
\begin{eqnarray}  
\begin{array}{l} \label{sync_condition}
\displaystyle \lim_{t \to \infty} \left\| x(t)-y(t) \right\| = 0,
\end{array}
\end{eqnarray}
holds, where $(x(t), y(t))$ is a solution of system $ (\ref{master_synch})+(\ref{slave_synch})$ with initial data $(x(0),y(0)) \in \mathscr{X} \times \mathscr{Y}.$ 

When the slave system is defined different than the master counterpart in the absence of driving, one can consider generalized synchronization. In this case, the dimensions of the master and slave systems are possibly different, say $d$ and $r$ respectively, and the definition of identical synchronization given by equation $(\ref{sync_condition})$ is replaced by
\begin{eqnarray}  
\begin{array}{l} \label{sync_condition2}
\displaystyle \lim_{t \to \infty} \left\| \psi[x(t)]-y(t) \right\| = 0,
\end{array}
\end{eqnarray}
where $\psi[x(t)]$ is a functional relation which determines the phase space trajectory $y(t)$ of the slave system from the trajectory $x(t)$ of the master. Here, it is crucial that the functional $\psi[x(t)]$ has to be the same for any initial data $(x(0),y(0))$ chosen from the set $\mathscr{X} \times \mathscr{Y},$ where the sets $\mathscr{X} \subset \mathbb R^d$ and $\mathscr{Y} \subset \mathbb R^r$ are defined in a similar way as in the case of identical synchronization. Generalized synchronization includes the identical synchronization as a particular case, and when $\psi[x(t)]$ is identity, one attains the equation $(\ref{sync_condition}).$

In our theoretical results of morphogenesis of chaos, we use coupled systems in which the generator system influences the replicator in a unidirectional way, that is the generator affects the behavior of the replicator, but not the converse. The possibility of making use of nonidentical systems in the replication of a known type of chaos furnished to the generator is an advantage of the procedure. On the other hand, contrary to the method that we present, in the synchronization of chaotic systems, one does not consider the type of the chaos that the master and slave systems admit. The problem that whether the synchronization of systems implies the same type of chaos for both master and slave has not been taken into account yet. 

In our results, we do not consider the synchronization problem as well the asymptotic properties, such as equations $(\ref{sync_condition})$ or $(\ref{sync_condition2}),$ for the systems; but we say that the type of the chaos is kept \textit{invariant} in the procedure.  That is why the classes which can be considered with respect to this \textit{invariance} is expectedly wider then those investigated for synchronization of chaos.  Since there is no strong relation and accordance between the solutions of the generator and the replicator in the asymptotic point of view, the terms \textit{master} and \textit{slave} as well as \textit{drive} and \textit{response} are not preferred to be used for the analyzed systems.

Nowadays,  one can  consider development  of a multidimensional chaos  from a low-dimensional one in different  ways. One of them is  chaotic itinerancy \cite{Kaneko2000}-\cite{Sauer2003}.
The itinerant motion among varieties of ordered states through high-dimensional chaotic motion can be observed and this behavior is named as chaotic itinerancy. In other words, chaotic itinerancy is a universal dynamics in high-dimensional dynamical systems, showing itinerant motion among varieties of low-dimensional ordered states through high-dimensional chaos. This  phenomenon occurs in different real world processes: optical turbulence \cite{Ikeda89}; globally coupled chaotic systems \cite{Kaneko90,Kaneko91}; non-equilibrium neural networks \cite{Tsuda91,Tsuda92}; analysis of brain activities \cite{Freeman94}, ecological systems \cite{Kim2007}. One can see that in its degenerated form chaotic itinerancy relates to intermittency \cite{Pomeau80, Moon}, since they both represent dynamical interchange of irregularity and regularity.

Likewise the itinerant chaos observed in brain activities, we have low-dimensional chaos in the subsystems considered and high-dimensional chaos is obtained when one considers all subsystems as a whole. The main difference compared to our technique is in the elapsed time for the occurrence of the process. In our discussions, no itinerant motion is observable and all resultant chaotic subsystems process simultaneously, whereas the low-dimensional chaotic motions take place as time elapses in the case of chaotic itinerancy. Knowledge of the type of chaos is another difference between chaotic itinerancy and our procedure. Possibly the present way  of replication of chaos will give a light to the solutions of problems about extension of irregular behavior (crises, collapses, etc.) in interrelated or multiple connected systems which can arise in problems of classical mechanics \cite{Moon} and electrical systems \cite{Chua94,Kapitaniak-Chua94}, economic theory \cite{Hanswalter89}, brain activity investigations \cite{Freeman94}.


In systems whose dimension is at least four, it is possible to observe chaotic attractors with at least two positive Lyapunov exponents and such systems are called hyperchaotic \cite{Sprott10}. An example of a four dimensional hyperchaotic system is discovered by R$\ddot{o}$ssler \cite{Rossler79}. Combining two or more chaotic, not necessarily identical, systems is a way of achieving hyperchaos \cite{Kapitaniak94}-\cite{Kapitaniak94b}. However, in the present paper, we take into account exactly one chaotic system with a known type of chaos, and use this system as the generator to reproduce the same type of chaos in other systems. On the other hand, the crucial phenomena in the hyperchaotic systems is the existence of two or more positive Lyapunov exponents and the type of chaos is not taken into account. In our way of morphogenesis, the critical situation is rather the replication of a known type of chaos.

\section{Preliminaries} \label{preliminaries}

Throughout the paper, the generator will be considered as a system of the form 
\begin{eqnarray}\label{1}
x'=F(t,x),
\end{eqnarray}
where $F:\mathbb R \times \mathbb R^{m}\to \mathbb R^{m},$ is a continuous function in all its arguments,  and the replicator is assumed to have the form 
\begin{eqnarray}\label{replicator}
y'=Ay+g(.,y),
\end{eqnarray}
where $g:\mathbb R^{m}\times \mathbb R^{n} \to \mathbb R^{n},$  is a continuous function in all its arguments, and the constant $n\times n$ real valued matrix $A$ has real parts of eigenvalues all negative.
To describe better how the replicator, $(\ref{replicator}),$ is involved in the morphogenesis process, we  use the dot, ``.", symbol in the system. To insert the replicator in the morphogenesis process we replace the dot, ``.",  by $x$ or a dependent variable of a replicator-predecessor and obtain the system  of the form
\begin{eqnarray}\label{2}
y'=Ay+g(x,y).
\end{eqnarray} 

Let us remind that in the morphogenesis process of our paper one generator and several replicators may be involved, but in future one can consider process of morphogenesis where several generators are used. 

In general, in our paper, the replicator system $(\ref{replicator})$ does not admit a chaos. That is, we say that the replicator is \textit{non-chaotic}. We understand the phrase ``non-chaotic" in the sense that if  ``." is, for example, replaced by a periodic function, then system $(\ref{replicator})$ does not produce a chaos, but admits a periodic solution and it is stable.  

We call the system which unites the generator system and several replicators, of type $(\ref{2}),$ in morphogenesis process as \textit{the result-system of morphogenesis} or in what follows just as \textit{the result-system}.

In what follows $ \mathbb R $ and $ \mathbb N$ denote the sets of real numbers and natural numbers, respectively, and the uniform norm $\left\|\Gamma\right\|=\displaystyle\sup_{\left\|v\right\|=1} \left\|\Gamma v \right\|$ for matrices is used in the paper. 

It is easy to verify the existence of positive real numbers $N$ and $\omega$ such that $\left\|e^{At}\right\| \leq Ne^{-\omega t},$ $t\geq 0,$ and these numbers will be used in the last assumption below.

The following assumptions are needed throughout the paper:
\begin{enumerate}
\item[\bf (A1)] There exists a positive real number $T$ such that the function $F(t,x)$ satisfies the periodicity condition $F(t+T,x)=F(t,x),$ for all $t\in \mathbb R,$ $x\in \mathbb R^m;$
\item[\bf (A2)] There exists a positive real number $L_0$ such that $\left\|F(t,x_1)-F(t,x_2)\right\| \leq L_0\left\|x_1-x_2\right\|,$ for all $t\in \mathbb R, x_1,x_2\in \mathbb R^m;$
\item[\bf (A3)] There exist a positive number $H_0<\infty$ such that $\displaystyle \sup_{t\in \mathbb R, x\in \mathbb R^m} \left\|F(t,x)\right\|=H_0;$
\item[\bf (A4)] There exists a positive real number $L_1$ such that 
$ \left\|g(x_1,y)-g(x_2,y)\right\| \geq L_1\left\|x_1-x_2\right\|,$ for all $x_1,x_2 \in \mathbb R^{m},$ $y \in \mathbb R^n;$
\item[\bf (A5)] There exist positive real numbers $L_2$ and $L_3$ such that 
$ 
\left\|g(x_1,y)-g(x_2,y)\right\| \leq L_2\left\|x_1-x_2\right\|,$ for all $x_1,x_2 \in \mathbb R^{m},$ $y\in \mathbb R^n,$ and $\left\|g(x,y_1)-g(x,y_2)\right\| \leq L_3\left\|y_1-y_2\right\|,$ for all $x\in \mathbb R^m,$ $y_1,y_2 \in \mathbb R^{n};$
\item[\bf (A6)] There exist a positive number $M_0 < \infty$ such that $\displaystyle \sup_{x\in \mathbb R^m, y\in \mathbb R^n} \left\|g(x,y)\right\|=M_0;$
\item[\bf (A7)] $NL_3-\omega<0.$
\end{enumerate} 

\begin{remark}
The results presented in the remaining parts are also true even if we replace the non-autonomous system $(\ref{1})$ by the autonomous equation 
\begin{eqnarray}\label{autonomus_master}
x'=\overline{F}(x),
\end{eqnarray}
where $\overline{F}:\mathbb R^{m}\to \mathbb R^{m},$ $m \geq 3,$ is continuous, and with conditions which are counterparts of $(A2)$ and $(A3).$
\end{remark}

Using the theory of quasilinear equations \cite{Hale80}, one can verify that for a given solution $x(t)$ of system (\ref{1}), there exists a unique bounded on $\mathbb R$ solution $y(t)$ of the system $y'=Ay+g(x(t),y),$ denoted by $y(t)=\phi_{x(t)}(t)$ throughout the paper, which satisfies the integral equation
\begin{eqnarray}\label{integral_eqns}
\begin{array}{l}
y(t)=\displaystyle\int^{t}_{-\infty} e^{A(t-s)}  g(x(s),y(s)) ds. \\
\end{array}
\end{eqnarray}

Our main assumption is the existence of a nonempty set $\mathscr{A}_x$ of all solutions of system (\ref{1}), uniformly bounded on $\mathbb R.$ That is, there exists a positive real number $H$ such that $\displaystyle \sup_{t \in \mathbb R} \left\|x(t)\right\| \leq H,$ for all $x(t) \in \mathscr{A}_x.$

Let us introduce the sets of functions
\begin{eqnarray}
\begin{array}{l} \label{A_y}
\mathscr{A}_y=\left\{\phi_{x(t)}(t) ~|~  x(t) \in \mathscr{A}_x \right\},
\end{array}
\end{eqnarray}
and 
\begin{eqnarray}
\begin{array}{l}
\mathscr{A}=\left\{(x(t),\phi_{x(t)}(t)) ~|~ x(t) \in \mathscr{A}_x \right\}.
\end{array}
\end{eqnarray}

We note that for all $y(t) \in \mathscr{A}_y$ one has $ \displaystyle \sup_{t \in \mathbb R} \left\|y(t)\right\| \leq M,$ where $M=\frac{NM_0}{\omega}.$

Next, we reveal that if the set $\mathscr{A}_x$ is an attractor with basin $\mathscr{U}_x,$ that is, for each $x(t)\in \mathscr{U}_x$ there exists $\overline{x}(t)\in \mathscr{A}_x$ such that $\left\|x(t)-\overline{x}(t)\right\| \to 0$ as $t \to \infty,$ then the set $\mathscr{A}_y$ is also an attractor in the same sense. In the next lemma we specify the basin of attraction of $\mathscr{A}_y.$
Define the set
\begin{eqnarray}
\begin{array}{l}
\mathscr{U}_y=\left\{y(t)~|~ y(t) ~\textrm{is a solution of the system}~ y'=Ay+g(x(t),y) ~\textrm{for some}~ x(t) \in \mathscr{U}_x  ~  \right\}. 
\end{array}
\end{eqnarray}
\begin{lemma}\label{attractor}
$\mathscr{U}_y$ is a basin of $\mathscr{A}_y.$ 
\end{lemma}

\noindent \textbf{Proof.}
Fix an arbitrary positive real number $\epsilon$ and let $y(t) \in \mathscr{U}_y$ be a given solution of the system $y'=Ay+g(x(t),y)$ for some $x(t) \in \mathscr{U}_x.$ In this case there exists $\overline{x}(t) \in \mathscr{A}_x$ such that $\left\|x(t)-\overline{x}(t)\right\| \to 0$ as $t \to \infty.$ Let $\alpha=\frac{\omega -NL_3}{\omega - NL_3 +NL_2}$ and $\overline{y}(t)= \phi_{\overline{x}(t)}(t).$ Condition $(A7)$ implies that the number $\alpha$ is positive. Under the circumstances, one can find $R_0(\epsilon)>0$ such that if $t \geq R_0$ then $\left\|x(t)-\overline{x}(t)\right\| < \alpha\epsilon$ and  
$N\left\|y(R_0)-\overline{y}(R_0)\right\|e^{(NL_3-\omega)t}<\alpha\epsilon.$
The functions $y(t)$ and $\overline{y}(t)$ satisfy the relations
\begin{eqnarray*}
y(t)= e^{A(t-R_0)}y(R_0)+ \displaystyle\int^{t}_{R_0}e^{A(t-s)}g(x(s),y(s))ds,  \nonumber
\end{eqnarray*}
and
\begin{eqnarray*}
\overline{y}(t)= e^{A(t-R_0)}\overline{y}(R_0)+ \displaystyle\int^{t}_{R_0}e^{A(t-s)}g(\overline{x}(s),\overline{y}(s))ds,  \nonumber
\end{eqnarray*}
respectively. Making use of these relations one has
\begin{eqnarray*}
&&y(t)-\overline{y}(t) = e^{A(t-R_0)} (y(R_0)-\overline{y}(R_0)) \nonumber\\
&& + \displaystyle\int^{t}_{R_0} e^{A(t-s)}  \left[g(x(s),y(s))-g(\overline{x}(s),\overline{y}(s))\right]  ds   \\
&& = e^{A(t-R_0)} (y(R_0)-\overline{y}(R_0)) \\
&& + \displaystyle\int^{t}_{R_0} e^{A(t-s)}  \left[g(x(s),y(s))-g(x(s),\overline{y}(s))\right]  ds \\
&& + \displaystyle\int^{t}_{R_0} e^{A(t-s)}  \left[g(x(s),\overline{y}(s))-g(\overline{x}(s),\overline{y}(s))\right]  ds.                               
\end{eqnarray*}

Therefore, we have

\begin{eqnarray*}
&&\left\|y(t)-\overline{y}(t)\right\| \leq \left\|e^{A(t-R_0)}\right\| \left\|y(R_0)-\overline{y}(R_0)\right\| \\
&& + \displaystyle\int^{t}_{R_0} \left\|e^{A(t-s)}\right\|  \left\|g(x(s),y(s))-g(x(s),\overline{y}(s))\right\|  ds  \nonumber\\
&& + \displaystyle\int^{t}_{R_0} \left\|e^{A(t-s)}\right\|  \left\|g(x(s),\overline{y}s))-g(\overline{x}(s),\overline{y}(s))\right\|  ds \\
&& \leq Ne^{-\omega(t-R_0)} \left\|y(R_0)-\overline{y}(R_0)\right\| \\
&& + \displaystyle\int^{t}_{R_0} Ne^{-\omega (t-s)} L_3 \left\|y(s)-\overline{y}(s)\right\| ds \\
&& + \displaystyle\int^{t}_{R_0} Ne^{-\omega (t-s)} L_2 \left\|x(s)-\overline{x}(s)\right\| ds \\
&& \leq Ne^{-\omega(t-R_0)} \left\|y(R_0)-\overline{y}(R_0)\right\| \\
&& + \frac{NL_2 \alpha \epsilon}{\omega}e^{-\omega t}\left(e^{\omega t}-e^{\omega R_0}\right) \\
&& + NL_3 \displaystyle\int^{t}_{R_0} e^{-\omega (t-s)} \left\|y(s)-\overline{y}(s)\right\| ds.
\end{eqnarray*}

Let $u:[R_0,\infty)\to [0, \infty)$ be a function defined as $u(t)=e^{\omega t}\left\|y(t)-\overline{y}(t)\right\|.$
Then we reach the inequality
\begin{eqnarray*}
u(t) \leq Ne^{\omega R_0}\left\|y(R_0)-\overline{y}(R_0)\right\| + \frac{NL_2\alpha\epsilon}{\omega}\left(e^{\omega t}-e^{\omega R_0}\right) + NL_3 \displaystyle\int^{t}_{R_0} u(s) ds.                         
\end{eqnarray*}

Now let $\psi(t)=\frac{NL_2\alpha\epsilon}{\omega}e^{\omega t}$ and $\phi(t)=\psi(t) + c$ where $c=Ne^{\omega R_0}\left\|y(R_0)-\overline{y}(R_0)\right\|-\frac{NL_2\alpha\epsilon}{\omega}e^{\omega R_0}.$
Using these functions we get
\begin{eqnarray*}
u(t) \leq \phi(t)+NL_3\displaystyle\int^{t}_{R_0} u(s) ds .   \nonumber                      
\end{eqnarray*}
Applying Lemma (Gronwall) \cite{Corduneanu77} to the last inequality for $t\geq R_0,$ we obtain
\begin{eqnarray*}
u(t) \leq c+\psi(t)+NL_3\displaystyle\int^{t}_{R_0} e^{NL_3(t-s)}c ds+NL_3\displaystyle\int^{t}_{R_0} e^{NL_3(t-s)}\psi(s) ds \nonumber
\end{eqnarray*}
and hence,
\begin{eqnarray*}
&&u(t) \leq c + \psi(t) + c\left(e^{NL_3(t-R_0)}-1\right) + \frac{N^{2}L_2L_3\alpha\epsilon}{\omega(\omega-NL_3)}e^{\omega t} \left(1-e^{(NL_3-\omega)(t-R_0)}\right) \nonumber\\
&& = \frac{NL_2\alpha\epsilon}{\omega}e^{\omega t} + N\left\|y(R_0)-\overline{y}(R_0)\right\|e^{\omega R_0}e^{NL_3(t-R_0)} \\
&& -\frac{NL_2\alpha\epsilon}{\omega}e^{\omega R_0}e^{NL_3(t-R_0)} + \frac{N^{2}L_2L_3\alpha\epsilon}{\omega(\omega-NL_3)}e^{\omega t} \left(1-e^{(NL_3-\omega)(t-R_0)}\right).
\end{eqnarray*}
Thus,
\begin{eqnarray*}
&& \left\|y(t)-\overline{y}(t)\right\|  \leq \frac{NL_2\alpha\epsilon}{\omega} + N\left\|y(R_0)-\overline{y}(R_0)\right\|e^{(NL_3-\omega)(t-R_0)} \nonumber\\
&& -\frac{NL_2\alpha\epsilon}{\omega}e^{(NL_3-\omega)(t-R_0)} + \frac{N^{2}L_2L_3\alpha\epsilon}{\omega(\omega-NL_3)} \left(1-e^{(NL_3-\omega)(t-R_0)}\right)\\
&& = N\left\|y(R_0)-\overline{y}(R_0)\right\|e^{(NL_3-\omega)(t-R_0)}\\
&& + \frac{NL_2\alpha\epsilon}{\omega}\left(1-e^{(NL_3-\omega)(t-R_0)}\right) + \frac{N^2L_2L_3\alpha\epsilon}{\omega(\omega-NL_3)}\left(1-e^{(NL_3-\omega)(t-R_0)}\right)\\
&& = N\left\|y(R_0)-\overline{y}(R_0)\right\|e^{(NL_3-\omega)(t-R_0)}+ \frac{NL_2\alpha\epsilon}{\omega-NL_3}\left(1-e^{(NL_3-\omega)(t-R_0)}\right)\\
&& < N\left\|y(R_0)-\overline{y}(R_0)\right\|e^{(NL_3-\omega)(t-R_0)}+ \frac{NL_2\alpha\epsilon}{\omega-NL_3}.
\end{eqnarray*}

Consequently, for $t \geq 2R_0,$ we have 
$\left\|y(t)-\overline{y}(t)\right\| < \left(1+ \frac{NL_2}{\omega-NL_3}\right) \alpha \epsilon = \epsilon,$
and hence $\left\|y(t)-\overline{y}(t)\right\| \to 0$ as $t \to \infty.$

The proof is completed. $\square$

Now, let us define the set
\begin{eqnarray*}
\mathscr{U}=\left\{(x(t),y(t))~|~ (x(t),y(t)) ~\textrm{is a solution of system}~ (\ref{1})+(\ref{2}) ~\textrm{such that}~  x(t) \in \mathscr{U}_x \right\}. 
\end{eqnarray*}
Next, we state the following corollary of Lemma \ref{attractor}.
\begin{corollary}
$\mathscr{U}$ is a basin of $\mathscr{A}.$ 
\end{corollary}

\noindent \textbf{Proof.}
Let $(x(t),y(t))\in \mathscr{U}$ be a given solution of system $(\ref{1})+(\ref{2}).$ According to Lemma \ref{attractor}, one can find $(\overline{x}(t),\overline{y}(t)) \in \mathscr{A}$ such that $\left\|x(t)-\overline{x}(t)\right\| \to 0$ as $t \to \infty$ and $\left\|y(t)-\overline{y}(t)\right\| \to 0$ as $t \to \infty.$ Consequently, $\left\|(x(t),y(t))-(\overline{x}(t),\overline{y}(t))\right\| \to 0$ as $t \to \infty.$ The proof is finalized. $\square$


\section{Description of chaotic sets of functions}\label{secchaos}

In this section of the paper, the descriptions for the chaotic sets of continuous functions will be introduced and the definitions of the chaotic features will be presented, both in the Devaney's sense and in the Li-Yorke sense.

Let us denote by 
\begin{eqnarray} \label{collection}
\mathscr{B}=\left\{\psi(t)~|~ \psi: \mathbb R \to K ~\textrm{is}  ~\textrm{continuous}  \right\} 
\end{eqnarray} 
a collection of functions, where $K \subset \mathbb R^q ,$ $q\in \mathbb N,$ is a bounded region.


We start with the description of chaotic set of functions in Devaney's sense and next continue with the Li-Yorke counterpart. 

\subsection{\textbf{Chaotic set of functions in Devaney's sense}}\label{subsecdevaney}
In this part, we shall elucidate the ingredients of the chaos in Devaney's sense for the set $\mathscr{B}$ and the first definition is about the sensitivity of chaotic set of functions.

\begin{definition}\label{dev1}
$\mathscr{B}$ is called sensitive if there exist positive numbers $\epsilon$ and $\Delta$ such that for every $\psi(t) \in \mathscr{B}$ and for arbitrary $\delta>0$ there exist $\overline{\psi}(t) \in \mathscr{B},$ $t_0\in \mathbb R$ and an interval $J \subset [t_0,\infty)$ with length not less than $\Delta$ such that $\left\|\psi(t_0)-\overline{\psi}(t_0)\right\|<\delta$ and $\left\|\psi(t)-\overline{\psi}(t)\right\| \geq \epsilon,$ for all $t\in J.$
\end{definition}

Definition \ref{dev1} considers the inequality $(\geq \epsilon)$ over the interval $J.$ In the Devaney's chaos definition for the map, the inequality is assumed for discrete moments. Let us reveal how one can extend the inequality from a discrete point to an interval by considering continuous flows.

In \cite{Dev90}, it is indicated that a continuous map $\varphi:\Lambda  \to \Lambda,$ with an invariant domain $\Lambda \subset \mathbb R^k, k\in \mathbb N,$ has sensitive dependence on initial conditions if there exists $\overline{\epsilon} >0$ such that for any $x \in \Lambda$ and any neighborhood $\mathscr{U}$ of $x,$ there exist $y \in \mathscr{U}$ and a natural number $n$ such that $\left\|\varphi^n(x)-\varphi^n(y)\right\| > \overline{\epsilon}.$

Suppose that the set $\mathscr{A}_x$ satisfies the definition of sensitivity in the following sense.
There exists $\overline{\epsilon} > 0$ such that for every $x(t) \in \mathscr{A}_x$ and arbitrary $\delta > 0,$ there exist $\overline{x}(t) \in \mathscr{A}_x,$ $t_0\in \mathbb R$ and a real number $\zeta \geq t_0$ such that $\left\|x(t_0)-\overline{x}(t_0)\right\| < \delta$ and $\left\|x(\zeta)-\overline{x}(\zeta)\right\| > \overline{\epsilon}.$

In this case, for given $x(t)\in \mathscr{A}_x$ and $\delta >0,$ one can find $\overline{x}(t) \in \mathscr{A}_x$ and $\zeta \geq t_0$ such that $\left\|x(t_0)-\overline{x}(t_0)\right\| < \delta $ and $\left\|x(\zeta)-\overline{x}(\zeta)\right\| > \overline{\epsilon}.$ Let $\Delta = \frac{\overline{\epsilon}}{8HL_0}$ and take a number $\Delta_1$ such that $\Delta \leq \Delta_1 \leq \frac{\overline{\epsilon}}{4HL_0}.$ Using the appropriate integral equations, for $t\in [\zeta,\zeta+\Delta_1],$ we have
\begin{eqnarray*}
&&\left\|x(t)-\overline{x}(t)\right\|  \geq \left\|x(\zeta)-\overline{x}(\zeta)\right\|- \left\|\displaystyle\int^{t}_{\zeta} [F(s,x(s))-F(s,\overline{x}(s))] ds\right\|  \nonumber \\
&& > \overline{\epsilon} - 2HL_0\Delta_1 \\
&& \geq \frac{\overline{\epsilon}}{2}.
\end{eqnarray*}

The last inequality confirms that $\mathscr{A}_x$ satisfies the Definition \ref{dev1} with $\epsilon = \frac{\overline{\epsilon}}{2}$ and $J=[\zeta,\zeta+\Delta_1].$
So the definition is a natural one. It provides more information then discrete moments and for us it is important that the extension on the interval is useful to prove the property for morphogenesis process.

In the next two definitions, we continue with the existence of a dense function in the set of chaotic functions followed by the transitivity property.

\begin{definition}\label{dev2}
$\mathscr{B}$ possesses a dense function $\psi^{*}(t)\in \mathscr{B}$ if for every function $\psi(t) \in \mathscr{B},$ arbitrary small $\epsilon>0$ and arbitrary large $E>0,$ there exist a number $\xi>0$ and an interval $J \subset \mathbb R$ with length $E$ such that $\left\|\psi(t)-\psi^*(t+\xi)\right\|< \epsilon,$ for all $t \in J.$
\end{definition}

\begin{definition}\label{dev3}
$\mathscr{B}$ is called transitive if it possesses a dense function.
\end{definition}

Now, let us recall the definition of transitivity for maps \cite{Dev90}. A continuous map $\varphi$ with an invariant domain $\Lambda \subset \mathbb R^k, k\in \mathbb N,$ possesses a dense orbit if there exists $c^* \in \Lambda$ such that for each $c \in \Lambda,$ arbitrary small real number $\epsilon >0$ and arbitrary large natural number $Q>0,$ there exist natural numbers $k_0$ and $l_0$ such that $\left\|\varphi^{n}(c)- \varphi^{n+k_0}(c^*)\right\| < \epsilon,$ for each integer $n$ between $l_0$ and $l_0+Q,$ and maps which have dense orbits are called transitive.

Suppose that $\mathscr{A}_x$ satisfies the transitivity property in the following sense. There exists $x^{*}(t)\in \mathscr{A}_x$ such that for each $x(t)\in \mathscr{A}_x,$ arbitrary small positive real number $\epsilon$ and arbitrary large natural number $Q,$ there exist natural numbers $k_0$ and $l_0$ such that $\left\|x(nT)-x^{*}((n+k_0)T)\right\| < \epsilon,$ for each integer $n$ between $l_0$ and $l_0+Q.$

Fix arbitrary $x(t)\in \mathscr{A}_x,$ $\epsilon >0$ and $Q \in \mathbb N.$ Under the circumstances, one can find $k_0,l_0 \in \mathbb N$ such that $\left\|x(nT)-x^{*}((n+k_0)T)\right\| < \epsilon e^{-L_0QT},$ for each integer $n=l_0,\ldots,l_0+Q.$

Using the condition $(A2)$ together with the convenient integral equations that $x(t)$ and $x^*(t)$ satisfy, it is easy to obtain for $t \in [l_0T,(l_0+Q)T]$ that

\begin{eqnarray*}
\left\|x(t)-x^{*}(t+k_0T)\right\| \leq \left\|x(l_0T)-x^{*}((l_0+k_0)T)\right\|+ \displaystyle\int^{t}_{l_0T} L_0 \left\|x(s)-x^{*}(s+k_0T)\right\| ds, \nonumber
\end{eqnarray*}
and by the help of the Gronwall-Bellman inequality \cite{Corduneanu08}, we get 
\begin{eqnarray*}
\left\|x(t)-x^{*}(t+k_0T)\right\| \leq \left\|x(l_0T)-x^{*}((l_0+k_0)T)\right\| e^{L_0(t-l_0T)}<\epsilon. \nonumber
\end{eqnarray*}
The last inequality shows that the set $\mathscr{A}_x$ satisfies the Definition \ref{dev2} with $\xi=k_0T$ and $E=QT,$ and is transitive according to Definition \ref{dev3}.

\begin{definition}\label{dev4}
$\mathscr{B}$ admits a dense countable collection $\mathscr{G} \subset \mathscr{B}$ of periodic functions if for every function $\psi(t) \in \mathscr{B},$ arbitrary small $\epsilon >0$ and arbitrary large $E>0,$ there exist $\widetilde{\psi}(t) \in \mathscr{G}$ and an interval $J\subset \mathbb R,$ with length $E,$ such that $\left\|\psi(t)-\widetilde{\psi}(t)\right\|<\epsilon,$ for all $t\in J.$ 
\end{definition}

Let us remind the definition of density of periodic orbits for maps \cite{Dev90}. The set of periodic orbits of a continuous map $\varphi$ with an invariant domain $\Lambda \subset \mathbb R^k, k\in \mathbb N,$ is called dense in $\Lambda$ if for each $c \in \Lambda,$ arbitrary small positive real number $\epsilon$ and  arbitrary large natural number $Q$ there exist a natural number $l_0$ and a point $\widetilde{c} \in \Lambda$ such that the sequence $\left\{\varphi^i(\widetilde{c})\right\}$ is periodic and $\left\|\varphi^n(c)-\varphi^n(\widetilde{c})\right\|< \epsilon,$ for each integer $n$ between $l_0$ and $l_0+Q.$

Let us denote by $\mathscr{G}_x$ the set of all periodic functions inside $\mathscr{A}_x.$ Suppose that $\mathscr{A}_x$ satisfies density of periodic solutions as follows. For each function $x(t) \in \mathscr{A}_x,$ arbitrary small $\epsilon >0$ and arbitrary large $Q\in \mathbb N,$ there exist $\widetilde{x}(t) \in \mathscr{G}_x$ and a number $l_0\in \mathbb N$  such that $\left\|x(nT)-\widetilde{x}(nT)\right\|<\epsilon,$ for each integer $n$ between $l_0$ and $l_0+Q.$ 

Let an arbitrary function $x(t)\in \mathscr{A}_x,$ arbitrary small $\epsilon >0$ and arbitrary large $Q \in \mathbb N$ be given. In that case, there exist $k_0,l_0 \in \mathbb N$ such that $\left\|x(nT)-\widetilde{x}(nT)\right\| < \epsilon e^{-L_0QT},$ for each integer $n=l_0,\ldots,l_0+Q.$

It can be easily verified that for $t \in [l_0T,(l_0+Q)T]$ the inequality
\begin{eqnarray*}
\left\|x(t)-\widetilde{x}(t)\right\| \leq \left\|x(l_0T)-\widetilde{x}(l_0T)\right\|+ \displaystyle\int^{t}_{l_0T} L_0 \left\|x(s)-\widetilde{x}(s)\right\| ds, \nonumber
\end{eqnarray*}
holds and therefore we have
\begin{eqnarray*}
\left\|x(t)-\widetilde{x}(t)\right\| \leq \left\|x(l_0T)-\widetilde{x}(l_0T)\right\| e^{L_0(t-l_0T)}<\epsilon. \nonumber
\end{eqnarray*}

Consequently, the set $\mathscr{A}_x$ satisfies Definition \ref{dev4} with $E=QT$ and $J=[l_0T,l_0T+E].$

\begin{definition}\label{dev5}
The collection of functions $\mathscr{B}$ is called a Devaney's chaotic set if
\begin{enumerate}
\item[\bf (D1)] $\mathscr{B}$ is sensitive;
\item[\bf (D2)] $\mathscr{B}$ is transitive; 
\item[\bf (D3)] $\mathscr{B}$ admits a dense countable collection of periodic functions. 
\end{enumerate} 
\end{definition}

In the next discussion, the chaotic properties on the set $\mathscr{B}$ will be imposed in Li-Yorke sense.

\subsection{\textbf{Chaotic set of functions in Li-Yorke sense}}\label{subsecliyorke}

The ingredients of Li-Yorke chaos for the collection of functions $\mathscr{B},$ which is defined by $(\ref{collection}),$ will be described in this part of the paper. Making use of discussions similar to the ones made in the previous subsection, we extend, below, the definitions for the ingredients of Li-Yorke chaos from maps \cite{Li75}-\cite{Akin03} to continuous flows and we just omit these indications here.

\begin{definition}\label{defliyorke1}
A couple of functions $ \left( \psi(t), \overline{\psi}(t) \right) \in \mathscr{B} \times \mathscr{B}$ is called proximal if for arbitrary $\epsilon>0, E>0$ there exist infinitely many disjoint intervals of length not less than $E$ such that $\left\|\psi(t)-\overline{\psi}(t)\right\| < \epsilon,$ for each $t$ from these intervals.
\end{definition}


\begin{definition}\label{defliyorke2}
A couple of functions $\left( \psi(t), \overline{\psi}(t) \right) \in \mathscr{B} \times \mathscr{B}$ is frequently $(\epsilon_0, \Delta)-$separated if there exist positive numbers $\epsilon_0, \Delta$ and infinitely many disjoint intervals of length not less than $\Delta$, such that $\left\|\psi(t)-\overline{\psi}(t)\right\| > \epsilon_0$ for each $t$ from these intervals.
\end{definition}

\begin{remark}
The numbers $\epsilon_0$ and $\Delta$ taken into account in Definition $\ref{defliyorke2}$ depend on the functions $\psi(t)$ and $\overline{\psi}(t).$
\end{remark}

%

\begin{definition}\label{defliyorke3}
A couple of functions $\left( \psi(t), \overline{\psi}(t) \right) \in \mathscr{B} \times \mathscr{B}$ is a Li$-$Yorke pair if they are proximal and frequently $(\epsilon_0, \Delta)-$separated for some positive numbers $\epsilon_0$ and $\Delta$. 
\end{definition}

\begin{definition}\label{defliyorke4}
An uncountable set $\mathscr{C} \subset \mathscr{B}$ is called a scrambled set if $\mathscr{C}$ does not contain any periodic functions and each couple of different functions inside $\mathscr{C} \times \mathscr{C}$ is a Li$-$Yorke pair.
\end{definition}

\begin{definition}\label{defliyorke5}
$\mathscr{B}$ is called a Li$-$Yorke chaotic set if 
\begin{description}
\item[(LY1)] There exists a positive real number $T_0$ such that  $\mathscr{B}$ admits a periodic function of period $kT_0,$ for any $k\in \mathbb N;$
\item[(LY2)] $\mathscr{B}$ possesses a scrambled set $\mathscr{C};$
\item[(LY3)] For any function $\psi(t)\in \mathscr{C}$ and any periodic function $\overline{\psi}(t)\in \mathscr{B},$ the couple $\left(\psi(t),\overline{\psi}(t)\right)$ is frequently $(\epsilon_0, \Delta)-$separated for some positive real numbers $\epsilon_0$ and $\Delta.$
\end{description}
\end{definition}


\section{Hyperbolic set of functions} \label{hyperbolic_sets}

The definitions of stable and unstable sets of hyperbolic periodic orbits of autonomous systems are given in \cite{Palmer2000}, and information about such sets of solutions of perturbed non-autonomous systems can be found in \cite{Lerman92}. Moreover, homoclinic structures in almost periodic systems were studied in \cite{Scheurle}-\cite{Palmer89}. In this section, we give definition for an hyperbolic collection of uniformly bounded functions and before this, we start with definitions of stable and unstable sets of a function.

We define the stable set of a function $\psi(t) \in \mathscr{B},$ where the collection $\mathscr{B}$ is defined by $(\ref{collection}),$ as the set of functions
\begin{eqnarray}
W^{s} \left( \psi(t) \right) = \left\{u(t) \in \mathscr{B} ~|~  \left\|u(t)-\psi(t)\right\| \to 0 ~\textrm{as}~ t \to \infty  \right\}, 
\end{eqnarray}
and, similarly, we define the unstable set of a function $\psi(t) \in \mathscr{B}$ as the set of functions
\begin{eqnarray}
W^{u} \left( \psi(t) \right) = \left\{v(t) \in \mathscr{B} ~|~  \left\|v(t)-\psi(t)\right\| \to 0 ~\textrm{as}~ t \to -\infty  \right\}. 
\end{eqnarray}

\begin{definition}
The set of functions $\mathscr{B}$ is called hyperbolic if each function inside this set has nonempty stable and unstable sets.
\end{definition}

\begin{theorem}\label{hyperbolicity}
If $\mathscr{A}_x $ is hyperbolic, then the same is true for $\mathscr{A}_y.$
\end{theorem}

\noindent \textbf{Proof.}
Fix arbitrary $\epsilon >0,$ $y(t)=\phi_{x(t)}(t)  \in \mathscr{A}_y,$ and let $\alpha=\frac{\omega -NL_3}{\omega - NL_3 +NL_2},$ $\beta=\frac{\omega-NL_3}{1+NL_2}.$ By condition $(A7),$ one can verify that the numbers $\alpha$ and $\beta$ are both positive. 

Due to hyperbolicity of $\mathscr{A}_x,$ the function $x(t)$ has a nonempty stable set $W^s(x(t))$ and a nonempty unstable set $W^u(x(t)).$

Take a function $u(t) \in W^s(x(t)).$ Since $\left\|x(t)-u(t)\right\| \to 0$ as $t \to \infty$ and $NL_3-\omega<0,$ there exists a positive number $R_1,$ which depends on $\epsilon,$ such that $\left\|x(t)-u(t)\right\| < \alpha \epsilon$ and $e^{(NL_3-\omega)t}<\frac{\omega \alpha \epsilon}{2M_0N},$ for $t \geq R_1.$

Let us denote $\overline{y}(t)=\phi_{u(t)}(t)$ and we shall prove that the function $\overline{y}(t)$ belongs to the stable set of $y(t).$

The bounded on $\mathbb R$ functions $y(t)$ and $\overline{y}(t)$ satisfy the relations
\begin{eqnarray*}
y(t) =  \displaystyle\int^{t}_{-\infty} e^{A(t-s)} g(x(s),y(s)) ds, \nonumber
\end{eqnarray*}
and
\begin{eqnarray*}
\overline{y}(t) =  \displaystyle\int^{t}_{-\infty} e^{A(t-s)} g(u(s),\overline{y}(s)) ds, \nonumber
\end{eqnarray*}
respectively, for $t\geq R_1.$

Therefore one can easily reach up the equation 
\begin{eqnarray*}
&& y(t)-\overline{y}(t) =  \displaystyle\int^{R_1}_{-\infty} e^{A(t-s)} [g(x(s),y(s))-g(u(s),\overline{y}(s))] ds \\
&& + \displaystyle\int^{t}_{R_1} e^{A(t-s)} \left\{[g(x(s),y(s))-g(x(s),\overline{y}(s))]+[g(x(s),\overline{y}(s))-g(u(s),\overline{y}(s))]\right\} ds,
\end{eqnarray*}
which implies that
\begin{eqnarray*}
&& \left\|y(t)-\overline{y}(t)\right\| \leq  \displaystyle\int^{R_1}_{-\infty} 2M_0N e^{-\omega(t-s)} ds \\
&& + \displaystyle\int^{t}_{R_1}   e^{-\omega(t-s)} \left(  NL_3\left\|y(s)-\overline{y}(s)\right\| +  NL_2\left\|x(s)-u(s)\right\| \right) ds \\
&& \leq \frac{2M_0N}{\omega} e^{-\omega(t-R_1)}  + \displaystyle\int^{t}_{R_1}  e^{-\omega(t-s)}  \left( NL_3 \left\|y(s)-\overline{y}(s)\right\| +NL_2\alpha \epsilon  \right) ds.
\end{eqnarray*}

Using the Gronwall type inequality indicated in \cite{Zadiraka68}, we attain the inequality

\begin{eqnarray*}
&& \left\|y(t)-\overline{y}(t)\right\| \leq \frac{2M_0N}{\omega} e^{(NL_3-\omega)(t-R_1)}  + \frac{NL_2\alpha\epsilon}{\omega-NL_3} [1-e^{(NL_3-\omega)(t-R_1)} ], ~~t \geq R_1.
\end{eqnarray*}

For that reason, for $t \geq 2R_1,$  one has

\begin{eqnarray*}
&& \left\|y(t)-\overline{y}(t)\right\| \leq \frac{2M_0N}{\omega} e^{(NL_3-\omega)R_1}  + \frac{NL_2\alpha\epsilon}{\omega-NL_3} \\
&& < \left( 1+\frac{NL_2}{\omega-NL_3} \right) \alpha \epsilon   \\
&& = \epsilon.
\end{eqnarray*}

The last inequality reveals that $\left\|y(t)-\overline{y}(t)\right\| \to 0$ as $t \to \infty,$ and hence $\overline{y}(t)$ belongs to the stable set $W^s(y(t))$ of $y(t).$

On the other hand, let $v(t)$ be a function inside the unstable set $W^u(x(t)).$

Since $\left\|x(t)-v(t)\right\| \to 0$ as $t \to -\infty,$ there exists a negative real number $R_2(\epsilon)$ such that $\left\|x(t)-v(t)\right\|<\beta \epsilon,$ for $t\leq R_2.$ Let $\widetilde{y}(t)=\phi_{v(t)}(t).$ Now, our purpose is to show that $\widetilde{y}(t)$ belongs to the unstable set of $y(t).$

By the help of the integral equations
\begin{eqnarray*}
y(t) =  \displaystyle\int^{t}_{-\infty} e^{A(t-s)} g(x(s),y(s)) ds, \nonumber
\end{eqnarray*}
and
\begin{eqnarray*}
\widetilde{y}(t) =  \displaystyle\int^{t}_{-\infty} e^{A(t-s)} g(v(s),\widetilde{y}(s)) ds, \nonumber
\end{eqnarray*}
we obtain that
\begin{eqnarray*}
&& y(t)-\widetilde{y}(t) =  \displaystyle\int^{t}_{-\infty} e^{A(t-s)} [g(x(s),y(s))-g(v(s),\widetilde{y}(s))] ds \\
&& = \displaystyle\int^{t}_{-\infty} e^{A(t-s)} [g(x(s),y(s))-g(v(s),y(s))] ds \\
&& + \displaystyle\int^{t}_{-\infty} e^{A(t-s)}  [g(v(s),y(s))-g(v(s),\widetilde{y}(s))] ds.
\end{eqnarray*}

Therefore, for $t\leq R_2,$ one has
\begin{eqnarray*}
&& \left\|y(t)-\widetilde{y}(t)\right\| \leq  \displaystyle\int^{t}_{-\infty} NL_2 e^{-\omega(t-s)} \left\|x(t)-v(t)\right\|  ds \\
&& + \displaystyle\int^{t}_{-\infty}   e^{-\omega(t-s)}  NL_3   \left\|y(s)-\widetilde{y}(s)\right\|   ds \\
&& \leq \frac{NL_2\beta\epsilon}{\omega} +\frac{NL_3}{\omega}  \sup_{t\leq R_2} \left\|y(t)-\widetilde{y}(t)\right\|.
\end{eqnarray*}
Hence,
\begin{eqnarray*}
\sup_{t\leq R_2} \left\|y(t)-\widetilde{y}(t)\right\| \leq \frac{NL_2\beta\epsilon}{\omega} +\frac{NL_3}{\omega} \sup_{t\leq R_2} \left\|y(t)-\widetilde{y}(t)\right\|
\end{eqnarray*}
and
\begin{eqnarray*}
\sup_{t\leq R_2} \left\|y(t)-\widetilde{y}(t)\right\| \leq \frac{NL_2\beta\epsilon}{\omega-NL_3}<\epsilon. 
\end{eqnarray*}
The last inequality confirms that $\left\|y(t)-\widetilde{y}(t)\right\| \to 0$ as $t \to -\infty.$ Therefore $\widetilde{y}(t) \in W^u(y(t)).$ 

Consequently, $\mathscr{A}_y$ is hyperbolic since $y(t)$ possesses both nonempty stable and unstable sets, denoted by $W^s(y(t))$ and $W^u(y(t)),$ respectively. 

The theorem is proved. $\square$

Theorem $\ref{hyperbolicity}$ implies the following corollary.

\begin{corollary}
 If $\mathscr{A}_x$ is hyperbolic, then the same is true for $\mathscr{A}.$ 
\end{corollary}

Next, we continue with another corollary of Theorem $(\ref{hyperbolicity}),$ following the definitions of homoclinic and heteroclinic functions.

A function $\psi_1(t) \in \mathscr{B}$ is said to be homoclinic to the function $\psi_0(t) \in \mathscr{B},$ $\psi_0(t) \neq \psi_1(t),$ if 
$\psi_1(t) \in  W^{s} \left( \psi_0(t) \right) \cap W^{u} \left( \psi_0(t) \right).$ 

On the other hand, a function $\psi_2(t) \in \mathscr{B}$ is called heteroclinic to the functions $\psi_0(t), \psi_1(t) \in \mathscr{B},$ $\psi_0(t) \neq \psi_2(t),$  $\psi_1(t) \neq \psi_2(t),$ if 
$\psi_2(t) \in  W^{s} \left( \psi_0(t) \right) \cap W^{u} \left( \psi_1(t) \right).$

\begin{corollary}
If $x_1(t) \in \mathscr{A}_x$ is homoclinic to the function $x_0(t) \in \mathscr{A}_x,$ $x_0(t) \neq x(t),$ then $\phi_{x_1(t)}(t) \in \mathscr{A}_y$ is homoclinic to the function $\phi_{x_0(t)}(t) \in \mathscr{A}_y.$

Similarly, if $x_2(t) \in \mathscr{A}_x$ is heteroclinic to the functions $x_0(t), x_1(t) \in \mathscr{A}_x,$ $x_0(t) \neq x_2(t),$  $x_1(t) \neq x_2(t),$ then $\phi_{x_2(t)}(t)$ is heteroclinic to the functions $\phi_{x_0(t)}(t),$ $\phi_{x_1(t)}(t)\in \mathscr{A}_y.$
\end{corollary}


\section{Self-replication of chaos} \label{self-replication}
In this section, which is the main part of our paper, we theoretically prove that the set $\mathscr{A}_y$ replicates the ingredients of chaos furnished to the set $\mathscr{A}_x$ and as a consequence the same is also valid for the set $\mathscr{A}.$ We start our discussions with chaos through period-doubling cascade, and continue with Devaney's and Li-Yorke's chaotic ingredients, respectively. 

The mechanism of morphogenesis was present before through the usage of the logistic map as an ``embryo" to generate the chaotic structure embedded in a system with an arbitrary finite dimension. Devaney's chaos \cite{Akh5}, Li-Yorke chaos \cite{Akh6}, chaos through period-doubling cascade \cite{Akh7} and intermittency \cite{Akh5} were provided in a quasilinear system of the type
\begin{eqnarray} 
\begin{array}{l}\label{previous}
z'(t)= Az(t) + f(t, z) + \nu(t,t_{0},\mu)  \\
z(t_{0}) = z_{0} , (t_{0},z_{0})\in [0,1] \times \mathbb R^{n},
\end{array}
\end{eqnarray}
where

\begin{eqnarray} \label{relay_function}
\mathcal \nu(t,t_0,\mu)=\left\{\begin{array}{ll} m_{0}, ~\textrm{if}  & \zeta_{2i}(t_0,\mu) < t  \leq \zeta_{2i+1}(t_0,\mu) \\
                                               m_{1}, ~\textrm{if}  & \zeta_{2i-1}(t_0,\mu) < t \leq \zeta_{2i}(t_0,\mu) , 
\end{array} \right.
\end{eqnarray} 
$i$ is an integer, $m_{0},m_{1} \in \mathbb R^{n},$ the sequence $\zeta(t_{0},\mu)=\{\zeta_{i}(t_{0},\mu)\}$ is defined through equations $\zeta_{i}(t_{0},\mu) =i+\kappa_{i}(t_{0},\mu),$  with   $\kappa_{i+1}(t_{0},\mu)=h(\kappa_{i}(t_{0},\mu),\mu),$ $\kappa_{0}(t_{0},\mu)=t_{0},$ and $h(x,\mu)=\mu x(1-x)$ is the logistic map. In the present paper, we do not consider an ``embryo" to generate the chaotic structure embedded in a system, but we make use of a chaotic system to replicate the same type of chaos.

\subsection{\textbf{Period-doubling cascade and morphogenesis}} \label{subsecperioddoubling}
We start this subsection by describing the chaos through period-doubling cascade for the set of functions $\mathscr{A}_x$ and deal with its replication by the set of functions $\mathscr{A}_y,$ which is defined by equation $(\ref{A_y}).$

Suppose that there exists a function $G: \mathbb R \times \mathbb R^m \times \mathbb R \to \mathbb R^m$  which is continuous in all of its arguments such that $G(t+T,x,\mu)=G(t,x,\mu),$ for all $t\in\mathbb R, x \in \mathbb R^m,\mu \in \mathbb R,$ and $F(t,x)=G(t,x,\mu_{\infty})$ for some finite value $\mu_{\infty}$ of the parameter $\mu,$ which will be explained below. 

To discuss chaos through period-doubling cascade, let us consider the system
\begin{eqnarray}
\begin{array}{l}
x'=G(t,x,\mu).  \label{perioddoubling1}
\end{array}
\end{eqnarray}

We say that the set $\mathscr{A}_x$ is chaotic through period-doubling cascade if there exists a natural number $k$ and a sequence of period-doubling bifurcation values $\left\{\mu_m\right\},$  $\mu_m \to \mu_{\infty}$ as $m \to \infty,$ such that for each $m\in \mathbb N$ as the parameter $\mu$ increases (or decreases) through $\mu_m,$ system $(\ref{perioddoubling1})$ undergoes a period-doubling bifurcation and a periodic solution with period $k2^mT$ appears. As a consequence, at $\mu=\mu_{\infty},$ there exist infinitely many periodic solutions of system $(\ref{perioddoubling1}),$ and hence of system $(\ref{1}),$ all lying in a bounded region.  In this case, the set $\mathscr{A}_x$ admits periodic functions of periods $kT, 2kT, 4kT, 8kT, \cdots.$

Now, making use of the equation (\ref{integral_eqns}), one can show that for any natural number $p,$ if $x(t) \in \mathscr{A}_x$ is a $pT-$periodic function then $\phi_{x(t)}(t) \in \mathscr{A}_y$ is also $pT-$periodic. Moreover, condition $(A4)$ implies that the converse is also true. Consequently, if the set $\mathscr{A}_x$ admits periodic functions of periods $kT, 2kT, 4kT, 8kT, \cdots,$ then the same is valid for $\mathscr{A}_y,$ with no additional periodic functions of any other period, and this provides us an opportunity to state the following theorem.

\begin{theorem}\label{morphogenesis_period-doubling_theorem}
If the set $\mathscr{A}_x$ is chaotic through period-doubling cascade, then the same is true for $\mathscr{A}_y.$
\end{theorem}

The following corollary states that the result-system $(\ref{1})+(\ref{2})$ is chaotic through the period-doubling cascade, provided the system $(\ref{1})$ is.

\begin{corollary}
If the set $\mathscr{A}_x$ is chaotic through period-doubling cascade, then the same is true for $\mathscr{A}.$
\end{corollary}

Our theoretical results show that the existing replicator systems, likewise the generator counterpart, undergoes period-doubling bifurcations as the parameter $\mu$ increases or decreases through the values $\mu_m, m\in \mathbb N.$ That is, the sequence $\left\{\mu_m\right\}$ of bifurcation parameters is exactly the same for both generator and replicator systems. Since the generator system necessarily obeys the Feigenbaum universality \cite{Th02,Feigenbaum80,Sch05,Zelinka}, one can conclude that the same is also true for the replicator. In other words, when $\lim_{m \to \infty } \frac{\mu_m-\mu_{m+1}}{\mu_{m+1}-\mu_{m+2}}$ is evaluated, the universal constant known as the Feigenbaum number $4.6692016\ldots$ is achieved  and this universal number is the same for both generator and replicators, and hence for the result-system. One can percieve this result as another aspect of  morphogenesis of chaos.

Next, let us illustrate by simulations, the morphogenesis of chaos through period-doubling cascade. In paper \cite{Sato83}, it is indicated that the Duffing's equation
\begin{eqnarray} 
\begin{array}{l}
x''+0.3x'+x^3=40 \cos t \label{per_doub_ex_1}
\end{array}
\end{eqnarray}
admits the chaos through period-doubling cascade. Defining the new variables $x_1=x$ and $x_2=x',$ equation $(\ref{per_doub_ex_1})$ can be rewritten as a system in the following form
\begin{eqnarray} 
\begin{array}{l}
x'_1=x_2 \\ \label{per_doub_ex_2}
x'_2=-0.3x_2-x_1^3+40 \cos t. \\ 
\end{array}
\end{eqnarray}
Making use of system $(\ref{per_doub_ex_2})$ as the generator, one can constitute the $8-$dimensional result-system
\begin{eqnarray} 
\begin{array}{l}
x'_1=x_2 \\
x'_2=-0.3x_2-x_1^3+40 \cos t \\ \label{per_doub_ex_3}
x'_3=2x_3-x_4+0.4\tan\left(  \frac{x_1+x_3}{10}  \right) \\
x'_4=17x_3-6x_4+x_2 \\
x'_5=-2x_5+0.5\sin x_6 -4x_4 \\
x'_6=-x_5-4x_6-\tan\left(  \frac{x_3}{2}  \right) \\
x'_7=2x_7+5x_8-0.0003(x_7-x_8)^3-1.5x_6 \\
x'_8=-5x_7-8x_8+4x_5,
\end{array}
\end{eqnarray}
which involves three consecutive $2-$dimensional replicator systems in addition to the generator, as the chain procedure indicated in Figure \ref{box1}.

We note that system $(\ref{per_doub_ex_3})$ exhibits a symmetry under the transformation
\begin{eqnarray}\label{transformation}
\mathscr{K}:(x_1,x_2,x_3,x_4,x_5,x_6,x_7,x_8,t) \to (-x_1,-x_2,-x_3,-x_4,-x_5,-x_6,-x_7,-x_8,t+\pi).
\end{eqnarray}

\begin{figure}[ht] 
\centering 
\includegraphics[width=14.0cm]{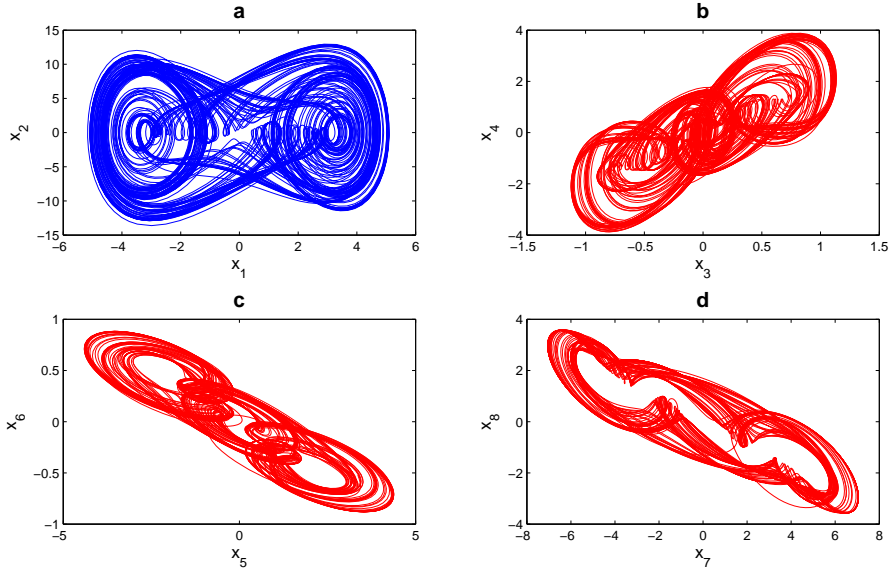} 
\caption{\footnotesize{$2-$dimensional projections of the chaotic attractor of the result-system $(\ref{per_doub_ex_3}).$ (a) Projection on the $x_1-x_2$ plane, (b) Projection on the $x_3-x_4$ plane, (c) Projection on the $x_5-x_6$ plane, (d) Projection on the $x_7-x_8$ plane. The picture in $(a)$ shows the attractor of the prior chaos produced by the generator system $(\ref{per_doub_ex_2})$ and in $(b)-(d)$ the chaotic attractors of the replicator systems are observable. The illustrations in $(b)-(d)$ repeated the structure of the attractor shown in $(a),$ and the similarity between these pictures is an indication of the morphogenesis of chaos. }} 
\label{per_doub_cas_fig1}
\end{figure}

According to our theoretical discussions, the result-system $(\ref{per_doub_ex_3})$ admits a chaotic attractor in the $8-$dimensional phase space which is obtained through period-doubling cascade. In Figure $\ref{per_doub_cas_fig1},(a)-(d),$ we illustrate the $2-$dimensional projections of the trajectory of system $(\ref{per_doub_ex_3}),$ with the initial data $x_1(0)=2.16,  x_2(0)=-9.28, x_3(0)=-0.21, x_4(0)=-2.03, x_5(0)=3.36, x_6(0)=-0.52, x_7(0)=3.07, x_8(0)=-0.32,$ on the planes $x_1-x_2,$ $x_3-x_4,$ $x_5-x_6,$ and $x_7-x_8,$ respectively. The picture in Figure $\ref{per_doub_cas_fig1}, (a),$ shows in fact the attractor of the prior chaos given by the system $(\ref{per_doub_ex_2})$ and similarly the illustrations in Figure $\ref{per_doub_cas_fig1}, (b)-(d)$ correspond to the chaotic attractors of the first, second and the third replicator systems, respectively. The resemblance between the shapes of the attractors of generator and replicator systems reflect the morphogenesis of chaos in the result-system $(\ref{per_doub_ex_3}).$

\begin{figure}[ht] 
\centering
\includegraphics[width=14.0cm]{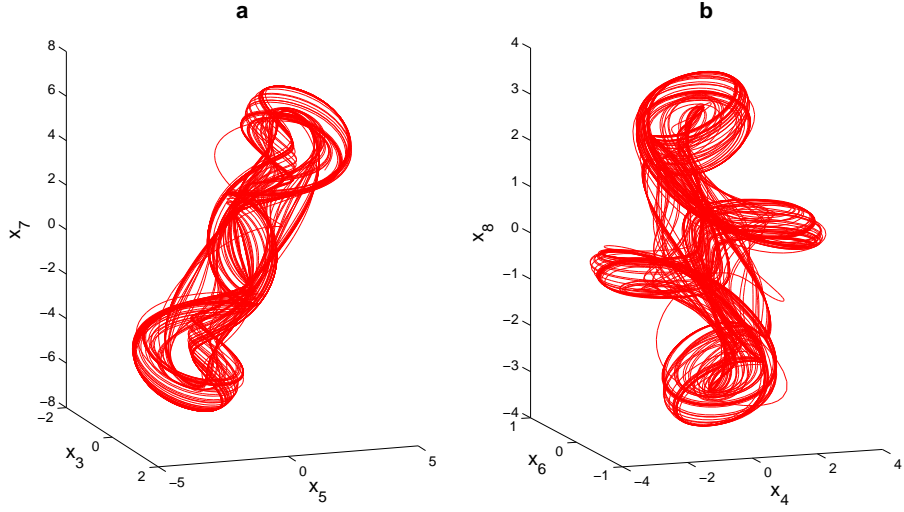} 
\caption{\footnotesize{$3-$dimensional projections of the chaotic attractor of the result-system $(\ref{per_doub_ex_3}).$ (a) Projection on the $x_3-x_5-x_7$ space, (b) Projection on the $x_4-x_6-x_8$ space. The illustrations presented in $(a)$ and $(b)$ give information about the impressive chaotic attractor in the $8-$dimensional space.}}
\label{per_doub_cas_fig2}
\end{figure}

To obtain better impression about the chaotic attractor of system $(\ref{per_doub_ex_3}),$ in Figure $\ref{per_doub_cas_fig2}$ we demonstrate the $3-$dimensional projections of the trajectory with the same initial data as above, on the $x_3-x_5-x_7$ and $x_4-x_6-x_8$ spaces.  Although we are restricted to make illustrations at most in $3-$dimensional spaces and not able to provide a picture of the whole chaotic attractor, the results shown both in Figure \ref{per_doub_cas_fig1} and Figure \ref{per_doub_cas_fig2} give us an idea about the spectacular chaotic attractor of system $(\ref{per_doub_ex_3}).$ We note that the presented attractors are symmetric around the origin due to the symmetry of the result-system $(\ref{per_doub_ex_3})$ under the transformation $\mathscr{K}$ introduced by $(\ref{transformation}).$

Now, let us show that the first replicator system which is included inside $(\ref{per_doub_ex_3})$ satisfies the condition $(A7).$ 

Let us denote by  $\left\|.\right\|$  the matrix norm which is induced by the usual Euclidean norm in $\mathbb R^l$. That is, 
\begin{eqnarray}
\left\| \Gamma \right\|= \max \left\{\sqrt{\varsigma} : \varsigma ~is ~an ~eigenvalue ~of ~\Gamma^{T}\Gamma \right\} \label{matrix_norm}
\end{eqnarray}
for any $l \times l$ matrix $\Gamma$ with real entries, and $\Gamma^T$ denotes the transpose of the matrix $\Gamma$ \cite{Horn92}.

When the system 
\begin{eqnarray} 
\begin{array}{l}
x'_3=2x_3-x_4+0.4\tan\left(  \frac{x_1+x_3}{10}  \right) \\ \label{per_doub_ex_4}
x'_4=17x_3-6x_4+x_2
\end{array}
\end{eqnarray}
is considered in the form of system $(\ref{2}),$ one can see that the matrix $A$ can be written in the form
$A=\left[
\begin {array}{ccc}
2&-1\\
\noalign{\medskip}
17&-6
\end {array}
\right],$
which admits the complex conjugate eigenvalues $-2 \mp i.$

The real Jordan form of the matrix $A$ is given by 
$J=\left[
\begin {array}{ccc}
-2 & -1 \\
\noalign{\medskip}
1 & -2
\end {array}
\right]$
and the identity $P^{-1}AP=J$ is satisfied where 
$P=\left[
\begin {array}{ccc}
0 & 1 \\
\noalign{\medskip}
-1 & 4
\end {array}
\right].$
Evaluating the exponential matrix $e^{At}$ we obtain that
\begin{eqnarray} 
e^{At}=e^{-2 t}P \left[
\begin {array}{ccc}  
cos(t) & -sin(t) \label{condition_evaluation_1}\\
\noalign{\medskip}
\sin(t) & cos(t)
\end {array}
\right]P^{-1}.
\end{eqnarray}

Taking $N=\left\|P\right\| \left\|P^{-1}\right\|<18$ and $\omega=2,$ one can see that the inequality $\left\|e^{At}\right\|\leq Ne^{-\omega t}$ holds for all $t \geq 0.$ The function $g:\mathbb R^2 \times \mathbb R^2 \to \mathbb R^2$ defined as 
\begin{eqnarray*}
g(x_1,x_2,x_3,x_4)=  \left( 0.4 \tan\left( \frac{x_1+x_3}{10} \right), x_2   \right)
\end{eqnarray*}
satisfies the conditions $(A4)$ and $(A5)$ with constants $L_1=\frac{\sqrt{2}}{50}, L_2=\sqrt{2}$ and $L_3=0.08$ since the chaotic attractor of system $(\ref{per_doub_ex_3})$ satisfies the inequalities $\left|x_1\right|\leq 6, \left|x_3\right| \leq \frac{3}{2},$ and consequently $\left| \frac{x_1+x_3}{10}\right|\leq \frac{3}{4}.$ Therefore, the condition $(A7)$ is satisfied. 

In a similar way, for the second replicator system, making use of $\left|x_3\right| \leq \frac{3}{2}$ once again, one can show that the function $h:\mathbb R^2 \times \mathbb R^2 \to \mathbb R^2$ defined as
\begin{eqnarray*}
h(x_3,x_4,x_5,x_6)=  \left( 0.5 \sin x_6 -4x_4, -\tan\left( \frac{x_3}{2}   \right)   \right)
\end{eqnarray*}
satisfies the counterparts of the conditions $(A4)$ and $(A5)$ with constants $L_1=\frac{1}{2\sqrt{2}}, L_2=4\sqrt{2}$ and $L_3=\frac{1}{2}.$

On the other hand, when one considers the third replicator, it is easy to show that the function $k:\mathbb R^2 \times \mathbb R^2 \to \mathbb R^2$ defined as 
\begin{eqnarray*}
k(x_5,x_6,x_7,x_8)=  \left( 0.0003(x_7-x_8)^3-1.5x_6, 4x_5   \right)
\end{eqnarray*}
satisfies the counterparts of the conditions $(A4)$ and $(A5)$ with constants $L_1=\frac{3\sqrt{2}}{4}, L_2=4\sqrt{2}$ and $L_3=0.19$ since the chaotic attractor of system $(\ref{per_doub_ex_3})$ satisfies the inequalities $\left|x_7\right|\leq 8, \left|x_8\right| \leq 4.$


\subsection{\textbf{Machinery of morphogenesis of Devaney's chaos}}\label{devaney}

In this part of our paper, we will prove theoretically that the ingredients of Devaney's chaos furnished to the set $\mathscr{A}_x$ are all replicated by the set $\mathscr{A}_y.$

\begin{lemma}\label{sensitivity}
Sensitivity of the set $\mathscr{A}_x$ implies the same feature for the set $\mathscr{A}_y.$ 
\end{lemma}

\noindent \textbf{Proof.}
Fix an arbitrary $\delta>0$ and let $y(t) \in \mathscr{A}_y$ be a given solution of system (\ref{2}). In this case, there exists $x(t)\in \mathscr{A}_x$ such that $y(t)=\phi_{x(t)}(t)$.

Let us choose an $\overline{\epsilon}=\overline{\epsilon}(\delta)>0$ small enough which satisfies the inequality $\left(1+\frac{NL_2}{\omega-NL_3}\right)\overline{\epsilon}<\delta.$ Then take $R=R(\overline{\epsilon})<0$ sufficiently large in absolute value such that $\frac{2M_0N}{\omega}e^{(\omega - NL_3)R}<\overline{\epsilon},$ and let $\delta_1=\delta_1(\overline{\epsilon},R)=\overline{\epsilon}e^{L_0R}.$
Since the set of functions $\mathscr{A}_x$ is sensitive, there exist positive real numbers $\epsilon_0$ and $\Delta$ such that the inequalities $\left\|x(t_0)-\overline{x}(t_0)\right\|<\delta_1$ and $\left\|x(t)-\overline{x}(t)\right\| \geq \epsilon_0,$ $t \in J,$ hold for some $\overline{x}(t)\in \mathscr{A}_x,$ $t_0 \in \mathbb R$ and for some interval $J \subset [t_0,\infty)$ whose length is not less than $\Delta.$

Using the couple of integral equations
\begin{eqnarray*}
x(t)=x(t_0)+\displaystyle \int_{t_0}^t F(s,x(s))ds,
\end{eqnarray*}
\begin{eqnarray*}
\overline{x}(t)=\overline{x}(t_0)+\displaystyle \int_{t_0}^t F(s,\overline{x}(s))ds
\end{eqnarray*}
together with condition $(A2),$ one see that the inequality
\begin{eqnarray*}
\left\|x(t)-\overline{x}(t)\right\| \leq \left\|x(t_0)-\overline{x}(t_0)\right\|+\left|\displaystyle \int_{t_0}^t L_0 \left\|x(s)-\overline{x}(s)\right\| ds \right|
\end{eqnarray*}
holds for $t \in [t_0+R, t_0].$
Applying Gronwall-Bellman inequality \cite{Corduneanu08}, we obtain
\begin{eqnarray*}
\left\|x(t)-\overline{x}(t)\right\| \leq \left\|x(t_0)-\overline{x}(t_0)\right\|e^{L_0\left|t-t_0\right|}
\end{eqnarray*} 
and therefore $\left\|x(t)-\overline{x}(t)\right\|<\overline{\epsilon},$ for $t \in [t_0+R, t_0].$ 

Suppose that $\overline{y}(t)=\phi_{\overline{x}(t)}(t) \in \mathscr{A}_y.$ First, we will show that $\left\|y(t_0)-\overline{y}(t_0)\right\|<\delta.$ 

The functions $y(t)$ and $\overline{y}(t)$ satisfy the integral equations 
\begin{eqnarray*} 
y(t) = \displaystyle\int^{t}_{-\infty} e^{A(t-s)}g(x(s),y(s))ds
\end{eqnarray*}
and 
\begin{eqnarray*} 
\overline{y}(t) = \displaystyle\int^{t}_{-\infty} e^{A(t-s)}g(\overline{x}(s),\overline{y}(s))ds,
\end{eqnarray*}
respectively. Therefore,

\begin{eqnarray*}
y(t)-\overline{y}(t)=\displaystyle\int^{t}_{-\infty} e^{A(t-s)} [g(x(s),y(s))-g(\overline{x}(s),\overline{y}(s))]ds
\end{eqnarray*}
and hence we obtain
\begin{eqnarray*}
&&\left\|y(t)-\overline{y}(t)\right\| \leq \displaystyle\int^{t}_{t_0+R} Ne^{-\omega(t-s)} \left\|g(x(s),y(s))-g(x(s),\overline{y}(s))\right\|ds \\
&& + \displaystyle\int^{t}_{t_0+R} Ne^{-\omega(t-s)} \left\|g(x(s),\overline{y}(s))-g(\overline{x}(s),\overline{y}(s))\right\|ds \\
&& + \displaystyle\int^{t_0+R}_{-\infty} Ne^{-\omega(t-s)} \left\|g(x(s),y(s))-g(\overline{x}(s),\overline{y}(s))\right\|ds. 
\end{eqnarray*}
Since $\left\|x(t)-\overline{x}(t)\right\|<\overline{\epsilon}$ for $t \in [t_0+R, t_0],$ one has
\begin{eqnarray*}
&& \left\|y(t)-\overline{y}(t)\right\| \leq NL_3\displaystyle\int^{t}_{t_0+R} e^{-\omega(t-s)} \left\|y(s)-\overline{y}(s)\right\|ds \\
&& + NL_2 \overline{\epsilon} \displaystyle\int^{t}_{t_0+R} e^{-\omega(t-s)}ds + 2M_0N\displaystyle\int^{t_0+R}_{-\infty}e^{-\omega(t-s)}ds \\
&& \leq NL_3\displaystyle\int^{t}_{t_0+R} e^{-\omega(t-s)} \left\|y(s)-\overline{y}(s)\right\|ds \\
&& + \frac{NL_2\overline{\epsilon}}{\omega}e^{-\omega t} (e^{\omega t}-e^{\omega (t_0+R)})+\frac{2M_0N}{\omega}e^{-\omega(t-t_0-R)}.
\end{eqnarray*}

Now, introduce the functions $u(t)=e^{\omega t}\left\|y(t)-\overline{y}(t)\right\|,$ $k(t)=\frac{NL_2\overline{\epsilon}}{\omega}e^{\omega t},$  and  $h(t)=c+ k(t)$ where $c=\left(\frac{2M_0N}{\omega}-\frac{NL_2\overline{\epsilon}}{\omega}\right)e^{\omega (t_0+R)}.$ 

These definitions give us the inequality 
\begin{eqnarray*}
u(t)\leq h(t)+\displaystyle\int^{t}_{t_0+R} N L_3 u(s)ds.
\end{eqnarray*}
Applying Lemma $2.2$ \cite{Bar70} to the last inequality, we get $u(t)\leq h(t)+NL_3\displaystyle\int^{t}_{t_0+R} e^{NL_3(t-s)}h(s)ds.$

Therefore, for $t\in [t_0+R, t_0]$ we have
\begin{eqnarray*}
&&u(t) \leq c+k(t)+c\left(e^{NL_3(t-t_0-R)}-1\right) + \frac{N^{2}L_2L_3\overline{\epsilon}}{\omega}e^{NL_3t} \displaystyle\int^{t}_{t_0+R} e^{(\omega -NL_3)s}ds \\
&& = k(t) +ce^{NL_3(t-t_0-R)} +\frac{N^{2}L_2L_3\overline{\epsilon}}{\omega(\omega-NL_3)}e^{\omega t} \left[1-e^{(NL_3-\omega)(t-t_0-R)}\right] \\
&& = \frac{NL_2\overline{\epsilon}}{\omega}e^{\omega t} + \left(\frac{2M_0N}{\omega}-\frac{NL_2\overline{\epsilon}}{\omega}\right)e^{\omega R}e^{NL_3(t-t_0-R)} \\
&& + \frac{N^2L_2L_3 \overline{\epsilon}}{\omega(\omega-NL_3)}e^{\omega t} \left[1-e^{(NL_3-\omega)(t-t_0-R)}\right].
\end{eqnarray*}

Hence,
\begin{eqnarray*}
&&\left\|y(t)-\overline{y}(t)\right\| \leq \frac{NL_2\overline{\epsilon}}{\omega} + \left(\frac{2M_0N}{\omega}-\frac{NL_2\overline{\epsilon}}{\omega}\right)e^{(NL_3-\omega)(t-t_0-R)} \\
&& + \frac{N^2L_2L_3\overline{\epsilon}}{\omega(\omega-NL_3)} \left[1-e^{(NL_3-\omega)(t-t_0-R)}\right] \\
&& = \frac{NL_2\overline{\epsilon}}{\omega-NL_3} + \left(\frac{2M_0N}{\omega}-\frac{NL_2\overline{\epsilon}}{\omega-NL_3}\right) e^{(NL_3-\omega)(t-t_0-R)} \\
&& = \frac{NL_2\overline{\epsilon}}{\omega-NL_3}\left[1-e^{(NL_3-\omega)(t-t_0-R)}\right] + \frac{2M_0N}{\omega}e^{(NL_3-\omega)(t-t_0-R)}.
\end{eqnarray*}

The last inequality leads to
\begin{eqnarray*}
\left\|y(t)-\overline{y}(t)\right\| \leq \frac{NL_2\overline{\epsilon}}{\omega-NL_3} + \frac{2M_0N}{\omega}e^{(NL_3-\omega)(t-t_0-R)}.
\end{eqnarray*}
Consequently,
\begin{eqnarray*}
&& \left\|y(t_0)-\overline{y}(t_0)\right\| \leq \frac{NL_2\overline{\epsilon}}{\omega-NL_3} + \frac{2M_0N}{\omega}e^{(\omega - NL_3)R} \\
&& < \left( 1+ \frac{NL_2}{\omega-NL_3} \right) \overline{\epsilon} \\
&& < \delta.
\end{eqnarray*}

In the remaining part of the proof, we will show the existence of a positive number $\epsilon_1$ and an interval $J^1 \subset J,$ with a fixed length which is independent of $y(t),\overline{y}(t)\in \mathscr{A}_y,$ such that the inequality $\left\|y(t)-\overline{y}(t)\right\|>\epsilon_1$ holds for all $t\in J^1.$

Suppose that
$
 g(x,y)
 = \left( \begin{array}{ccc}
 g_1(x,y)   \\
 g_2(x,y) \\
\vdots \\
 g_n(x,y)
\end{array} \right),
$
where each $g_j,$ $1 \leq j \leq n,$ is a real valued function.

Since for each $x(t)\in \mathscr{A}_x$ the functions $x'(t)$ and $\phi'_{x(t)}(t)$  lie inside the tubes with radii $H_0$ and $\left\|A\right\|M+M_0,$ respectively, one can conclude that the set of functions $\mathscr{A}_x$ and $\mathscr{A}_y$ are both equicontinuous on $\mathbb R.$ Making use of the uniform continuity of the function $\overline{g}: \mathbb R^m \times \mathbb R^m \times \mathbb R^n \to \mathbb R^n$ defined as $\overline{g}(x_1,x_2,x_3)=g(x_1,x_3)-g(x_2,x_3)$ on the compact region 
\begin{eqnarray*}
\mathscr{D}=\left\{(x_1,x_2,x_3) \in \mathbb R^m \times \mathbb R^m \times \mathbb R^n ~|~ \left\|x_1\right\| \leq H, \left\|x_2\right\| \leq H, \left\|x_3\right\|\leq M \right\}, \nonumber 
\end{eqnarray*}
together with the equicontinuity of the sets  $\mathscr{A}_x$ and $\mathscr{A}_y,$ one can easily verify that the set 
\begin{eqnarray}\label{equicontinuous}
\mathscr{F}=\left\{g_j(x(t),\phi_{x(t)}(t))-g_j(\overline{x}(t),\phi_{x(t)}(t))~|~1\leq j\leq n,~  x(t)\in \mathscr{A}_x,~ \overline{x}(t) \in \mathscr{A}_x \right\} \end{eqnarray} 
is an equicontinuous family on $\mathbb R.$ 

Therefore, there exists a positive real number $\tau<\Delta,$ independent of the functions $x(t),\overline{x}(t)\in \mathscr{A}_x,$ $y(t),\overline{y}(t)\in \mathscr{A}_y,$ such that for any $t_1,t_2\in \mathbb R$ with $\left|t_1-t_2\right|<\tau$ the inequality 
\begin{eqnarray}\label{sensitivity_eqn_1}
\left| \left(g_j\left(x(t_1),y(t_1)\right) - g_j\left(\overline{x}(t_1),y(t_1)\right)  \right) - \left(g_j\left(x(t_2),y(t_2)\right) - g_j\left(\overline{x}(t_2),y(t_2)\right)  \right)   \right|<\frac{L_1\epsilon_0}{2n}
\end{eqnarray}
holds, for all $1\leq j \leq n.$

Condition $(A4)$ implies that, for all $t \in J,$ the inequality $\left\|g(x(t),y(t))-g(\overline{x}(t),y(t))  \right\| \geq L_1 \left\|x(t)-\overline{x}(t)\right\|$ holds. Therefore, for each $t\in J,$ there exists an integer $j_0=j_0(t),$ $1 \leq j_0 \leq n,$ such that 
\begin{eqnarray}
\begin{array}{l}
\left|g_{j_0}(x(t),y(t))-g_{j_0}(\overline{x}(t),y(t))\right|   \geq \frac{L_1}{n} \left\|x(t)-\overline{x}(t)\right\| \nonumber.
\end{array}
\end{eqnarray}
Otherwise, if there exists $s\in J$ such that for all $1\leq j\leq n,$ the inequality  
\begin{eqnarray}
\begin{array}{l}
\left|g_{j} \left(x\left(s\right) ,y\left(s\right) \right)-g_{j}(\overline{x}(s),y(s)) \right|< \frac{L_1}{n} \left\|x(s)-\overline{x}(s)\right\| \nonumber
\end{array}
\end{eqnarray}
holds, then one encounters with a contradiction since
\begin{eqnarray*}
&& \left\|g(x(s),y(s))-g(\overline{x}(s),y(s))  \right\|   \nonumber\\
&& \leq \sum_{j=1}^{n}\left| g_{j}(x(s),y(s))-g_{j}(\overline{x}(s),y(s)) \right| \nonumber\\
&& < L_1 \left\|x(s)-\overline{x}(s)\right\|. 
\end{eqnarray*}

Now, let $s_0$ be the midpoint of the interval $J$ and $\theta=s_0-\frac{\tau}{2}.$ One can find an integer $j_0=j_0(s_0),$  $1 \leq j_0 \leq n,$ such that 
\begin{eqnarray}
\begin{array}{l} \label{sensitivity_eqn_2}
\left|g_{j_0}(x(s_0),y(s_0))-g_{j_0}(\overline{x}(s_0),y(s_0))\right|   \geq \frac{L_1}{n} \left\|x(s_0)-\overline{x}(s_0)\right\| \geq \frac{L_1\epsilon_0}{n}. 
\end{array}
\end{eqnarray}

On the other hand, making use of inequality $(\ref{sensitivity_eqn_1}),$ for all $t \in \left[\theta, \theta+\tau\right]$ we have
\begin{eqnarray*}
&& \left|g_{j_0}\left(x(s_0),y(s_0)\right) - g_{j_0}\left(\overline{x}(s_0),y(s_0)\right) \right| - \left|g_{j_0}\left(x(t),y(t)\right) - g_{j_0}\left(\overline{x}(t),y(t)\right) \right| \\
&& \leq \left| \left(g_{j_0}\left(x(t),y(t)\right) - g_{j_0}\left(\overline{x}(t),y(t)\right)  \right) - \left(g_{j_0}\left(x(s_0),y(s_0)\right) - g_{j_0}\left(\overline{x}(s_0),y(s_0)\right)  \right)   \right| \\
&&<\frac{L_1\epsilon_0}{2n}
\end{eqnarray*}
and therefore by means of $(\ref{sensitivity_eqn_2}),$ we obtain that the inequality
\begin{eqnarray} \label{sensitivity_eqn_3}
 \left|g_{j_0}\left(x(t),y(t)\right) - g_{j_0}\left(\overline{x}(t),y(t)\right) \right| 
 > \left|g_{j_0}\left(x(s_0),y(s_0)\right) - g_{j_0}\left(\overline{x}(s_0),y(s_0)\right) \right|  - \frac{L_1\epsilon_0}{2n}
 \geq \frac{L_1\epsilon_0}{2n}
\end{eqnarray}
holds for all $t \in \left[\theta, \theta+\tau\right].$

By applying the mean value theorem for integrals, one can find $s_1, s_2, \ldots, s_n \in [\theta,\theta+\tau]$ such that
\begin{eqnarray*}
&& \left\|\displaystyle\int^{\theta + \tau}_{\theta} \left[g(x(s),y(s))-g(\overline{x}(s),y(s))\right] ds \right\| \nonumber\\
&& = \left\|\left( \begin{array}{ccc}
\displaystyle\int^{\theta + \tau}_{\theta} \left[g_1(x(s),y(s))-g_1(\overline{x}(s),y(s))\right] ds  \\
\displaystyle\int^{\theta + \tau}_{\theta} \left[g_2(x(s),y(s))-g_2(\overline{x}(s),y(s))\right] ds \\
\vdots \\
\displaystyle\int^{\theta + \tau}_{\theta} \left[g_n(x(s),y(s))-g_n(\overline{x}(s),y(s))\right] ds \\
\end{array} \right)\right\| \\
&& = \left\|\left( \begin{array}{ccc}
\tau \left[g_1(x(s_1),y(s_1))-g_1(\overline{x}(s_1),y(s_1))\right]   \\
\tau \left[g_2(x(s_2),y(s_2))-g_2(\overline{x}(s_2),y(s_2))\right]  \\
\vdots \\
\tau \left[g_n(x(s_n),y(s_n))-g_n(\overline{x}(s_n),y(s_n))\right] 
\end{array} \right)\right\|.
\end{eqnarray*}

Thus, using $(\ref{sensitivity_eqn_3}),$ one can obtain that
\begin{eqnarray*}
&& \left\|\displaystyle\int^{\theta + \tau}_{\theta} \left[g(x(s),y(s))-g(\overline{x}(s),y(s))\right] ds \right\| \nonumber\\
&& \geq \tau \left|g_{j_0}(x(s_{j_0}),y(s_{j_0}))-g_{j_0}(\overline{x}(s_{j_0}),y(s_{j_0}))\right| \\
&& > \frac{\tau  L_1 \epsilon_0}{2n}.
\end{eqnarray*}

It is clear that, for $t\in [\theta,\theta+\tau],$ $y(t)$ and $\overline{y}(t)$ satisfy the integral equations
\begin{eqnarray*}
y(t)= y(\theta)+ \displaystyle\int^{t}_{\theta} Ay(s) ds + \displaystyle\int^{t}_{\theta} g(x(s),y(s)) ds,  \nonumber
\end{eqnarray*}
and
\begin{eqnarray*}
\overline{y}(t)= \overline{y}(\theta)+ \displaystyle\int^{t}_{\theta} A\overline{y}(s) ds + \displaystyle\int^{t}_{\theta} g(\overline{x}(s),\overline{y}(s)) ds,  \nonumber
\end{eqnarray*}
respectively, and herewith the equation
\begin{eqnarray*}
&& y(t)-\overline{y}(t) = (y(\theta)-\overline{y}(\theta))  + \displaystyle\int^{t}_{\theta} A(y(s)-\overline{y}(s)) ds \nonumber\\
&& + \displaystyle\int^{t}_{\theta}  [g(x(s),y(s))-g(\overline{x}(s),y(s))]  ds \\
&& + \displaystyle\int^{t}_{\theta}  [g(\overline{x}(s),y(s))-g(\overline{x}(s),\overline{y}(s))]  ds
\end{eqnarray*}
is achieved.
Hence, we have the inequality
\begin{eqnarray} \label{sensitivity_eqn_4}
\begin{array}{l}
\left\|y(\theta+\tau)-\overline{y}(\theta+\tau)\right\| \geq  \left\|\displaystyle\int^{\theta+\tau}_{\theta}  [g(x(s),y(s))-g(\overline{x}(s),y(s))]ds\right\| \\
- \left\|y(\theta)-\overline{y}(\theta)\right\|  \\
-  \displaystyle\int^{\theta+\tau}_{\theta} \left\|A\right\|\left\|y(s)-\overline{y}(s)\right\| ds  \\
- \displaystyle\int^{\theta+\tau}_{\theta}  L_3\left\|y(s)-\overline{y}(s)\right\| ds. \\
\end{array}
\end{eqnarray}

Now, assume that $\displaystyle \max_{t\in [\theta,\theta+\tau]}\left\|y(t)-\overline{y}(t)\right\| \leq \frac{\tau L_1 \epsilon_0}{2n[2+\tau (L_3 + \left\|A\right\|)]}.$
In the present case, one encounters with a contradiction since, by means of the inequalities $(\ref{sensitivity_eqn_3})$ and $(\ref{sensitivity_eqn_4}),$ we have 
\begin{eqnarray*}
&& \displaystyle \max_{t\in [\theta,\theta+\tau]}\left\|y(t)-\overline{y}(t)\right\| \geq \left\|y(\theta+\tau)-\overline{y}(\theta+\tau)\right\| \\
&& > \frac{\tau L_1 \epsilon_0}{2n} - [1+ \tau(L_3 + \left\|A\right\|)] \displaystyle \max_{t\in [\theta,\theta+\tau]}\left\|y(t)-\overline{y}(t)\right\| \\
&& \geq \frac{\tau L_1 \epsilon_0}{2n} - [1+ \tau(L_3 + \left\|A\right\|)] \frac{\tau L_1 \epsilon_0}{2n[2+\tau (L_3 + \left\|A\right\|)]} \\
&& = \frac{\tau L_1 \epsilon_0}{2n} \left(1-\frac{1+ \tau(L_3 + \left\|A\right\|)}{2+ \tau(L_3 + \left\|A\right\|)}\right) \\
&& = \frac{\tau L_1 \epsilon_0}{2n[2+\tau (L_3 + \left\|A\right\|)]}.
\end{eqnarray*}
Therefore, one can see that the inequality $\displaystyle \max_{t\in [\theta,\theta+\tau]}\left\|y(t)-\overline{y}(t)\right\| > \frac{\tau L_1 \epsilon_0}{2n[2+\tau (L_3 + \left\|A\right\|)]}$ is valid.

Suppose that at the point $\eta,$ the real valued function $\left\|y(t)-\overline{y}(t)\right\|$ takes its maximum on the interval $[\theta,\theta+\tau],$  that is, 
\begin{eqnarray*}
\displaystyle \max_{t \in [\theta,\theta+\tau]} \left\|y(t)-\overline{y}(t)\right\| = \left\|y(\eta)-\overline{y}(\eta)\right\| \nonumber
\end{eqnarray*} 
for some $\theta \leq \eta \leq \theta+\tau.$

For $t\in [\theta,\theta+\tau],$ by favour of the integral equations
\begin{eqnarray*}
y(t)= y(\eta)+ \displaystyle\int^{t}_{\eta} Ay(s) ds + \displaystyle\int^{t}_{\eta} g(x(s),y(s)) ds,  \nonumber
\end{eqnarray*}
and
\begin{eqnarray*}
\overline{y}(t)= \overline{y}(\eta)+ \displaystyle\int^{t}_{\eta} A\overline{y}(s) ds + \displaystyle\int^{t}_{\eta} g(\overline{x}(s),\overline{y}(s)) ds,  \nonumber
\end{eqnarray*}
we obtain
\begin{eqnarray*}
&& y(t)-\overline{y}(t) = (y(\eta)-\overline{y}(\eta))  + \displaystyle\int^{t}_{\eta} A(y(s)-\overline{y}(s)) ds \nonumber\\
&& + \displaystyle\int^{t}_{\eta}  [g(x(s),y(s))-g(\overline{x}(s),\overline{y}(s))]  ds.
\end{eqnarray*}
Define
$\tau^1=\displaystyle \min \left\{ \frac{\tau}{2}, \frac{\tau L_1 \epsilon_0}{8n(M\left\|A\right\|+M_0)[2+\tau (L_3 + \left\|A\right\|)]}   \right\}$
and let
$
\theta^1=\left\{\begin{array}{ll} \eta, & ~\textrm{if}~  \eta \leq \theta + \frac{\tau}{2}   \\
\eta - \tau^1, & ~\textrm{if}~  \eta > \theta + \frac{\tau}{2}  \\
\end{array} \right. .\nonumber
$ 
We note that the interval $J^1=[\theta^1, \theta^1+\tau^1]$ is a subset of $[\theta,\theta+\tau]$ and hence of $J.$

For $t\in J^1,$ we have
\begin{eqnarray} 
\begin{array}{l}
\left\|y(t)-\overline{y}(t)\right\| \geq  \left\|y(\eta)-\overline{y}(\eta)\right\| - \left|  \displaystyle\int^{t}_{\eta} \left\|A\right\|\left\|y(s)-\overline{y}(s)\right\| ds   \right| \nonumber  \\
-\left|  \displaystyle\int^{t}_{\eta} \left\|  g(x(s),y(s))-g(\overline{x}(s),\overline{y}(s))  \right\| ds  \right| \\
 > \frac{\tau L_1 \epsilon_0}{2n[2+\tau (L_3 + \left\|A\right\|)]} -2M\left\|A\right\|\tau^1-2M_0\tau^1 \\
 = \frac{\tau L_1 \epsilon_0}{2n[2+\tau (L_3 + \left\|A\right\|)]} -2\tau^1(M\left\|A\right\|+M_0) \\
 \geq \frac{\tau L_1 \epsilon_0}{4n[2+\tau (L_3 + \left\|A\right\|)]}.
\end{array}
\end{eqnarray}

Consequently, we get $\left\|y(t)-\overline{y}(t)\right\| > \epsilon_1,$ $t\in J^1,$ where $\epsilon_1=\frac{\tau L_1 \epsilon_0}{4n[2+\tau (L_3 + \left\|A\right\|)]}$ and the length of the interval $J^1$ does not depend on the functions $x(t),\overline{x}(t) \in \mathscr{A}_x.$

The proof of the lemma is finalized. $\square$

We say that the set $\mathscr{A}$ is sensitive if both $\mathscr{A}_x$ and $\mathscr{A}_y$ are sensitive. According to this definition, if $\mathscr{A}_x$ is sensitive, then Lemma \ref{sensitivity} implies that the set $\mathscr{A}$ is also sensitive.

Now, let us illustrate the replication of sensitivity through an example. It is known that the Lorenz system  
\begin{eqnarray} 
\begin{array}{l}
x'_1=\sigma \left(  -x_1 + x_2 \right) \\  \label{Lorenz_system}
x'_2=-x_2 +rx_1 -x_1x_3 \\
x'_3=-bx_3+x_1x_2, \\
\end{array}
\end{eqnarray}
with the coefficients $\sigma=10, b=8/3, r=28$ admits sensitivity \cite{Lorenz63}. We use system (\ref{Lorenz_system}) with the specified coefficients as the master system and constitute the $6-$dimensional result-system 
\begin{eqnarray} 
\begin{array}{l}
x'_1=10(-x_1+x_2) \\ \label{sensitivity_example}
x'_2=-x_2+28x_1-x_1x_3 \\
x'_3=-\frac{8}{3}x_3+x_1x_2 \\
x'_4=-5x_4+x_3 \\
x'_5=-2x_5+0.0002(x_2-x_5)^3+4x_2 \\
x'_6=-3x_6-3x_1.
\end{array}
\end{eqnarray}

When system $(\ref{sensitivity_example})$ is considered in the form of system $(\ref{1})+(\ref{2}),$ one can see that the diagonal matrix $A$ whose entries on the diagonal are $-5, -2, -3$ satisfies that $\left\|e^{At}\right\| \leq Ne^{-\omega t},$ where $N=1$ and $\omega=2.$ We note that the function $g:\mathbb R^3 \times \mathbb R^3 \to \mathbb R^3$ defined as 
\begin{eqnarray*} 
g(x_1,x_2,x_3,x_4,x_5,x_6)= \left(  x_3,0.0002(x_2-x_5)^3+4x_2,-3x_1 \right) 
\end{eqnarray*}  
provides the conditions $(A4)$ and $(A5)$ with constants $L_1=\frac{1}{\sqrt{3}}, L_2=\frac{11 \sqrt{3}}{2}$ and $L_3=\frac{3}{2}$ since the chaotic attractor of system $(\ref{sensitivity_example})$ is inside a compact region such that $\left|x_2\right|\leq 30$ and $\left|x_5\right|\leq 50.$ Consequently, system $(\ref{sensitivity_example})$ satisfies the condition $(A7).$

\begin{figure}[ht] 
\centering
\includegraphics[width=14.0cm]{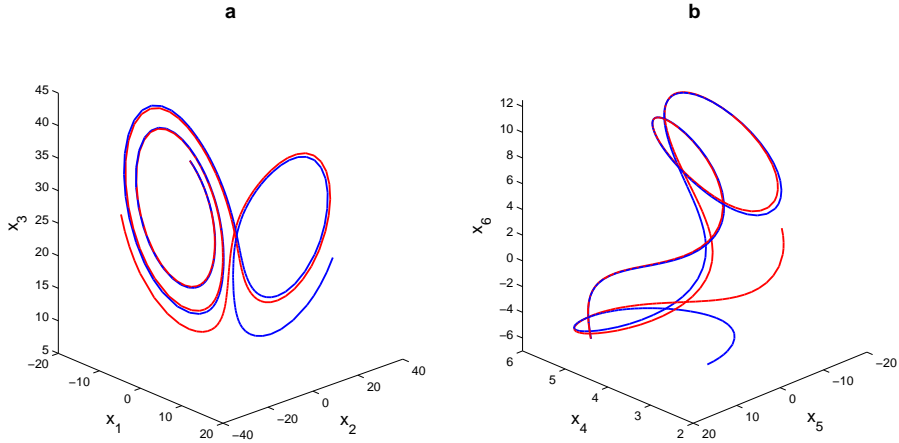}
\caption{\footnotesize{Replication of sensitivity in the result-system $(\ref{sensitivity_example}).$  (a)  $3-$dimensional projection in the $x_1-x_2-x_3$ space. (b) $3-$dimensional projection in the $x_4-x_5-x_6$ space.  The sensitivity property is observable both in $(a)$ and $(b)$ such that the trajectories presented by blue and red colors move together in the first stage and then diverge. In other words, the sensitivity property of the generator system is mimicked by the replicator counterpart.}}
\label{sensitivity_figure}
\end{figure}

In Figure $\ref{sensitivity_figure},$ one can see the $3-$dimensional projections in the $x_1-x_2-x_3$ and $x_4-x_5-x_6$ spaces of two different trajectories of the result-system $(\ref{sensitivity_example})$ with adjacent initial conditions, such that one of them is in blue color and the other in red color. For the trajectory with blue color, we make use of the initial data $x_1(0)=-8.57,  x_2(0)=-2.39,  x_3(0)=33.08,  x_4(0)=5.32,  x_5(0)=10.87,  x_6(0)=-6.37$ and for the one with red color, we use the initial data $x_1(0)=-8.53,  x_2(0)=-2.47,  x_3(0)=33.05,  x_4(0)=5.33,  x_5(0)=10.86,  x_6(0)=-6.36.$ In the simulation, the trajectories move on the time interval $[0,3].$ The results seen in Figure \ref{sensitivity_figure} supports our theoretical results indicated in Lemma \ref{sensitivity} such that the replicator system, likewise the generator counterpart, admits the sensitivity feature. That is, the solutions of both the generator and the replicator given by blue and red colors diverge, even though they start and move close to each other in the first stage.

\begin{lemma} \label{transitivity}
Transitivity of $\mathscr{A}_x$ implies the same feature for $\mathscr{A}_y.$ 
\end{lemma}

\noindent \textbf{Proof.}
Fix arbitrary small $\epsilon>0,$ arbitrary large $E>0$ and let $y(t)\in \mathscr{A}_y$ be a given function. Arising from the definition, introduced by $(\ref{A_y}),$ of the set $\mathscr{A}_y$, there exists a function $x(t)\in \mathscr{A}_x$ such that $y(t)=\phi_{x(t)}(t).$  Let $\gamma=\frac{\omega(\omega-NL_3)}{2M_0N(\omega-NL_3)+NL_2\omega}.$ Condition $(A7)$ guarantees that $\gamma$ is positive.
Since there exists a solution $x^*(t) \in \mathscr{A}_x,$ which is dense in $\mathscr{A}_x,$ one can find $\xi>0$ and an interval $J \subset \mathbb R$ with length $E$ such that $\left\|x(t)-x^*(t+\xi)\right\| < \gamma \epsilon,$ for all $t\in J.$ Without loss of generality, assume that $J$ is a closed interval, that is, $J=[a,a+E]$ for some real number $a.$ 

Let $y^*(t)=\phi_{x^*(t)}(t).$ 
For $t\in J,$ the bounded on $\mathbb R$ functions $y(t)$ and $y^*(t)$ must satisfy the relations
\begin{eqnarray*}
y(t) =  \displaystyle\int^{t}_{-\infty} e^{A(t-s)} g(x(s),y(s)) ds, 
\end{eqnarray*}
and
\begin{eqnarray*}
y^*(t) =  \displaystyle\int^{t}_{-\infty} e^{A(t-s)} g(x^*(s),y^*(s)) ds,
\end{eqnarray*}
respectively. 
The second equation above implies that
\begin{eqnarray*}
y^*(t+\xi) =  \displaystyle\int^{t+\xi}_{-\infty} e^{A(t+\xi-s)} g(x^*(s),y^*(s)) ds.
\end{eqnarray*}
Using the transformation $\overline{s}=s-\xi,$ and replacing $\overline{s}$ by $s$ again, we get
\begin{eqnarray*}
y^*(t+\xi) =  \displaystyle\int^{t}_{-\infty} e^{A(t-s)} g(x^*(s+\xi),y^*(s+\xi)) ds.
\end{eqnarray*}

Therefore, for $t\in J,$ we have 
\begin{eqnarray*}
&& y(t)-y^*(t+\xi) =  \displaystyle\int^{a}_{-\infty} e^{A(t-s)} [g(x(s),y(s))-g(x^*(s+\xi),y^*(s+\xi))] ds \\
&& + \displaystyle\int^{t}_{a} e^{A(t-s)} [g(x(s),y(s))-g(x(s),y^*(s+\xi))] ds \\
&& + \displaystyle\int^{t}_{a} e^{A(t-s)} [g(x(s),y^*(s))-g(x^*(s+\xi),y^*(s+\xi))] ds,
\end{eqnarray*}
which implies the inequality
\begin{eqnarray*}
&& \left\|y(t)-y^*(t+\xi)\right\| \leq  \displaystyle\int^{a}_{-\infty} 2M_0N e^{-\omega(t-s)} ds \\
&& + \displaystyle\int^{t}_{a} NL_3  e^{-\omega(t-s)}  \left\|y(s)-y^*(s+\xi)\right\| ds \\
&& + \displaystyle\int^{t}_{a} NL_2  e^{-\omega(t-s)}  \left\|x(s)-x^*(s+\xi)\right\| ds \\
&& \leq \frac{2M_0N}{\omega} e^{-\omega(t-a)} + \frac{NL_2\gamma\epsilon}{\omega} e^{-\omega t} \left(e^{\omega t} - e^{\omega a}\right) \\
&& + \displaystyle\int^{t}_{a} NL_3  e^{-\omega(t-s)}  \left\|y(s)-y^*(s+\xi)\right\| ds.
\end{eqnarray*}  
Hence, we have
\begin{eqnarray*}
&& e^{\omega t}\left\|y(t)-y^*(t+\xi)\right\| \leq \frac{2M_0N}{\omega} e^{\omega a} + \frac{NL_2\gamma\epsilon}{\omega}  \left(e^{\omega t} - e^{\omega a}\right) \\
&& + \displaystyle\int^{t}_{a} NL_3  e^{\omega s}  \left\|y(s)-y^*(s+\xi)\right\| ds.
\end{eqnarray*}  
Through the implementation of Lemma 2.2 \cite{Bar70} to the last inequality, we attain 
\begin{eqnarray*}
&& e^{\omega t}\left\|y(t)-y^*(t+\xi)\right\| \leq \frac{2M_0N}{\omega} e^{\omega a} + \frac{NL_2\gamma\epsilon}{\omega}  \left(e^{\omega t} - e^{\omega a}\right) \\
&& + \displaystyle\int^{t}_{a} NL_3 \left[\frac{2M_0N}{\omega} e^{\omega a} + \frac{NL_2\gamma\epsilon}{\omega}  \left(e^{\omega s} - e^{\omega a}\right)\right] e^{NL_3(t-s)}  ds \\
&& = \frac{NL_2\gamma\epsilon}{\omega}e^{\omega t} + \left(\frac{2M_0N}{\omega}-\frac{NL_2\gamma\epsilon}{\omega}\right) e^{\omega a}e^{NL_3(t-a)} \\
&& + \frac{N^2L_2L_3\gamma\epsilon}{\omega(\omega-NL_3)} e^{NL_3 t} \left(e^{(\omega-NL_3)t}-e^{(\omega-NL_3)a}\right).
\end{eqnarray*}  
Multiplying both sides by $e^{-\omega t},$ we achieve that
\begin{eqnarray*}
&& \left\|y(t)-y^*(t)\right\| \leq \frac{2M_0N}{\omega} e^{(NL_3-\omega)(t-a)} \nonumber \\
&& + \left( \frac{NL_2\gamma\epsilon}{\omega} + \frac{N^2L_2L_3\gamma\epsilon}{\omega(\omega-NL_3)} \right) \left(1-e^{(NL_3-\omega)(t-a)}\right) \\
&& = \frac{2M_0N}{\omega} e^{(NL_3-\omega)(t-a)} +\frac{NL_2\gamma\epsilon}{\omega-NL_3} \left(1-e^{(NL_3-\omega)(t-a)}\right).
\end{eqnarray*}  
Now, suppose that $E>\frac{2}{\omega -NL_3} \ln \left(\frac{1}{\gamma\epsilon}\right).$ If $t \in \left[ a+\frac{E}{2}, a+E \right],$ then it is true that $e^{(NL_3-\omega)(t-a)} \leq e^{(NL_3-\omega)\frac{E}{2}}<\gamma\epsilon.$
As a result, we have
\begin{eqnarray*}
\left\|y(t)-y^*(t+\xi)\right\| < \left[\frac{2M_0N}{\omega}  + \frac{NL_2}{\omega-NL_3}\right] \gamma\epsilon =\epsilon, 
\end{eqnarray*}  
for $t\in J_1= \left[ a_1, a_1+E_1 \right],$ where $a_1=a+\frac{E}{2}$ and $E_1=\frac{E}{2}.$ Consequently, the set $\mathscr{A}_y$ is transitive in compliance with Definition $\ref{dev3}.$

The lemma is proved. $\square$

\begin{lemma}
If $\mathscr{A}_x$ admits a dense countable collection of periodic functions, then the same is true for $\mathscr{A}_y.$
\end{lemma}

\noindent \textbf{Proof.}
Fix $y(t)=\phi_{x(t)}(t) \in \mathscr{A}_y,$ arbitrary small $\epsilon>0$ and arbitrary large $E>0.$   
Assume that $\mathscr{A}_x$ admits a dense countable collection $\mathscr{G}_x$ of periodic functions.  Let $\gamma=\frac{\omega(\omega-NL_3)}{2M_0N(\omega-NL_3)+NL_2\omega}.$ Note that $\gamma$ is a positive number by condition $(A7).$ In this case, there exist $\widetilde{x}(t)\in \mathscr{G}_x$ and an interval $J$ with length $E$ such that $\left\|x(t)-\widetilde{x}(t)\right\| < \gamma\epsilon$ for $t\in J.$ Without loss of generality, assume that $J$ is a closed interval, that is, $J=[a,a+E]$ for some $a \in \mathbb R.$ 

We note that by condition $(A4)$ there is a one-to-one correspondence between the sets $\mathscr{G}_x$ and $\mathscr{G}_y = \left\{  \phi_{x(t)}(t) ~|~ x(t) \in \mathscr{G}_x  \right\},$ and $x(t) \in \mathscr{G}_x$ is $pT-$periodic if and only if $\phi_{x(t)}(t) \in \mathscr{G}_y$ is $pT-$periodic. Therefore $\mathscr{G}_y \subset \mathscr{A}_y$ is a countable collection of periodic functions and our aim is to show that the set $\mathscr{G}_y$ is dense in $\mathscr{A}_y$ in accordance with Definition $\ref{dev4}.$

Let $\widetilde{y}(t)=\phi_{\widetilde{x}(t)}(t),$ which clearly belongs to the set $\mathscr{G}_y.$
Making use of the relations
\begin{eqnarray*}
y(t) =  \displaystyle\int^{t}_{-\infty} e^{A(t-s)} g(x(s),y(s)) ds, 
\end{eqnarray*}
and
\begin{eqnarray*}
\widetilde{y}(t) =  \displaystyle\int^{t}_{-\infty} e^{A(t-s)} g(\widetilde{x}(s),\widetilde{y}(s)) ds,
\end{eqnarray*}
for $t\in J,$ we attain 
\begin{eqnarray*}
&& y(t)-\widetilde{y}(t) =  \displaystyle\int^{a}_{-\infty} e^{A(t-s)} [g(x(s),y(s))-g(\widetilde{x}(s),\widetilde{y}(s))] ds \\
&& + \displaystyle\int^{t}_{a} e^{A(t-s)} [g(x(s),y(s))-g(x(s),\widetilde{y}(s))] ds \\
&& + \displaystyle\int^{t}_{a} e^{A(t-s)} [g(x(s),\widetilde{y}(s))-g(\widetilde{x}(s),\widetilde{y}(s))] ds.
\end{eqnarray*}
The last equation implies that
\begin{eqnarray*}
&& \left\|y(t)-\widetilde{y}(t)\right\| \leq  \displaystyle\int^{a}_{-\infty} 2M_0N e^{-\omega(t-s)} ds \\
&& + \displaystyle\int^{t}_{a} NL_3  e^{-\omega(t-s)}  \left\|y(s)-\widetilde{y}(s)\right\| ds \\
&& + \displaystyle\int^{t}_{a} NL_2  e^{-\omega(t-s)}  \left\|x(s)-\widetilde{x}(s)\right\| ds \\
&& \leq \frac{2M_0N}{\omega} e^{-\omega(t-a)} + \frac{NL_2\gamma\epsilon}{\omega} e^{-\omega t} \left(e^{\omega t} - e^{\omega a}\right) \\
&& + \displaystyle\int^{t}_{a} NL_3  e^{-\omega(t-s)}  \left\|y(s)-\widetilde{y}(s)\right\| ds.
\end{eqnarray*}  
Hence, we achive that
\begin{eqnarray*}
&& e^{\omega t}\left\|y(t)-\widetilde{y}(t)\right\| \leq \frac{2M_0N}{\omega} e^{\omega a} + \frac{NL_2\gamma\epsilon}{\omega}  \left(e^{\omega t} - e^{\omega a}\right) \\
&& + \displaystyle\int^{t}_{a} NL_3  e^{\omega s}  \left\|y(s)-\widetilde{y}(s)\right\| ds.
\end{eqnarray*}  
Application of Lemma 2.2 \cite{Bar70} to the last inequality provides us
\begin{eqnarray*}
&& e^{\omega t}\left\|y(t)-\widetilde{y}(t)\right\| \leq \frac{2M_0N}{\omega} e^{\omega a} + \frac{NL_2\gamma\epsilon}{\omega}  \left(e^{\omega t} - e^{\omega a}\right) \\
&& + \displaystyle\int^{t}_{a} NL_3 \left[\frac{2M_0N}{\omega} e^{\omega a} + \frac{NL_2\gamma\epsilon}{\omega}  \left(e^{\omega s} - e^{\omega a}\right)\right] e^{NL_3(t-s)}  ds \\
&& = \frac{NL_2\gamma\epsilon}{\omega}e^{\omega t} + \left(\frac{2M_0N}{\omega}-\frac{NL_2\gamma\epsilon}{\omega}\right) e^{\omega a}e^{NL_3(t-a)} \\
&& + \frac{N^2L_2L_3\gamma\epsilon}{\omega(\omega-NL_3)} e^{NL_3 t} \left(e^{(\omega-NL_3)t}-e^{(\omega-NL_3)a}\right).
\end{eqnarray*}  
Multiplying both sides by $e^{-\omega t},$ we obtain
\begin{eqnarray*}
&& \left\|y(t)-\widetilde{y}(t)\right\| \leq \frac{2M_0N}{\omega} e^{(NL_3-\omega)(t-a)} \nonumber \\
&& + \left( \frac{NL_2\gamma\epsilon}{\omega} + \frac{N^2L_2L_3\gamma\epsilon}{\omega(\omega-NL_3)} \right) \left(1-e^{(NL_3-\omega)(t-a)}\right) \\
&& = \frac{2M_0N}{\omega} e^{(NL_3-\omega)(t-a)} +\frac{NL_2\gamma\epsilon}{\omega-NL_3} \left(1-e^{(NL_3-\omega)(t-a)}\right).
\end{eqnarray*}  
Suppose that $E>\frac{2}{\omega -NL_3} \ln \left(\frac{1}{\gamma\epsilon}\right).$ If $a+\frac{E}{2}\leq t \leq a+E ,$ then one has $e^{(NL_3-\omega)(t-a)} \leq e^{(NL_3-\omega)\frac{E}{2}}<\gamma\epsilon.$
Consequently, the inequality
\begin{eqnarray*}
\left\|y(t)-\widetilde{y}(t)\right\| < \left[\frac{2M_0N}{\omega}  + \frac{NL_2}{\omega-NL_3}\right] \gamma\epsilon =\epsilon, 
\end{eqnarray*}  
holds for $t\in J_1= \left[ a_1, a_1+E_1 \right],$ where $a_1=a+\frac{E}{2}$ and $E_1=\frac{E}{2}.$

The proof of the Lemma is completed. $\square$

\begin{theorem}\label{morphogenesis_devaney_theorem}
If the set $\mathscr{A}_x$ is Devaney's chaotic, then the same is true for the set $\mathscr{A}_y.$
\end{theorem}


\subsection{\textbf{Machinery of Li-Yorke chaos extension}}\label{secliyorke}

In this part of our paper, the extension of the chaos in the Li$-$Yorke sense is taken into account.
Our aim is to prove that if $\mathscr{A}_x$ is chaotic in the sense of Li-Yorke, then the same is true for the set $\mathscr{A}_y$ and hence for the set $\mathscr{A}.$

We start by presenting the following lemma which states the replication of proximality property.

\begin{lemma} \label{proximality}
If a couple of functions $\left(x(t),\overline{x}(t)\right) \in \mathscr{A}_x \times  \mathscr{A}_x $ is proximal, then the same is true for the couple $\left(\phi_{x(t)}(t),\phi_{\overline{x}(t)}(t)\right) \in \mathscr{A}_y \times  \mathscr{A}_y .$
\end{lemma}
\noindent \textbf{Proof.}
Fix arbitrary positive real numbers $\epsilon$ and $E.$ Let us define $\gamma=\frac{\omega(\omega-NL_3)}{2M_0N(\omega-NL_3)+NL_2\omega}$ which is a positive number by condition $(A7).$ Suppose that a given  couple of functions $\left(x(t),\overline{x}(t)\right) \in \mathscr{A}_x \times  \mathscr{A}_x $ is proximal. Therefore, one can find a sequence of real numbers $\left\{E_i\right\}$ satisfying  $E_i \geq E$ for each $i\in \mathbb N,$ and a sequence $\left\{a_i\right\},$ $a_i \to \infty$ as $i \to \infty,$ such that we have $\left\|x(t)-\overline{x}(t)\right\|<\gamma\epsilon,$ for each $t$ from the intervals $J_i=[a_i,a_i+E_i],$ $i\in \mathbb N,$ and $J_i \cap J_j = \emptyset$ whenever $i \neq j.$ 

Now, we shall continue our proof by fixing an arbitrary natural number $i.$ Since the functions $y(t)=\phi_{x(t)}(t) \in \mathscr{A}_y$ and $\overline{y}(t)=\phi_{\overline{x}(t)}(t) \in \mathscr{A}_y$  satisfy the relations
\begin{eqnarray*}
y(t) =  \displaystyle\int^{t}_{-\infty} e^{A(t-s)} g(x(s),y(s)) ds, \nonumber
\end{eqnarray*}
and
\begin{eqnarray*}
\overline{y}(t) =  \displaystyle\int^{t}_{-\infty} e^{A(t-s)} g(\overline{x}(s),\overline{y}(s)) ds, \nonumber
\end{eqnarray*}
respectively, for $t\in J_i,$ we have 
\begin{eqnarray*}
&& y(t)-\overline{y}(t) =  \displaystyle\int^{a_i}_{-\infty} e^{A(t-s)} [g(x(s),y(s))-g(\overline{x}(s),\overline{y}(s))] ds \\
&& + \displaystyle\int^{t}_{a_i} e^{A(t-s)} [g(x(s),y(s))-g(x(s),\overline{y}(s))] ds \\
&& + \displaystyle\int^{t}_{a_i} e^{A(t-s)} [g(x(s),\overline{y}(s))-g(\overline{x}(s),\overline{y}(s))] ds.
\end{eqnarray*}
This implies that the inequality
\begin{eqnarray*}
&& \left\|y(t)-\overline{y}(t)\right\| \leq  \displaystyle\int^{a_i}_{-\infty} 2M_0N e^{-\omega(t-s)} ds \\
&& + \displaystyle\int^{t}_{a_i} NL_3  e^{-\omega(t-s)}  \left\|y(s)-\overline{y}(s)\right\| ds \\
&& + \displaystyle\int^{t}_{a_i} NL_2  e^{-\omega(t-s)}  \left\|x(s)-\overline{x}(s)\right\| ds \\
&& \leq \frac{2M_0N}{\omega} e^{-\omega(t-a)} + \frac{NL_2\gamma\epsilon}{\omega} e^{-\omega t} \left(e^{\omega t} - e^{\omega a}\right) \\
&& + \displaystyle\int^{t}_{a_i} NL_3  e^{-\omega(t-s)}  \left\|y(s)-\overline{y}(s)\right\| ds
\end{eqnarray*}  
is valid. Hence, we achieve that
\begin{eqnarray*}
&& e^{\omega t}\left\|y(t)-\overline{y}(t)\right\| \leq \frac{2M_0N}{\omega} e^{\omega a_i} + \frac{NL_2\gamma\epsilon}{\omega}  \left(e^{\omega t} - e^{\omega a_i}\right) \\
&& + \displaystyle\int^{t}_{a_i} NL_3  e^{\omega s}  \left\|y(s)-\overline{y}(s)\right\| ds.
\end{eqnarray*}  
Implementing Lemma 2.2 \cite{Bar70} to the last inequality, we obtain
\begin{eqnarray*}
&& e^{\omega t}\left\|y(t)-\overline{y}(t)\right\| \leq \frac{2M_0N}{\omega} e^{\omega a_i} + \frac{NL_2\gamma\epsilon}{\omega}  \left(e^{\omega t} - e^{\omega a_i}\right) \\
&& + \displaystyle\int^{t}_{a} NL_3 \left[\frac{2M_0N}{\omega} e^{\omega a_i} + \frac{NL_2\gamma\epsilon}{\omega}  \left(e^{\omega s} - e^{\omega a_i}\right)\right] e^{NL_3(t-s)}  ds \\
&& = \frac{NL_2\gamma\epsilon}{\omega}e^{\omega t} + \left(\frac{2M_0N}{\omega}-\frac{NL_2\gamma\epsilon}{\omega}\right) e^{\omega a_i}e^{NL_3(t-a_i)} \\
&& + \frac{N^2L_2L_3\gamma\epsilon}{\omega(\omega-NL_3)} e^{NL_3 t} \left(e^{(\omega-NL_3)t}-e^{(\omega-NL_3)a_i}\right).
\end{eqnarray*}  
By multiplying both sides by $e^{-\omega t},$ one can easily verify that
\begin{eqnarray*}
&& \left\|y(t)-\overline{y}(t)\right\| \leq \frac{2M_0N}{\omega} e^{(NL_3-\omega)(t-a_i)} \nonumber \\
&& + \left( \frac{NL_2\gamma\epsilon}{\omega} + \frac{N^2L_2L_3\gamma\epsilon}{\omega(\omega-NL_3)} \right) \left(1-e^{(NL_3-\omega)(t-a_i)}\right) \\
&& = \frac{2M_0N}{\omega} e^{(NL_3-\omega)(t-a_i)} +\frac{NL_2\gamma\epsilon}{\omega-NL_3} \left(1-e^{(NL_3-\omega)(t-a_i)}\right).
\end{eqnarray*}  
If $E>\frac{2}{\omega -NL_3} \ln \left(\frac{1}{\gamma\epsilon}\right)$ and $t$ belongs to the interval $ \left[ a_i+\frac{E_i}{2}, a_i+E_i \right],$ then one has
\begin{eqnarray*}
&& e^{(NL_3-\omega)(t-a_i)} < e^{(NL_3-\omega)\frac{E_i}{2}} \\
&& \leq e^{(NL_3-\omega)\frac{E}{2}} \\
&& < \gamma\epsilon.
\end{eqnarray*}
Since the natural number $i$ was arbitrarily chosen, for each $i \in \mathbb N,$ we have
\begin{eqnarray*}
\left\|y(t)-\overline{y}(t)\right\| < \left[\frac{2M_0N}{\omega}  + \frac{NL_2}{\omega-NL_3}\right] \gamma \epsilon = \epsilon, 
\end{eqnarray*}  
for each $t\in \widetilde{J}_i= \left[ \widetilde{a}_i, \widetilde{a}_i + \widetilde{E}_i \right],$ where $\widetilde{a}_i=a_i+\frac{E_i}{2}$ and $\widetilde{E}_i=\frac{E_i}{2}.$ Note that $\widetilde{J}_i \subset \mathbb R$ is an interval with length not less than $\widetilde{E}=\frac{E}{2}.$ As a consequence, the couple of functions $\left(\phi_{x(t)}(t),\phi_{\overline{x}(t)}(t)\right) \in \mathscr{A}_y \times  \mathscr{A}_y$ is proximal according to the Definition $\ref{defliyorke1}.$

The proof of the Lemma is accomplished. $\square$

\begin{lemma} \label{li-yorke_separated}
If a couple of functions $\left(x(t),\overline{x}(t)\right) \in \mathscr{A}_x \times  \mathscr{A}_x $ is frequently $(\epsilon_0,\Delta)-$separated for some positive real numbers $\epsilon_0$ and $\Delta$, then the couple of functions  $\left(\phi_{x(t)}(t),\phi_{\overline{x}(t)}(t) \right) \in \mathscr{A}_y \times  \mathscr{A}_y $ is frequently $(\epsilon_1,\overline{\Delta})-$separated for some positive real numbers $\epsilon_1$ and $\overline{\Delta}.$
\end{lemma}

\noindent \textbf{Proof.} Suppose that a given couple of functions $\left( x(t), \overline{x}(t) \right) \in \mathscr{A}_x \times  \mathscr{A}_x $  is frequently $(\epsilon_0,\Delta)$ separated, for some $\epsilon_0>0$ and $\Delta>0.$ In this case, there exist infinitely many disjoint intervals, each with length not less than $\Delta,$ such that $\left\|x(t)-\overline{x}(t)\right\| > \epsilon_0,$ for each $t$ from these intervals. Without loss of generality, assume that these intervals are all closed subsets of $\mathbb R$. In that case, one can find a sequence $\left\{\Delta_i\right\}$ satisfying  $\Delta_i \geq \Delta,$ $i\in \mathbb N,$ and a sequence $\left\{d_i\right\},$ $d_i \to \infty$ as $i \to \infty,$ such that for each $i\in \mathbb N$ the inequality $\left\|x(t)-\overline{x}(t)\right\|>\epsilon_0$ holds for $t\in J_i=[d_i,d_i+\Delta_i],$ and $J_i \cap J_j = \emptyset$ whenever $i \neq j.$ 

Throughout the proof, let us denote $y(t)=\phi_{x(t)}(t) \in \mathscr{A}_y$ and $\overline{y}(t)=\phi_{\overline{x}(t)}(t)\in \mathscr{A}_y.$

Our aim is to show the existence of positive real numbers $\epsilon_1, \overline{\Delta}$ and infinitely many disjoint intervals $\overline{J}_i \subset J_i, i\in \mathbb{N},$ each with length $\overline{\Delta},$ such that the inequality $\left\|y(t)-\overline{y}(t)\right\|>\epsilon_1$ holds for each $t$ from the intervals $ \overline{J}_i, i\in \mathbb N.$

As in the proof of Lemma $\ref{sensitivity},$ we again suppose that
$
 g(x,y)
 = \left( \begin{array}{ccc}
 g_1(x,y)   \\
 g_2(x,y) \\
\vdots \\
 g_n(x,y)
\end{array} \right),
$
where each $g_j,$ $1 \leq j \leq n,$ is a real valued function.

Using the equicontinuity  on $\mathbb R$ of the family $\mathscr{F},$ defined by equation $(\ref{equicontinuous}),$
one can find a  positive real number $\tau<\Delta,$ independent of the functions $x(t),\overline{x}(t)\in \mathscr{A}_x,$ $y(t),\overline{y}(t)\in \mathscr{A}_y,$ such that for any $t_1,t_2\in \mathbb R$ with $\left|t_1-t_2\right|<\tau$ the inequality 
\begin{eqnarray}\label{liyorke_1}
\left| \left(g_j\left(x(t_1),y(t_1)\right) - g_j\left(\overline{x}(t_1),y(t_1)\right)  \right) - \left(g_j\left(x(t_2),y(t_2)\right) - g_j\left(\overline{x}(t_2),y(t_2)\right)  \right)   \right|<\frac{L_1\epsilon_0}{2n}
\end{eqnarray}
holds, for all $1\leq j \leq n.$

Suppose that the sequence $\left\{s_i\right\}$ denote the midpoints of the intervals $J_i,$ that is, $s_i=d_i+\frac{\Delta_i}{2},$ for each $i\in \mathbb N.$  Let us define a sequence $\left\{\theta_i\right\}$ through the equation $\theta_i=s_i-\frac{\tau}{2}.$ 

Now, let us fix an arbitrary natural number $i.$ 

In a similar way to the method specified in the proof of Lemma $\ref{sensitivity},$ one can show the existence of an integer $j_i=j_i(s_i),$  $1 \leq j_i \leq n,$ such that 
\begin{eqnarray}
\begin{array}{l} \label{liyorke_2}
\left|g_{j_i}(x(s_i),y(s_i))-g_{j_i}(\overline{x}(s_i),y(s_i))\right|   \geq \frac{L_1}{n} \left\|x(s_i)-\overline{x}(s_i)\right\| \geq \frac{L_1\epsilon_0}{n}. 
\end{array}
\end{eqnarray}

On the other hand, making use of the inequality $(\ref{liyorke_1}),$ it is easy to verify that
\begin{eqnarray*}
&& \left|g_{j_i}\left(x(s_i),y(s_i)\right) - g_{j_i}\left(\overline{x}(s_i),y(s_i)\right) \right| - \left|g_{j_i}\left(x(t),y(t)\right) - g_{j_i}\left(\overline{x}(t),y(t)\right) \right| \\
&& \leq \left| \left(g_{j_i}\left(x(t),y(t)\right) - g_{j_i}\left(\overline{x}(t),y(t)\right)  \right) - \left(g_{j_i}\left(x(s_i),y(s_i)\right) - g_{j_i}\left(\overline{x}(s_i),y(s_i)\right)  \right)   \right| \\
&&<\frac{L_1\epsilon_0}{2n},
\end{eqnarray*}
for all $t \in \left[\theta_i, \theta_i+\tau\right]$
and therefore by favour of $(\ref{liyorke_2}),$ we obtain that the inequality
\begin{eqnarray} \label{liyorke_3}
 \left|g_{j_i}\left(x(t),y(t)\right) - g_{j_i}\left(\overline{x}(t),y(t)\right) \right| 
 > \left|g_{j_i}\left(x(s_i),y(s_i)\right) - g_{j_i}\left(\overline{x}(s_i),y(s_i)\right) \right|  - \frac{L_1\epsilon_0}{2n}
 \geq \frac{L_1\epsilon_0}{2n}
\end{eqnarray}
is valid on the same interval.

Using the mean value theorem for integrals, it is possible to find real numbers $s^i_1, s^i_2, \ldots, s^i_n \in [\theta_i,\theta_i+\tau]$ such that
\begin{eqnarray*}
&& \left\|\displaystyle\int^{\theta_i + \tau}_{\theta_i} \left[g(x(s),y(s))-g(\overline{x}(s),y(s))\right] ds \right\| \nonumber\\
&& = \left\|\left( \begin{array}{ccc}
\displaystyle\int^{\theta_i + \tau}_{\theta_i} \left[g_1(x(s),y(s))-g_1(\overline{x}(s),y(s))\right] ds  \\
\displaystyle\int^{\theta_i + \tau}_{\theta_i} \left[g_2(x(s),y(s))-g_2(\overline{x}(s),y(s))\right] ds \\
\vdots \\
\displaystyle\int^{\theta_i + \tau}_{\theta_i} \left[g_n(x(s),y(s))-g_n(\overline{x}(s),y(s))\right] ds \\
\end{array} \right)\right\| \\
&& = \left\|\left( \begin{array}{ccc}
\tau \left[g_1(x(s^i_1),y(s^i_1))-g_1(\overline{x}(s^i_1),y(s^i_1))\right]   \\
\tau \left[g_2(x(s^i_2),y(s^i_2))-g_2(\overline{x}(s^i_2),y(s^i_2))\right]  \\
\vdots \\
\tau \left[g_n(x(s^i_n),y(s^i_n))-g_n(\overline{x}(s^i_n),y(s^i_n))\right] 
\end{array} \right)\right\|.
\end{eqnarray*}

Hence, the inequality $(\ref{liyorke_3})$ gives us that
\begin{eqnarray*}
&& \left\|\displaystyle\int^{\theta_i + \tau}_{\theta_i} \left[g(x(s),y(s))-g(\overline{x}(s),y(s))\right] ds \right\| \nonumber\\
&& \geq \tau \left|g_{j_i}(x(s^i_{j_i}),y(s^i_{j_i}))-g_{j_i}(\overline{x}(s^i_{j_i}),y(s^i_{j_i}))\right| \\
&& > \frac{\tau  L_1 \epsilon_0}{2n}.
\end{eqnarray*}

For $t\in [\theta_i,\theta_i+\tau],$ the functions $y(t)\in \mathscr{A}_y$ and $\overline{y}(t)\in \mathscr{A}_y$ satisfy the relations
\begin{eqnarray*}
y(t)= y(\theta_i)+ \displaystyle\int^{t}_{\theta_i} Ay(s) ds + \displaystyle\int^{t}_{\theta_i} g(x(s),y(s)) ds,  \nonumber
\end{eqnarray*}
and
\begin{eqnarray*}
\overline{y}(t)= \overline{y}(\theta_i)+ \displaystyle\int^{t}_{\theta_i} A\overline{y}(s) ds + \displaystyle\int^{t}_{\theta_i} g(\overline{x}(s),\overline{y}(s)) ds,  \nonumber
\end{eqnarray*}
respectively, and herewith the equation
\begin{eqnarray*}
&& y(t)-\overline{y}(t) = (y(\theta_i)-\overline{y}(\theta_i))  + \displaystyle\int^{t}_{\theta_i} A(y(s)-\overline{y}(s)) ds \nonumber\\
&& + \displaystyle\int^{t}_{\theta_i}  [g(x(s),y(s))-g(\overline{x}(s),y(s))]  ds \\
&& + \displaystyle\int^{t}_{\theta_i}  [g(\overline{x}(s),y(s))-g(\overline{x}(s),\overline{y}(s))]  ds
\end{eqnarray*}
is achieved.
Taking $t=\theta_i+\tau$ in the last equation, we attain the inequality
\begin{eqnarray} \label{liyorke_4}
\begin{array}{l}
\left\|y(\theta_i+\tau)-\overline{y}(\theta_i+\tau)\right\| \geq  \left\|\displaystyle\int^{\theta+\tau}_{\theta}  [g(x(s),y(s))-g(\overline{x}(s),y(s))]ds\right\| \\
- \left\|y(\theta_i)-\overline{y}(\theta_i)\right\| -  \displaystyle\int^{\theta_i+\tau}_{\theta_i} \left(\left\|A\right\| +L_3 \right) \left\|y(s)-\overline{y}(s)\right\| ds  \\
\end{array}
\end{eqnarray}

Now, assume that $\displaystyle \max_{t\in [\theta_i,\theta_i+\tau]}\left\|y(t)-\overline{y}(t)\right\| \leq \frac{\tau L_1 \epsilon_0}{2n[2+\tau (L_3 + \left\|A\right\|)]}.$
In this case, one arrives at a contradiction since, by means of the inequalities $(\ref{liyorke_3})$ and $(\ref{liyorke_4}),$ we have 
\begin{eqnarray*}
&& \displaystyle \max_{t\in [\theta_i,\theta_i+\tau]}\left\|y(t)-\overline{y}(t)\right\| \geq \left\|y(\theta_i+\tau)-\overline{y}(\theta_i+\tau)\right\| \\
&& > \frac{\tau L_1 \epsilon_0}{2n} - [1+ \tau(L_3 + \left\|A\right\|)] \displaystyle \max_{t\in [\theta_i,\theta_i+\tau]}\left\|y(t)-\overline{y}(t)\right\| \\
&& \geq \frac{\tau L_1 \epsilon_0}{2n} - [1+ \tau(L_3 + \left\|A\right\|)] \frac{\tau L_1 \epsilon_0}{2n[2+\tau (L_3 + \left\|A\right\|)]} \\
&& = \frac{\tau L_1 \epsilon_0}{2n} \left(1-\frac{1+ \tau(L_3 + \left\|A\right\|)}{2+ \tau(L_3 + \left\|A\right\|)}\right) \\
&& = \frac{\tau L_1 \epsilon_0}{2n[2+\tau (L_3 + \left\|A\right\|)]}.
\end{eqnarray*}
Therefore, it is true that $\displaystyle \max_{t\in [\theta_i,\theta_i+\tau]}\left\|y(t)-\overline{y}(t)\right\| > \frac{\tau L_1 \epsilon_0}{2n[2+\tau (L_3 + \left\|A\right\|)]}.$

Suppose that  the real valued function $\left\|y(t)-\overline{y}(t)\right\|$ assumes its maximum on the interval $[\theta_i,\theta_i+\tau],$ at a point $\eta_i.$ In other words, for some $ \eta_i \in \left[\theta_i,\theta_i+\tau\right],$ we have
\begin{eqnarray*}
\displaystyle \max_{t \in [\theta_i,\theta_i+\tau]} \left\|y(t)-\overline{y}(t)\right\| = \left\|y(\eta_i)-\overline{y}(\eta_i)\right\|. \nonumber
\end{eqnarray*}

Making use of the integral equations
\begin{eqnarray*}
y(t)= y(\eta_i)+ \displaystyle\int^{t}_{\eta_i} Ay(s) ds + \displaystyle\int^{t}_{\eta_i} g(x(s),y(s)) ds,  \nonumber
\end{eqnarray*}
and
\begin{eqnarray*}
\overline{y}(t)= \overline{y}(\eta_i)+ \displaystyle\int^{t}_{\eta_i} A\overline{y}(s) ds + \displaystyle\int^{t}_{\eta_i} g(\overline{x}(s),\overline{y}(s)) ds,  \nonumber
\end{eqnarray*}
on the time interval $[\theta_i,\theta_i+\tau],$ one can obtain that
\begin{eqnarray*}
&& y(t)-\overline{y}(t) = (y(\eta_i)-\overline{y}(\eta_i))  + \displaystyle\int^{t}_{\eta_i} A(y(s)-\overline{y}(s)) ds \nonumber\\
&& + \displaystyle\int^{t}_{\eta_i}  [g(x(s),y(s))-g(\overline{x}(s),\overline{y}(s))]  ds.
\end{eqnarray*}
Define
$\overline{\Delta}=\displaystyle \min \left\{ \frac{\tau}{2}, \frac{\tau L_1 \epsilon_0}{8n(M\left\|A\right\|+M_0)[2+\tau (L_3 + \left\|A\right\|)]}   \right\}$
and let
$
\theta_i^1=\left\{\begin{array}{ll} \eta_i, & ~\textrm{if}~  \eta \leq \theta_i + \frac{\tau}{2}   \\
\eta_i - \tau^1, & ~\textrm{if}~  \eta_i > \theta_i + \frac{\tau}{2}  \\
\end{array} \right. .\nonumber
$

For each $t\in [\theta_i^1, \theta_i^1+\overline{\Delta}],$ we have
\begin{eqnarray} 
\begin{array}{l}
\left\|y(t)-\overline{y}(t)\right\| \geq  \left\|y(\eta_i)-\overline{y}(\eta_i)\right\| - \left|  \displaystyle\int^{t}_{\eta_i} \left\|A\right\|\left\|y(s)-\overline{y}(s)\right\| ds   \right| \nonumber  \\
-\left|  \displaystyle\int^{t}_{\eta_i} \left\|  g(x(s),y(s))-g(\overline{x}(s),\overline{y}(s))  \right\| ds  \right| \\
 > \frac{\tau L_1 \epsilon_0}{2n[2+\tau (L_3 + \left\|A\right\|)]} -2M\left\|A\right\|\tau^1-2M_0\tau^1 \\
 = \frac{\tau L_1 \epsilon_0}{2n[2+\tau (L_3 + \left\|A\right\|)]} -2\tau^1(M\left\|A\right\|+M_0) \\
 \geq \frac{\tau L_1 \epsilon_0}{4n[2+\tau (L_3 + \left\|A\right\|)]}.
\end{array}
\end{eqnarray}

The information mentioned above is true  for arbitrarily chosen the natural number $i.$ Therefore, for each $i\in \mathbb N,$ the interval $\overline{J}_i=[\theta_i^1, \theta_i^1+\overline{\Delta}]$ is a subset of $[\theta_i,\theta_i+\tau],$ and hence of $J_i.$ Moreover, for any $i \in \mathbb N,$ we have $\left\|y(t)-\overline{y}(t)\right\| > \epsilon_1,$ $t\in \overline{J}_i,$ where $\epsilon_1=\frac{\tau L_1 \epsilon_0}{4n[2+\tau (L_3 + \left\|A\right\|)]}.$ 

Consequently, according to Definition $\ref{defliyorke2},$ the couple of functions  $\left(\phi_{x(t)}(t),\phi_{\overline{x}(t)}(t) \right) \in \mathscr{A}_y \times  \mathscr{A}_y $ is frequently $(\epsilon_1,\overline{\Delta})-$separated.

The proof of the lemma is finalized. $\square$

We end up the present subsection by stating the following theorem and its corollary.
\begin{theorem}\label{morphogenesis_li-yorke_theorem}
If the set $\mathscr{A}_x$ is Li-Yorke chaotic, then the same is true for the set $\mathscr{A}_y.$
\end{theorem}

\noindent \textbf{Proof.} Assume that the set $\mathscr{A}_x$ is Li-Yorke chaotic. As we indicated in Subsection \ref{subsecperioddoubling}, it is easy to show that for any natural number $k,$ $x(t) \in \mathscr{G}_x$ is a $kT-$periodic function if and only if $\phi_{x(t)}(t) \in \mathscr{G}_y$ is also $kT-$periodic, where $\mathscr{G}_x$ and $\mathscr{G}_y$ denote the sets of all periodic functions inside $\mathscr{A}_x$ and $\mathscr{A}_y,$ respectively. Therefore, the set $\mathscr{A}_y$ admits a $kT-$periodic function for any $k\in \mathbb N.$

Now, suppose that the set  $\mathscr{C}_x$ is a scrambled set inside $\mathscr{A}_x$ and define the set 
\begin{eqnarray}
\begin{array}{l}
\mathscr{C}_y = \left\{  \phi_{x(t)}(t) ~|~ x(t) \in \mathscr{C}_x  \right\}.
\end{array}
\end{eqnarray}
Condition $(A4)$ implies that there is one-to-one correspondence between the sets $\mathscr{C}_x$ and $\mathscr{C}_y.$ Since the scrambled set $\mathscr{C}_x$ is uncountable, it is clear that the set $\mathscr{C}_y$ is also uncountable. Moreover, using the same condition one can show that no periodic functions exist inside $\mathscr{C}_y$ since no such functions exist inside $\mathscr{C}_x.$ That is, $\mathscr{C}_y \cap \mathscr{G}_y = \emptyset.$

Since each couple of functions inside $\mathscr{C}_x \times  \mathscr{C}_x $ is proximal, Lemma $\ref{proximality}$ implies the same feature for each couple of functions inside $\mathscr{C}_y \times  \mathscr{C}_y .$

Similarly, Lemma $\ref{li-yorke_separated}$ implies that if each couple of functions $\left(x(t),\overline{x}(t)\right) \in \mathscr{C}_x \times  \mathscr{C}_x $ $  \left(\mathscr{C}_x \times  \mathscr{G}_x \right)$ is frequently $(\epsilon_0,\Delta)-$separated for some positive real numbers $\epsilon_0$ and $\Delta$, then each couple of functions  $\left(y(t),\overline{y}(t) \right) \in \mathscr{C}_y \times  \mathscr{C}_y $ $  \left(\mathscr{C}_y \times  \mathscr{G}_y \right)$ is frequently $(\epsilon_1,\overline{\Delta})-$separated for some positive real numbers $\epsilon_1$ and $\overline{\Delta}.$ Consequently, the set $\mathscr{C}_y$ is a scrambled set inside $\mathscr{A}_x,$ and according to Definition \ref{defliyorke5}, $\mathscr{A}_y$ is Li-Yorke chaotic.

The proof of the theorem is accomplished. $\square$

An immediate corollary of Theorem $\ref{morphogenesis_li-yorke_theorem}$ is the following.

\begin{corollary}
If the set $\mathscr{A}_x$ is Li-Yorke chaotic, then the set $\mathscr{A}$ is chaotic in the same way.
\end{corollary}


\subsection{\textbf{Consecutive replications of chaos}}\label{consecutive_replications}

In this part of the paper, in addition to the system $(\ref{1}) + (\ref{2})$, we take into account the system 
\begin{eqnarray} \label{3}
z'=Bz+h(y,z),
\end{eqnarray}
where $h:\mathbb R^{n}\times \mathbb R^{l} \to \mathbb R^{l}$  is a continuous function in all of its arguments, and the constant $l\times l$ real valued matrix $B$ has real parts of eigenvalues all negative.

It is easy to verify that there exist positive real numbers $\widetilde{N}$ and $\widetilde{\omega}$ such that $\left\|e^{Bt}\right\| \leq \widetilde{N}e^{-\widetilde{\omega} t},$ for all $t\geq 0.$

In our next theoretical discussions, the system $(\ref{3})$ will serve as the second replicator system in the mechanism presented by Figure \ref{box1}, and we need the following assumptions which are counterparts of the conditions $(A4)-(A7)$ presented in Section $\ref{preliminaries}.$

\begin{enumerate} 
\item[\bf (A8)] There exists a positive real number $\widetilde{L}_1$ such that 
$ \left\|h(y_1,z)-h(y_2,z)\right\| \geq \widetilde{L}_1\left\|y_1-y_2\right\|,$ for all $y_1,y_2 \in \mathbb R^{n},$ $z \in \mathbb R^l;$
\item[\bf (A9)] There exist positive real numbers $\widetilde{L}_2$ and $\widetilde{L}_3$ such that 
$ 
\left\|h(y_1,z)-h(y_2,z)\right\| \leq \widetilde{L}_2\left\|y_1-y_2\right\|,$ for all $y_1,y_2 \in \mathbb R^{n},$ $z\in \mathbb R^l,$ and $\left\|h(y,z_1)-h(y,z_2)\right\| \leq \widetilde{L}_3\left\|z_1-z_2\right\|,$ for all $y\in \mathbb R^n,$ $z_1,z_2 \in \mathbb R^{l};$
\item[\bf (A10)] There exist a positive number $K_0 < \infty$ such that $\displaystyle \sup_{y\in \mathbb R^n, z\in \mathbb R^l} \left\|h(y,z)\right\|=K_0;$
\item[\bf (A11)] $\widetilde{N}\widetilde{L}_3-\widetilde{\omega} < 0.$
\end{enumerate} 


Likewise the definition for the set of functions $\mathscr{A}_y,$ given by $(\ref{A_y})$,  let us denote by $\mathscr{A}_z$ the set of all bounded on $\mathbb R$ solutions of system $z'=Bz+h(y(t),z),$ for any $y(t) \in \mathscr{A}_y.$

In a similar way to the Lemma (\ref{attractor}), one can show that the set
\begin{eqnarray}
\begin{array}{l}
\mathscr{U}_z=\left\{z(t)~|~ z(t) ~\textrm{is a solution of the system}~ z'=Az+g(y(t),z) ~\textrm{for some}~ y(t) \in \mathscr{U}_y  ~  \right\}. 
\end{array}
\end{eqnarray}
is a basin of $\mathscr{A}_z.$

Similar results of Theorem $\ref{hyperbolicity}$ introduced in Section $\ref{hyperbolic_sets},$ and of the Theorems $\ref{morphogenesis_period-doubling_theorem}$, $\ref{morphogenesis_devaney_theorem}$ and $\ref{morphogenesis_li-yorke_theorem}$ presented in Subsections $\ref{subsecperioddoubling},$ $\ref{devaney}$ and $\ref{secliyorke},$ respectively, hold also for the set $\mathscr{A}_z,$ and we state this result in the next theorem.

We note that, in the presence of arbitrary finite number of consecutive replicator systems, which obey conditions that are counterparts of $(A4)-(A7),$ one can prove that a similar result of the next theorem also holds.

\begin{theorem}\label{second_slave_theorem}
If the set $\mathscr{A}_x$ is chaotic through period-doubling cascade or Devaney chaotic or Li-Yorke chaotic, then the set $\mathscr{A}_z$ is chaotic in the same way as both $\mathscr{A}_x$ and $\mathscr{A}_y.$
\end{theorem}

\noindent \textbf{Proof.} 
In the proof, we will show that for each $z(t) \in \mathscr{A}_z$ and arbitrary $\delta >0,$ there exist $\overline{z}(t) \in \mathscr{A}_z$ and $t_0 \in \mathbb R$ such that $\left\|z(t_0)-\overline{z}(t_0)\right\|<\delta,$ which is needed to show sensitivity of $\mathscr{A}_z.$ The remaining parts of the proof can be performed in a similar way to the proofs presented in Subsections $\ref{subsecperioddoubling},$ $\ref{devaney}$ and $\ref{secliyorke},$ and therefore are omitted.

Suppose that the set $\mathscr{A}_x$ is sensitive. Fix an arbitrary $\delta>0$ and let $z(t) \in \mathscr{A}_z$ be a given solution of system $(\ref{3}).$ In this case, there exists $y(t)=\phi_{x(t)}(t) \in \mathscr{A}_y,$ where $x(t) \in \mathscr{A}_x,$ such that $z(t)$ is the unique bounded on $\mathbb R$ solution of the system $z'=Bz+h(y(t),z).$

Let us choose a positive number $\overline{\epsilon}=\overline{\epsilon}(\delta)$ small enough such that the inequality 
\begin{eqnarray*}
\left( 1+ \frac{\widetilde{N}\widetilde{L}_2}{\widetilde{\omega}-\widetilde{N}\widetilde{L}_3} \right) \left(1+\frac{NL_2}{\omega-NL_3}\right)\overline{\epsilon} <\delta
\end{eqnarray*}
holds, and denote $\epsilon_1=\left(1+\frac{NL_2}{\omega-NL_3}\right)\overline{\epsilon}.$ Now, take $R=R(\overline{\epsilon})<0$ sufficiently large in absolute value such that both the inequalities $\frac{2M_0N}{\omega}e^{-(NL_3-\omega)\frac{R}{2}} \leq \overline{\epsilon}$ and $\frac{2\widetilde{M}_0\widetilde{N}}{\widetilde{\omega}}e^{-( \widetilde{N}\widetilde{L}_3-\widetilde{\omega})\frac{R}{2}} \leq \epsilon_1$ are valid, and let $\delta_1=\delta_1(\overline{\epsilon},R)=\overline{\epsilon}e^{L_0 R}.$
Since the set $\mathscr{A}_x$ is sensitive, one can find $\overline{x}(t)\in \mathscr{A}_x$ and $t_0 \in \mathbb R$ such that the inequality $\left\|x(t_0)-\overline{x}(t_0)\right\|<\delta_1$ holds.

As shown in the proof of Lemma $\ref{sensitivity},$ for $t \in [t_0+R, t_0],$ one has  
\begin{eqnarray*}
\left\|x(t)-\overline{x}(t)\right\|<\overline{\epsilon},
\end{eqnarray*}
and
\begin{eqnarray*}
\left\|y(t)-\overline{y}(t)\right\| \leq \frac{NL_2\overline{\epsilon}}{\omega-NL_3} + \frac{2M_0N}{\omega}e^{(NL_3-\omega)(t-t_0-R)}.
\end{eqnarray*}

As a consequence of the last inequality, we have 
\begin{eqnarray*}
\left\|y(t)-\overline{y}(t)\right\| \leq \epsilon_1,
\end{eqnarray*}
for $t\in [t_0+ \frac{R}{2}, t_0].$ 

Suppose that $\overline{z}(t)$ is the unique bounded on $\mathbb R$ solution of the system $z'=Bz+h(\overline{y}(t),z).$ One can see that the relations 
\begin{eqnarray*} 
z(t) = \displaystyle\int^{t}_{-\infty} e^{B(t-s)}h(y(s),z(s))ds
\end{eqnarray*}
and 
\begin{eqnarray*} 
\overline{z}(t) = \displaystyle\int^{t}_{-\infty} e^{B(t-s)}h(\overline{y}(s),\overline{z}(s))ds,
\end{eqnarray*}
are valid. 

Therefore,
\begin{eqnarray*}
z(t)-\overline{z}(t)=\displaystyle\int^{t}_{-\infty} e^{B(t-s)} [h(y(s),z(s))-h(\overline{y}(s),\overline{z}(s))]ds
\end{eqnarray*}
and hence we obtain
\begin{eqnarray*}
&&\left\|z(t)-\overline{z}(t)\right\| \leq \displaystyle\int^{t}_{t_0+\frac{R}{2}} \widetilde{N} e^{-\widetilde{\omega}(t-s)} \left\|h(y(s),z(s))-h(y(s),\overline{z}(s))\right\|ds \\
&& + \displaystyle\int^{t}_{t_0+\frac{R}{2}} \widetilde{N} e^{-\widetilde{\omega}(t-s)} \left\|h(y(s),\overline{z}(s))-h(\overline{y}(s),\overline{z}(s))\right\|ds \\
&& + \displaystyle\int^{t_0+\frac{R}{2}}_{-\infty} \widetilde{N}e^{-\widetilde{\omega}(t-s)} \left\|h(y(s),z(s))-h(\overline{y}(s),\overline{z}(s))\right\|ds. 
\end{eqnarray*}
Since $\left\|y(t)-\overline{y}(t)\right\|<\epsilon_1$ for $t \in [t_0+\frac{R}{2}, t_0],$ one has
\begin{eqnarray*}
&& \left\|z(t)-\overline{z}(t)\right\| \leq \widetilde{N}\widetilde{L}_3\displaystyle\int^{t}_{t_0+\frac{R}{2}} e^{-\widetilde{\omega}(t-s)} \left\|z(s)-\overline{z}(s)\right\|ds \\
&& + \widetilde{N}\widetilde{L}_2 \epsilon_1 \displaystyle\int^{t}_{t_0+\frac{R}{2}} e^{-\widetilde{\omega}(t-s)}ds + 2\widetilde{M}_0\widetilde{N}\displaystyle\int^{t_0+\frac{R}{2}}_{-\infty}e^{-\widetilde{\omega}(t-s)}ds \\
&& \leq \widetilde{N}\widetilde{L}_3\displaystyle\int^{t}_{t_0+\frac{R}{2}} e^{-\widetilde{\omega}(t-s)} \left\|z(s)-\overline{z}(s)\right\|ds \\
&& + \frac{\widetilde{N}\widetilde{L}_2\epsilon_1}{\widetilde{\omega}}e^{-\widetilde{\omega} t} (e^{\widetilde{\omega} t}-e^{\widetilde{\omega} (t_0+\frac{R}{2})})+\frac{2\widetilde{M}_0\widetilde{N}}{\widetilde{\omega}}e^{-\widetilde{\omega}(t-t_0-\frac{R}{2})}.
\end{eqnarray*}

Now, introduce the functions $u(t)=e^{\widetilde{\omega} t}\left\|z(t)-\overline{z}(t)\right\|,$ $k(t)=\frac{\widetilde{N}\widetilde{L}_2\epsilon_1}{\widetilde{\omega}}e^{\widetilde{\omega} t},$  and  $v(t)=c+ k(t)$ where $c=\left(\frac{2\widetilde{M}_0\widetilde{N}}{\widetilde{\omega}}-\frac{\widetilde{N}\widetilde{L}_2\epsilon_1}{\widetilde{\omega}}\right)e^{\widetilde{\omega} (t_0+\frac{R}{2})}.$ 

These definitions give us the inequality 
\begin{eqnarray*}
u(t)\leq v(t)+\displaystyle\int^{t}_{t_0+\frac{R}{2}} \widetilde{N} \widetilde{L}_3 u(s)ds.
\end{eqnarray*}
Applying Lemma $2.2$ \cite{Bar70} to the last inequality, one can verify that 
\begin{eqnarray*}
u(t)\leq v(t)+\widetilde{N}\widetilde{L}_3\displaystyle\int^{t}_{t_0+\frac{R}{2}} e^{\widetilde{N}\widetilde{L}_3(t-s)}h(s)ds.
\end{eqnarray*}

Therefore, for $t\in [t_0+\frac{R}{2}, t_0]$ we have
\begin{eqnarray*}
&&u(t) \leq c+k(t)+c\left(e^{\widetilde{N}\widetilde{L}_3(t-t_0-\frac{R}{2})}-1\right) + \frac{N^{2}\widetilde{L}_2\widetilde{L}_3\epsilon_1}{\widetilde{\omega}}e^{\widetilde{N}\widetilde{L}_3t} \displaystyle\int^{t}_{t_0+\frac{R}{2}} e^{(\widetilde{\omega} -\widetilde{N}\widetilde{L}_3)s}ds \\
&& = k(t) +ce^{\widetilde{N}\widetilde{L}_3(t-t_0-\frac{R}{2})} +\frac{\widetilde{N}^{2}\widetilde{L}_2\widetilde{L}_3\epsilon_1}{\widetilde{\omega}(\widetilde{\omega}-\widetilde{N}\widetilde{L}_3)}e^{\widetilde{\omega} t} \left[1-e^{(\widetilde{N}\widetilde{L}_3-\widetilde{\omega})(t-t_0-\frac{R}{2})}\right] \\
&& = \frac{\widetilde{N}\widetilde{L}_2\epsilon_1}{\widetilde{\omega}}e^{\widetilde{\omega} t} + \left(\frac{2\widetilde{M}_0N}{\widetilde{\omega}}-\frac{\widetilde{N}\widetilde{L}_2\epsilon_1}{\widetilde{\omega}}\right)e^{\widetilde{\omega} T}e^{\widetilde{N}\widetilde{L}_3(t-t_0-\frac{R}{2})} \\
&& + \frac{\widetilde{N}^2\widetilde{L}_2\widetilde{L}_3 \epsilon_1}{\widetilde{\omega}(\widetilde{\omega}-\widetilde{N}\widetilde{L}_3)}e^{\widetilde{\omega} t} \left[1-e^{(\widetilde{N}\widetilde{L}_3-\widetilde{\omega})(t-t_0-\frac{R}{2})}\right].
\end{eqnarray*}

Hence,
\begin{eqnarray*}
&&\left\|z(t)-\overline{z}(t)\right\| \leq \frac{\widetilde{N}\widetilde{L}_2\epsilon_1}{\widetilde{\omega}} + \left(\frac{2\widetilde{M}_0\widetilde{N}}{\widetilde{\omega}}-\frac{\widetilde{N}\widetilde{L}_2\epsilon_1}{\widetilde{\omega}}\right)e^{(\widetilde{N}\widetilde{L}_3-\widetilde{\omega})(t-t_0-\frac{R}{2})} \\
&& + \frac{\widetilde{N}^2\widetilde{L}_2\widetilde{L}_3\epsilon_1}{\widetilde{\omega}(\widetilde{\omega}-\widetilde{N}\widetilde{L}_3)} \left[1-e^{(\widetilde{N}\widetilde{L}_3-\widetilde{\omega})(t-t_0-\frac{R}{2})}\right] \\
&& = \frac{\widetilde{N}\widetilde{L}_2\epsilon_1}{\widetilde{\omega}-\widetilde{N}\widetilde{L}_3} + \left(\frac{2\widetilde{M}_0N}{\widetilde{\omega}}-\frac{\widetilde{N}\widetilde{L}_2\epsilon_1}{\widetilde{\omega}-\widetilde{N}\widetilde{L}_3}\right) e^{(\widetilde{N}\widetilde{L}_3-\widetilde{\omega})(t-t_0-\frac{T}{2})} \\
&& = \frac{\widetilde{N}\widetilde{L}_2\epsilon_1}{\widetilde{\omega}-\widetilde{N}\widetilde{L}_3}\left[1-e^{(\widetilde{N}\widetilde{L}_3-\widetilde{\omega})(t-t_0-\frac{R}{2})}\right] + \frac{2\widetilde{M}_0N}{\widetilde{\omega}}e^{(\widetilde{N}\widetilde{L}_3-\widetilde{\omega})(t-t_0-\frac{R}{2})}.
\end{eqnarray*}

Consequently,
\begin{eqnarray*}
&& \left\|z(t_0)-\overline{z}(t_0)\right\| \leq \frac{\widetilde{N}\widetilde{L}_2\epsilon_1}{\widetilde{\omega}-\widetilde{N}\widetilde{L}_3} + \frac{2\widetilde{M}_0\widetilde{N}}{\widetilde{\omega}}e^{(\widetilde{\omega}-\widetilde{N}\widetilde{L}_3)\frac{R}{2}} \\
&& < \left( 1+ \frac{\widetilde{N}\widetilde{L}_2}{\widetilde{\omega}-\widetilde{N}\widetilde{L}_3} \right) \epsilon_1 \\
&& < \delta.
\end{eqnarray*}

The theorem is proved. $\square$

\section{Controlling morphogenesis of chaos}
In previous section we have theoretically proved self-replication of chaos for specific types. Controlling the self-replicated chaos is another interesting problem and in this section we give an answer. The next theorem and its corollary indicates a method to control the chaos of the replicator system $(\ref{2})$ and the result-system $(\ref{1})+(\ref{2}),$ respectively,  and reveals that controlling the chaos of system $(\ref{1})$ is sufficient for this.  

\begin{theorem}\label{control_of_chaos}
Assume that for arbitrary $\epsilon >0,$ a periodic solution $x_p(t) \in \mathscr{A}_x$ is stabilized such that for any solution $x(t)$ of system $(\ref{1})$ there exist real numbers $a$ and $E>0$ such that the inequality $\left\|x(t)-x_p(t)\right\| < \epsilon$ holds for $t\in[a,a+E].$  

Then, the periodic solution $\phi_{x_p(t)}(t) \in \mathscr{A}_y$ is stabilized such that for any solution $y(t)$ of system $(\ref{2})$ there exists a number $b\geq a$ such that the inequality $\left\|y(t)-\phi_{x_p(t)}(t)\right\|<\left(1+\frac{NL_2}{\omega-NL_3}\right)\epsilon$ holds for $t\in[b,a+E],$ provided that $E$ is sufficiently large.
\end{theorem}

\noindent \textbf{Proof.} Fix an arbitrary solution $y(t)$ of system $y'=Ay+g(x(t),y)$ for some solution $x(t)$ of system $(\ref{1}).$ According to our assumption,  there exist numbers $a$ and $E>0$ such that the inequality $\left\|x(t)-x_p(t)\right\| < \epsilon$ holds for $t\in[a,a+E].$  
Let us denote $y_p(t)=\phi_{x_p(t)}(t) \in \mathscr{A}_y.$ It is clear that the function $y_p(t)$ is periodic with the same period as $x_p(t).$
Since $y(t)$ and $y_p(t)$ satisfy the integral equations
\begin{eqnarray*}
y(t)= e^{A(t-a)}y(a)+ \displaystyle\int^{t}_{a}e^{A(t-s)}g(x(s),y(s))ds,  \nonumber
\end{eqnarray*}
and
\begin{eqnarray*}
y_p(t)= e^{A(t-a)}y_p(a)+ \displaystyle\int^{t}_{a}e^{A(t-s)}g(x_p(s),y_p(s))ds,  \nonumber
\end{eqnarray*}
respectively, one has
\begin{eqnarray*}
&&y(t)-y_p(t) = e^{A(t-a)} (y(a)-y_p(a)) \nonumber\\
&& + \displaystyle\int^{t}_{a} e^{A(t-s)}  \left[g(x(s),y(s))-g(x_p(s),y_p(s))\right]  ds   \\
&& = e^{A(t-a)} (y(a)-y_p(a)) \\
&& + \displaystyle\int^{t}_{a} e^{A(t-s)}  \left[g(x(s),y(s))-g(x(s),y_p(s))\right]  ds \\
&& + \displaystyle\int^{t}_{a} e^{A(t-s)}  \left[g(x(s),y_p(s))-g(x_p(s),y_p(s))\right]  ds.                               
\end{eqnarray*}

By the help of the last equation, we have

\begin{eqnarray*}
&&\left\|y(t)-y_p(t)\right\| \leq \left\|e^{A(t-a)}\right\| \left\|y(a)-y_p(a)\right\| \\
&& + \displaystyle\int^{t}_{a} \left\|e^{A(t-s)}\right\|  \left\|g(x(s),y(s))-g(x(s),y_p(s))\right\|  ds  \nonumber\\
&& + \displaystyle\int^{t}_{a} \left\|e^{A(t-s)}\right\|  \left\|g(x(s),y_p(s))-g(x_p(s),y_p(s))\right\|  ds \\
&& \leq Ne^{-\omega(t-a)} \left\|y(a)-y_p(a)\right\| \\
&& + \displaystyle\int^{t}_{a} Ne^{-\omega (t-s)} L_3 \left\|y(s)-y_p(s)\right\| ds \\
&& + \displaystyle\int^{t}_{a} Ne^{-\omega (t-s)} L_2 \left\|x(s)-x_p(s)\right\| ds \\
&& \leq Ne^{-\omega(t-a)} \left\|y(a)-y_p(a)\right\| \\
&& + \frac{NL_2\epsilon}{\omega}e^{-\omega t}\left(e^{\omega t}-e^{\omega a}\right) \\
&& + NL_3 \displaystyle\int^{t}_{a} e^{-\omega (t-s)} \left\|y(s)-y_p(s)\right\| ds.
\end{eqnarray*}

Let $u:[a,a+E]\to \mathbb R_+$ be a function defined as $u(t)=e^{\omega t}\left\|y(t)-y_p(t)\right\|.$
Then we reach the inequality
\begin{eqnarray*}
u(t) \leq Ne^{\omega a}\left\|y(a)-y_p(a)\right\| + \frac{NL_2\epsilon}{\omega}\left(e^{\omega t}-e^{\omega a}\right) + NL_3 \displaystyle\int^{t}_{a} u(s) ds.                         
\end{eqnarray*}

Now, define $\psi:[a,a+E]\to \mathbb R_+$ as $\psi(t)=\frac{NL_2\epsilon}{\omega}e^{\omega t}$ and similarly $\phi:[a,a+E]\to \mathbb R_+$ as $\phi(t)=\psi(t) + c$ where $c=Ne^{\omega a}\left\|y(a)-y_p(a)\right\|-\frac{NL_2\epsilon}{\omega}e^{\omega a}.$
By means of these definitions we get
\begin{eqnarray*}
u(t) \leq \phi(t)+NL_3\displaystyle\int^{t}_{a} u(s) ds .   \nonumber                      
\end{eqnarray*}
Implementation of Lemma $2.2$ \cite{Bar70} to the last inequality, where $t\in [a,a+E],$ provides us the inequality
\begin{eqnarray*}
u(t) \leq c+\psi(t)+NL_3\displaystyle\int^{t}_{a} e^{NL_3(t-s)}c ds+NL_3\displaystyle\int^{t}_{a} e^{NL_3(t-s)}\psi(s) ds \nonumber
\end{eqnarray*}
and therefore,
\begin{eqnarray*}
&&u(t) \leq c + \psi(t) + c\left(e^{NL_3(t-a)}-1\right) + \frac{N^{2}L_2L_3\epsilon}{\omega(\omega-NL_3)}e^{\omega t} \left(1-e^{(NL_3-\omega)(t-a)}\right) \nonumber\\
&& = \frac{NL_2\epsilon}{\omega}e^{\omega t} + N\left\|y(a)-y_p(a)\right\|e^{\omega a}e^{NL_3(t-a)} \\
&& -\frac{NL_2\epsilon}{\omega}e^{\omega a}e^{NL_3(t-a)} + \frac{N^{2}L_2L_3\epsilon}{\omega(\omega-NL_3)}e^{\omega t} \left(1-e^{(NL_3-\omega)(t-a)}\right).
\end{eqnarray*}
Consequently,
\begin{eqnarray*}
&& \left\|y(t)-y_p(t)\right\|  \leq \frac{NL_2\epsilon}{\omega} + N\left\|y(a)-y_p(a)\right\|e^{(NL_3-\omega)(t-a)} \nonumber\\
&& -\frac{NL_2\epsilon}{\omega}e^{(NL_3-\omega)(t-a)} + \frac{N^{2}L_2L_3\epsilon}{\omega(\omega-NL_3)} \left(1-e^{(NL_3-\omega)(t-a)}\right)\\
&& = N\left\|y(a)-y_p(a)\right\|e^{(NL_3-\omega)(t-a)}\\
&& + \frac{NL_2\epsilon}{\omega}\left(1-e^{(NL_3-\omega)(t-a)}\right) + \frac{N^2L_2L_3\epsilon}{\omega(\omega-NL_3)}\left(1-e^{(NL_3-\omega)(t-a)}\right)\\
&& = N\left\|y(a)-y_p(a)\right\|e^{(NL_3-\omega)(t-a)}+ \frac{NL_2\epsilon}{\omega-NL_3}\left(1-e^{(NL_3-\omega)(t-a)}\right)\\
&& < N\left\|y(a)-y_p(a)\right\|e^{(NL_3-\omega)(t-a)}+ \frac{NL_2\epsilon}{\omega-NL_3}.
\end{eqnarray*}

If $y(a)=y_p(a),$ then clearly $\left\|y_p(t)-y(t)\right\|<\left(1+\frac{NL_2}{\omega-NL_3}\right)\epsilon,$ $t \in [a,a+E]$. Suppose that $y(a)\neq y_p(a).$ For $t \geq a + \frac{1}{NL_3-\omega} \ln\left(\frac{\epsilon}{N\left\|y(a)-y_p(a)\right\|}\right),$ the inequality $e^{(NL_3-\omega)(t-a)} \leq \frac{\epsilon}{N\left\|y(a)-y_p(a)\right\|}$ is satisfied. Assume that the number $E$ is sufficiently large so that $ E>\frac{1}{NL_3-\omega}\ln\left(\frac{\epsilon}{N\left\|y(a)-y_p(a)\right\|}\right).$  Thus, taking
$ b=\max\left\{a,a+\frac{1}{NL_3-\omega}\ln\left(\frac{\epsilon}{N\left\|y(a)-y_p(a)\right\|}\right)\right\} $
and 
$\widetilde{E}=\min\left\{E,E-\frac{1}{NL_3-\omega}\ln\left(\frac{\epsilon}{N\left\|y(a)-y_p(a)\right\|}\right)\right\}$
one attains that $\left\|y(t)-y_p(t)\right\| < \left(\frac{\omega-NL_3+NL_2}{\omega-NL_3}\right)\epsilon,$ for $t \in [b,b+\widetilde{E}].$
Here the number $\widetilde{E}$ stands for the duration of control for system $(\ref{2}).$ We note that $b \geq a,$ $0<\widetilde{E} \leq E$ and $b+\widetilde{E}=a+E.$

Hence $\left\|y(t)-y_p(t)\right\| \leq \left(1+\frac{NL_2}{\omega-NL_3}\right)\epsilon,$ for $t \in [b,a+E].$

The proof is finalized. $\square$

An immediate corollary of Theorem \ref{control_of_chaos} is the following. 

\begin{corollary}
Assume that the conditions of Theorem $\ref{control_of_chaos}$ hold. In this case, the periodic solution $z_p(t)=\left(x_p(t),\phi_{x_p(t)}(t)\right) \in \mathscr{A}$ is stabilized such that for any solution $z(t)$ of system $(\ref{1})+(\ref{2})$ there exists a number $b\geq a$ such that the inequality $\left\|z_p(t)-z(t)\right\|<\left(2+\frac{NL_2}{\omega-NL_3}\right)\epsilon$ holds for $t\in[b,a+E],$ provided that $E$ is sufficiently large. 
\end{corollary}
\noindent \textbf{Proof.}
On account of the inequality $ \left\|z(t)-z_p(t)\right\|  \leq \left\|x(t) - x_p(t)\right\| + \left\| y(t) - \phi_{x_p(t)}(t)  \right\|,$ and using the conclusion of Theorem \ref{control_of_chaos} one can show that $\left\|z_p(t)-z(t)\right\|<\left(2+\frac{NL_2}{\omega-NL_3}\right)\epsilon$ holds for $t\in[b,a+E]$ and for some $b \geq a.$

\begin{remark}
As a conclusion of the Theorem \ref{control_of_chaos}, the transient time for control to take effect may increase and the duration of control may decrease as the number of consecutive replicator systems increase.
\end{remark}

In the remaining part of this section, our aim is to present an illustration which confirms the results of Theorem \ref{control_of_chaos}, and for our purposes, we will make use of the Pyragas control method \cite{Pyragas92}. Therefore, primarily, we continue with a brief explanation of this method.

A delayed feedback control method for the stabilization of unstable periodic orbits of a chaotic system was proposed by Pyragas \cite{Pyragas92}. In this method, one considers a system of the form
\begin{eqnarray}
\begin{array}{l}
x'=H(x,q), \label{Duffing_pyragas1} 
\end{array}
\end{eqnarray}
where $q=q(t)$ is an externally controllable parameter and for $q=0$ it is assumed that the system $(\ref{Duffing_pyragas1})$ is in the chaotic state of interest, whose periodic orbits are to be stabilized \cite{Gon04,Zelinka,Pyragas92,Fra07}. According to Pyragas method, an unstable periodic solution with period $\xi$ of the system $(\ref{Duffing_pyragas1})$ with $q=0,$ can be stabilized by the control law $q(t)=C \left[ s\left(t- \xi\right) -s(t) \right],$ where the parameter $C$ represents the strength of the perturbation and $s(t)= \sigma \left[x(t)\right]$ is a scalar signal given by some function of the state of the system.

It is indicated in \cite{Gon04} that in order to apply the Pyragas control method to the chaotic Duffing oscillator
\begin{eqnarray}
\begin{array}{l}
x''+0.10 x'- \frac{1}{2} x \left( 1 - x^2 \right)=0.24 \displaystyle \sin(t),  \label{Duffing_pyragas3} \\
\end{array}
\end{eqnarray}
one can make use of the new variables $x_1=x, x_2=x', x_3=t,$ to  construct the system
\begin{eqnarray}
\begin{array}{l}
x_1'=x_2 \\  \label{Duffing_pyragas4}
x_2'=-0.10 x_2 + \frac{1}{2} x_1 \left( 1-x_1^2  \right) + 0.24 \displaystyle \sin(x_3) + C \left[ x_2(t-\xi) -x_2(t)  \right] \\
x_3'=1,
\end{array}
\end{eqnarray}
where $q(t)=C \left[ x_2(t-\xi) -x_2(t)  \right]$ is the control law and the less unstable $2\pi-$periodic solution can be stabilized by choosing the parameter values $C=0.36$ and $\xi=2\pi.$

Now, we shall indicate an implementation of Theorem  \ref{control_of_chaos} through simulations. Let us combine system (\ref{Duffing_pyragas4})  with two consecutive replicator systems and set up the $9-$dimensional result-system

\begin{eqnarray}
\begin{array}{l}
x_1'=x_2 \\  \label{Duffing_pyragas5}
x_2'=-0.10 x_2 + \frac{1}{2} x_1 \left( 1-x_1^2  \right) + 0.24 \displaystyle \sin(x_3) + C \left[ x_2(t-\xi) -x_2(t)  \right] \\
x_3'=1  \\
x_4'=x_5-0.1x_1 \\
x_5'=-3x_4-2x_5-0.008x_4^3+1.6x_2 \\
x_6'=-x_6+x_3 \\ 
x_7'=x_8+0.6x_4 \\
x_8'=-3.1x_7-2.1x_8-0.007x_7^3+2.5x_5 \\
x_9'=-x_9+0.2x_6.
\end{array}
\end{eqnarray}

Since our procedure of morphogenesis is valid for specific types of chaos such as in Devaney's and Li-Yorke sense and through period-doubling cascade, we expect that our procedure is also applicable to any other chaotic system with an unspecified type of chaos. Accordingly, the system $(\ref{Duffing_pyragas5})$ is chaotic for the parameter value $C=0,$ since the system (\ref{Duffing_pyragas4}) admits the chaos for the same value of the parameter.

\begin{figure}[ht] 
\centering
\includegraphics[width=13.5cm]{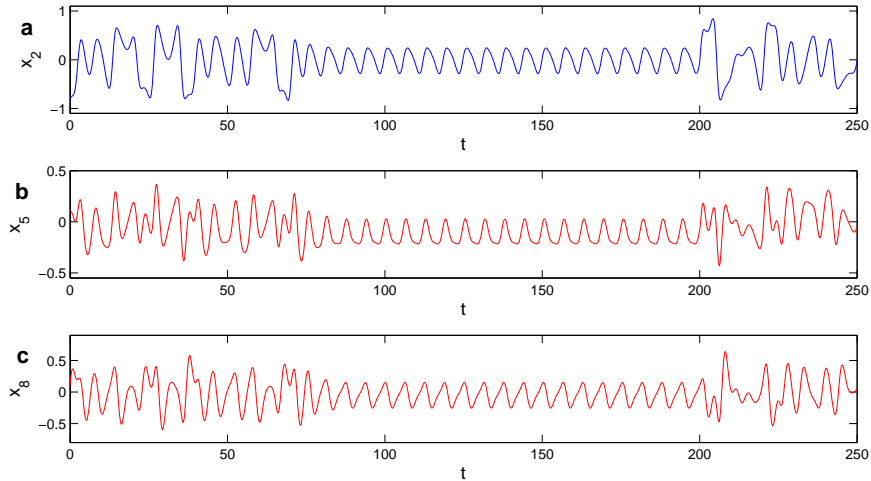}
\caption{\footnotesize{Pyragas control method  applied to the result-system $(\ref{Duffing_pyragas5}).$ (a)  The graph of $x_2-$coordinate. (b) The graph of  $x_5-$coordinate. (c) The graph of  $x_8-$coordinate. The result of Pyragas control method applied to the generator system $(\ref{Duffing_pyragas4})$ is seen in $(a).$ Through this method, the $2\pi-$periodic solution of the generator and accordingly the $2\pi-$periodic solutions of the first and the second replicator systems are stabilized. In other words, the morphogenesis of chaos is controlled. For the coordinates $x_1,x_4$ and $x_7$ we have similar results which are not just pictured here. The control starts at $t=70$ and ends at $t=200,$ after which emergence of the chaos is observable again.}}
\label{pyragas}
\end{figure}

Let us consider a solution of system $(\ref{Duffing_pyragas5})$ with the initial data $x_1(0)=x_5(0)=x_8(0)=0.1, x_2(0)=-0.8, x_3(0)=0, x_4(0)=-0.5, x_6(0)=-0.1, x_7(0)=-0.2, x_9(0)=-0.3.$ We let the system evolve freely taking $C=0$ until $t=70,$ and at that moment  we switch on the control  by taking $C=0.36.$ At time $t=200,$ we switch off the control and start to use the value of the parameter $C=0$ again. In Figure $\ref{pyragas}$ one can see the graphs of the $x_2, x_5, x_8$ coordinates of the solution. After switching off the control, the $2\pi-$periodic solution loses its stability and chaos emerges again.

\section{Discussion}

In this part of our paper, we intend to consider not rigorously proved, but interesting phenomena which can be considered in the framework morphogenesis of chaos. So we shall discuss again obtained above results and say about the possibility of the morphogenesis for other type of chaos, for intermittency, for relay systems, and discuss the possibility of quasiperiodic motions as an infinite basis of chaos.

The type of chaos for the double-scroll Chua circuit is proposed in paper \cite{Chua86}. It is an interesting problem to prove that this type of chaos can be replicated through the method discussed in our paper. Nevertheless, let us show by simulations that the regular behavior in replicator systems can be also seen. This means that next special investigation has to be done. 

Moreover, this will show how one can use morphogenesis not only for chaos, but also for Chua circuits by uniting them in complexes in electrical (physical) sense, and observing same properties as unique separated Chua circuit admits. This is an interesting problem which can give a light for the complex behavior of huge electrical circuits.

We start our discussions with morphogenesis of intermittency.

\subsection{\textbf{Morphogenesis of intermittency}}\label{intermittency}

In Section \ref{self-replication}, we have rigorously proved self-replication of specific types of chaos such as period-doubling cascade, Devaney's and Li-Yorke chaos. Consequently, one can expect that the same procedure also works for the intermittency route.

\begin{figure}[ht] 
\centering
\includegraphics[width=13.7cm]{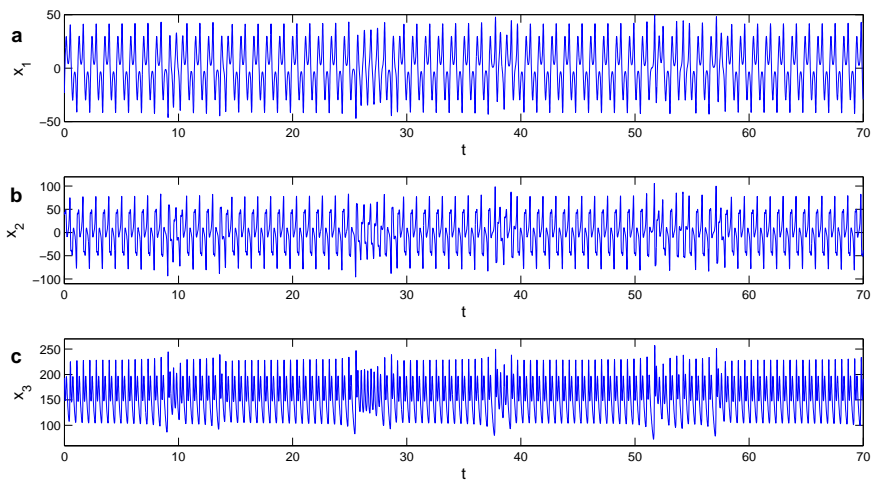}
\caption{\footnotesize{Intermittency in the Lorenz system (\ref{Lorenz_system}), where $\sigma=10,$ $b=8/3$ and $r=166.25.$ (a) The graph of the $x_1-$coordinate, (b) The graph of the $x_2-$coordinate, (c) The graph of the $x_3-$coordinate.}}
\label{intermittency_fig1}
\end{figure}

Pomeau and Manneville \cite{Pomeau80} observed chaos through intermittency in the Lorenz system $(\ref{Lorenz_system}),$ with the coefficients $\sigma=10,$ $b=8/3$ and values of $r$ slightly larger than the critical value $r_c\approx166.06.$ To observe intermittent behavior in the Lorenz system, let us consider a solution of system $(\ref{Lorenz_system})$ together with the coefficients $\sigma=10,$ $b=8/3,$ $r=166.25$ using the initial data $x_1(0)=-23.3,$ $x_2(0)= 38.3$ and $x_3(0)= 193.4.$  The trajectories of the $x_1,$ $x_2$ and $x_3$ coordinates of the solution are indicated in Figure $\ref{intermittency_fig1},$ where one can see that regular oscillations are interrupted by irregular ones.

\begin{figure}[ht] 
\centering
\includegraphics[width=13.5cm]{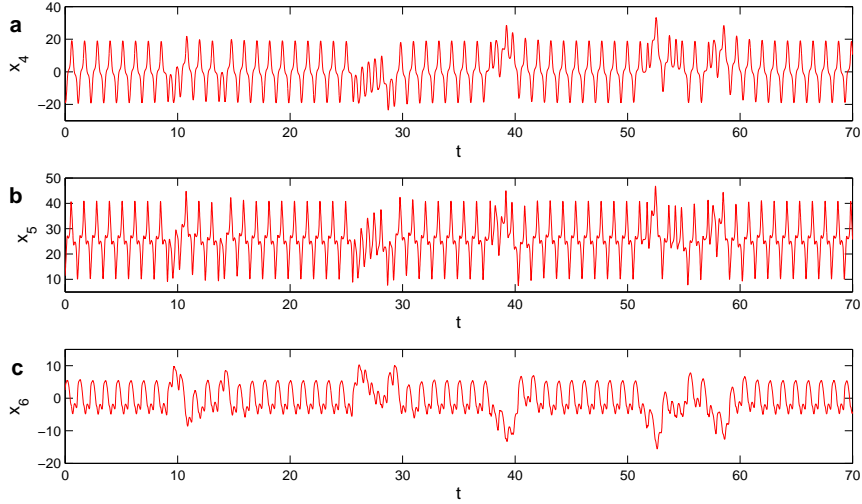}
\caption{\footnotesize{Intermittency in the replicator system.  (a) The graph of the $x_4-$coordinate, (b) The graph of the $x_5-$coordinate, (c) The graph of the $x_6-$coordinate. The analogy between the time-series of the generator and the replicator systems indicates the morphogenesis of intermittency.}}
\label{intermittency_fig2}
\end{figure}

To perform morphogenesis of intermittency, let us consider the Lorenz system $(\ref{Lorenz_system})$ as a generator and set up the $6-$dimensional result-system
\begin{eqnarray} 
\begin{array}{l}
x'_1=\sigma \left(  -x_1 + x_2 \right) \\  \label{replication_of_intermittency}
x'_2=-x_2 +rx_1 -x_1x_3 \\
x'_3=-bx_3+x_1x_2 \\
x'_4=-x_4+4x_1 \\
x'_5=x_6+2x_2 \\
x'_6=-3x_5-2x_6-0.00005x_5^3+0.5x_4,
\end{array}
\end{eqnarray}
again with the coefficients $\sigma=10,$ $b=8/3$ and $r=166.25.$ It can be easily verified that condition $(A7)$ is valid for system $(\ref{replication_of_intermittency}).$ We consider the trajectory of system $(\ref{replication_of_intermittency})$ corresponding to the initial data $x_1(0)=-23.3,$   $x_2(0)=38.3,$  $x_3(0)=193.4,$  $x_4(0)=-17.7,$   $x_5(0)=11.4,$  and  $x_6(0)=2.5,$ and represent the graphs for the $x_4,x_5$ and $x_6$ coordinates in Figure $\ref{intermittency_fig2}$  such that the intermittent behavior in the replicator system is observable. The similarity between the time-series of the generator and the replicator counterpart shows the morphogenesis of intermittency.

In next our discussion, we will emphasize by means of simulations the morphogenesis of the double-scroll Chua's attractor in a unidirectionally coupled open chain of Chua circuits. Approaches for the generation of hyperchaotic systems have already been discussed  making use of Chua circuits which are all chaotic \cite{Kapitaniak94b,Anishchenko94}. It deserves to remark that to create hyperchaotic attractors in previous papers, others consider both involved interacting systems chaotic, but in our case only the generator is chaotic and all other Chua systems considered as replicators are non-chaotic.

\subsection{\textbf {Morphogenesis of the double-scroll Chua's attractor}} \label{chua_system}

There is well known result of the chaoticity based on the double-scroll Chua's attractor \cite{Chua85}. It was proven first in the paper \cite{Chua86} rigorously and the proof is based on the Shil'nikov's theorem \cite{Shilnikov65}. Since the Chua circuit and its chaotic behavior is of extreme importance from the theoretical point of view and its usage area in electrical circuits by radio physicists and nonlinear scientists from other disciplines, one can suppose that morphogenesis of the chaos  will also be  of a practical and a theoretical interest.

We just take into account a simulation result which supports that morphogenesis idea can be developed also from this point of view.

Let us consider the dimensionless form of the Chua's oscillator given by the system
\begin{eqnarray} 
\begin{array}{l}
x_1'=k \alpha [x_2-x_1 -f(x_1)] \\ \label{Chua1}
x_2'= k (x_1-x_2+x_3) \\
x_3'=k (-\beta x_2-\gamma x_3) \\
f(x)=bx+0.5(a-b)\left(  \left| x+1 \right|+ \left|x-1\right|  \right)
\end{array}
\end{eqnarray}
where $\alpha,\beta,\gamma, a, b$ and $k$ are constants.

In paper \cite{Chua93}, it is indicated that system $(\ref{Chua1})$ with the coefficients $\alpha=\frac{21.32}{5.75},$ $\beta=7.8351,$ $\gamma=\frac{1.38166392}{12},$ $a=-1.8459,$ $b=-0.86604$ and $k=1$ admits a stable equilibrium.

In what follows, as the generator, we make use of the system $(\ref{Chua1})$ together with the coefficients $\alpha=15.6, \beta=25.58, \gamma=0, a=-\frac{8}{7}, b=-\frac{5}{7}$ and $k=1$ such that a double-scroll Chua's attractor takes place \cite{Alligood}, and consider the $12-$dimensional result-system
\begin{eqnarray} 
\begin{array}{l}
x_1'=15.6 [x_2-\frac{2}{7}x_1+\frac{3}{14} \left(  \left| x_1+1 \right|+ \left|x_1-1\right|  \right) ] \\ \label{Chua2}
x_2'= x_1-x_2+x_3 \\
x_3'=-25.58 x_2\\
x_4'=\frac{21.32}{5.75} [x_5-0.13396x_4+0.48993\left(  \left| x_4+1 \right|+ \left|x_4-1\right|  \right) ] +2x_1\\ 
x_5'= x_4-x_5+x_6 +5x_2\\
x_6'=-7.8351 x_5-\frac{1.38166392}{12} x_6+2x_3\\
x_7'=\frac{21.32}{5.75} [x_8-0.13396x_7+0.48993\left(  \left| x_7+1 \right|+ \left|x_7-1\right|  \right) ] +2x_4\\ 
x_8'= x_7-x_8+x_9 + 3x_5\\
x_9'=-7.8351 x_8-\frac{1.38166392}{12} x_9-0.001x_6\\
x_{10}'=\frac{21.32}{5.75} [x_{11}-0.13396x_{10}+0.48993\left(  \left| x_{10}+1 \right|+ \left|x_{10}-1\right|  \right) ] +4x_7\\ 
x_{11}'= x_{10}-x_{11}+x_{12} -0.1x_8\\
x_{12}'=-7.8351 x_{11}-\frac{1.38166392}{12} x_{12}+2x_9.
\end{array}
\end{eqnarray}

\begin{figure}[ht] 
\centering
\includegraphics[width=14.5cm]{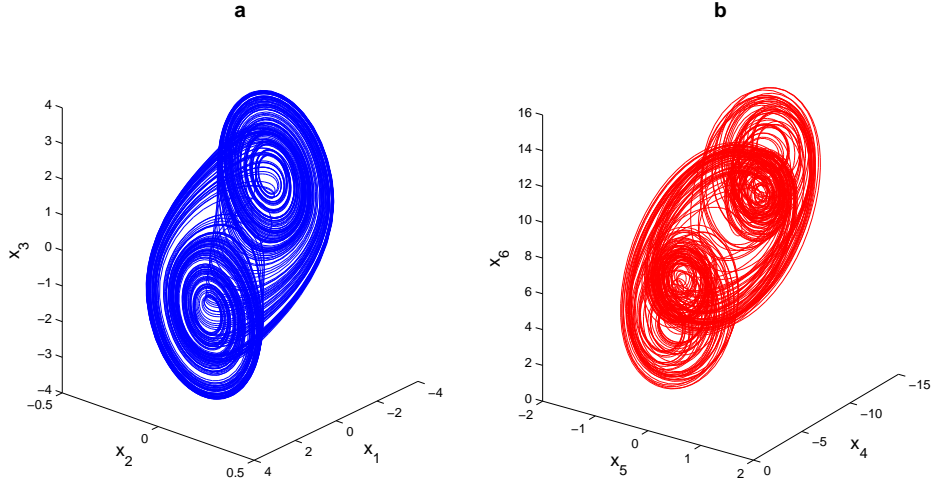}
\caption{\footnotesize{$3-$dimensional projections of the chaotic attractor of the result-system (\ref{Chua2}). (a) Projection on the $x_1-x_2-x_3$ space, (b) Projection on the $x_4-x_5-x_6$ space. The picture in $(a)$ shows the attractor of the original prior chaos of the generator system (\ref{Chua1}) and $(b)$ represents the attractor of the first replicator. The resemblance between shapes of the attractors of the generator and the replicator systems makes the morphogenesis of chaos apparent. }}
\label{Chua_double1}
\end{figure}

In Figure \ref{Chua_double1}, we simulate the $3-$dimensional projections on the $x_1-x_2-x_3$ and $x_4-x_5-x_6$ spaces of the trajectory of the result-system $(\ref{Chua2})$ with the initial data $x_1(0)=0.634,$  $x_2(0)=-0.093,$  $x_3(0)=-0.921,$  $x_4(0)=-8.013,$  $x_5(0)=0.221,$  $x_6(0)=6.239,$ $x_7(0)=-50.044,$  $x_8(0)=-0.984,$  $x_9(0)=48.513,$  $x_{10}(0)=-256.325,$ $x_{11}(0)=7.837,$  $x_{12}(0)=264.331.$ The projection on the $x_1-x_2-x_3$ space shows the  double-scroll Chua's attractor produced by the generator system $(\ref{Chua1}),$ and projection on the $x_4-x_5-x_6$ space represents the chaotic attractor of the first replicator.

\begin{figure}[ht] 
\centering
\includegraphics[width=14.5cm]{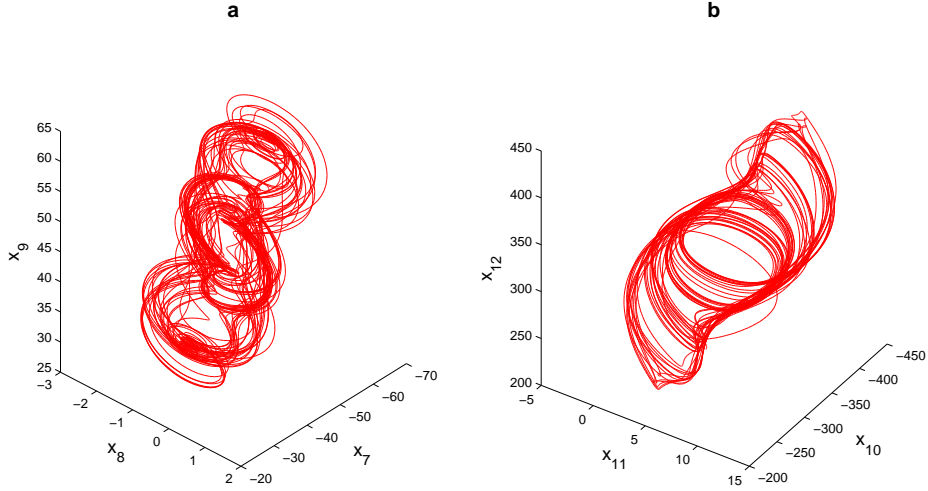}
\caption{\footnotesize{$3-$dimensional projections of the chaotic attractor of the result-system (\ref{Chua2}). (a) Projection on the $x_7-x_8-x_9$ space, (b) Projection on the $x_{10}-x_{11}-x_{12}$ space. $(a)$ and $(b)$ demonstrates the attractors generated by the second and the third replicator systems, respectively.}}
\label{Chua_double2}
\end{figure}

\begin{figure}[ht] 
\centering
\includegraphics[width=13.5cm]{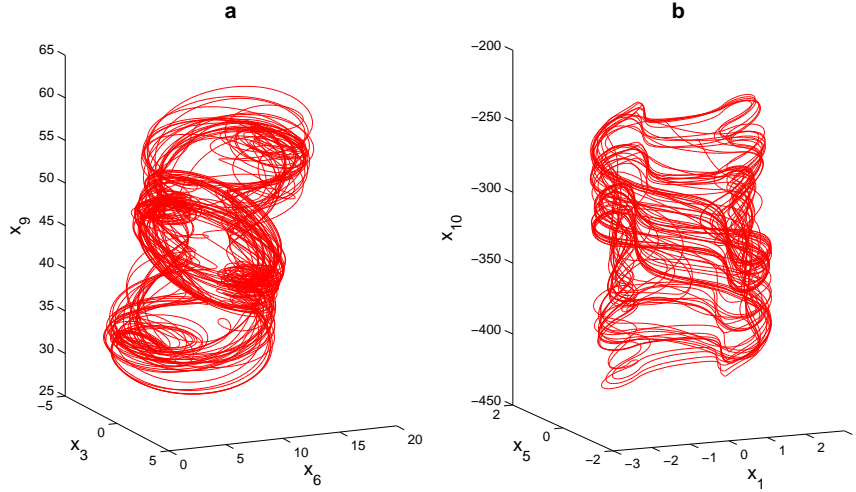}
\caption{\footnotesize{$3-$dimensional projections of the chaotic attractor of the result-system (\ref{Chua2}). (a) Projection on the $x_3-x_6-x_9$ space, (b) Projection on the $x_{1}-x_{5}-x_{10}$ space. Since we are not able to provide a picture of the whole chaotic attractor in the $12-$dimensional space, to obtain better impression,  the projections shown in $(a)$ and $(b)$ are presented.  These pictures give us an idea about the fascinating chaotic attractor of the result-system, which can possibly be called as 12D Chua's attractor. }}
\label{Chua_double3}
\end{figure}

In a similar way, we display the projections of the same trajectory on the $x_7-x_8-x_9$ and $x_{10}-x_{11}-x_{12}$ spaces, which correspond to the attractors of the second and the third replicator systems, in Figure $\ref{Chua_double2};$ and the projections on the $x_3-x_6-x_9$ and $x_{1}-x_{5}-x_{10}$ spaces in Figure $\ref{Chua_double3}.$ The illustrations shown in Figure $\ref{Chua_double1},$ Figure $\ref{Chua_double2}$ and  Figure $\ref{Chua_double3}$ indicate the morphogenesis double-scroll Chua attractor.

Possibly the result-system $(\ref{Chua2})$ produces a double-scroll Chua's attractor with hyperchaos where the number of positive Lyapunov exponents are more than one and even four.

\subsection{\textbf{Morphogenesis and the logistic map}} \label{relay_system_section}

Since the logistic map is very convenient to verify Devaney's, Li-Yorke chaos and also chaos through period-doubling cascade, one can expect that it can be used as a generator. This is true in the theoretical sense, but not for simulations since only chaos through period-doubling route can be presented through simulations.

In paper \cite{Akh7} it is rigorously proved that the Duffing equation perturbed with a pulse function 
\begin{eqnarray}
\begin{array}{l}
x''+0.18x'+2x+0.00004x^3=0.02\displaystyle \cos(2\pi t)+\nu(t,t_0,\mu_{\infty}), \label{forced_Duffing}
\end{array}
\end{eqnarray}
with the coefficients $m_0=2, m_1=1$ and $\mu_{\infty}=3.8,$ where the relay function $\nu(t,t_0,\mu_{\infty})$ is described by equation $(\ref{relay_function}),$ admits the chaos through period-doubling cascade on the time interval $[0, \infty)$ and obeys the Feigenbaum universal behavior \cite{Feigenbaum80}. 

By favour of the new variables $x_1=x$ and $x_2=x',$ equation $(\ref{forced_Duffing})$ can be reduced to the system
\begin{eqnarray}
\begin{array}{l}
x_{1}'=x_2  \label{forced_Duffing_system} \\
x_{2}'=-0.18x_2-2x_1-0.00004x_1^3+0.02\displaystyle\cos(2\pi t)+\nu(t,t_0,\mu_{\infty}).
\end{array}
\end{eqnarray}
One can see that system $(\ref{forced_Duffing_system})$ is in the form of system $(\ref{previous}).$ For the illustration of morphogenesis of chaos, we will make use of the relay-system $(\ref{forced_Duffing_system})$ as the generator, in the role of the core as displayed in Figure \ref{box2}, and attach three replicator systems to obtain the $8-$dimensional result-relay-system 
\begin{eqnarray}
\begin{array}{l}
x_1'=x_2  \label{relay_system} \\
x_2'=-0.18x_2-2x_1-0.00004x_1^3+0.02\displaystyle\cos(2\pi t)+\nu(t,t_0,\mu_{\infty}) \\
x_3'=x_4-0.1x_1 \\
x_4'=-10x_3-6x_4-0.03x_3^3+4x_2 \\
x_5'=x_6+2x_1 \\
x_6'=-2x_5-2x_6+0.007x_5^3+0.6x_2 \\
x_7'=x_8-0.5x_2 \\
x_8'=-5x_7-4x_8-0.05x_7^3+2.5x_1,
\end{array}
\end{eqnarray}
where again $m_0=2, m_1=1$ and $\mu_{\infty}=3.8.$

\begin{figure}[ht] 
\centering
\includegraphics[width=14.5cm]{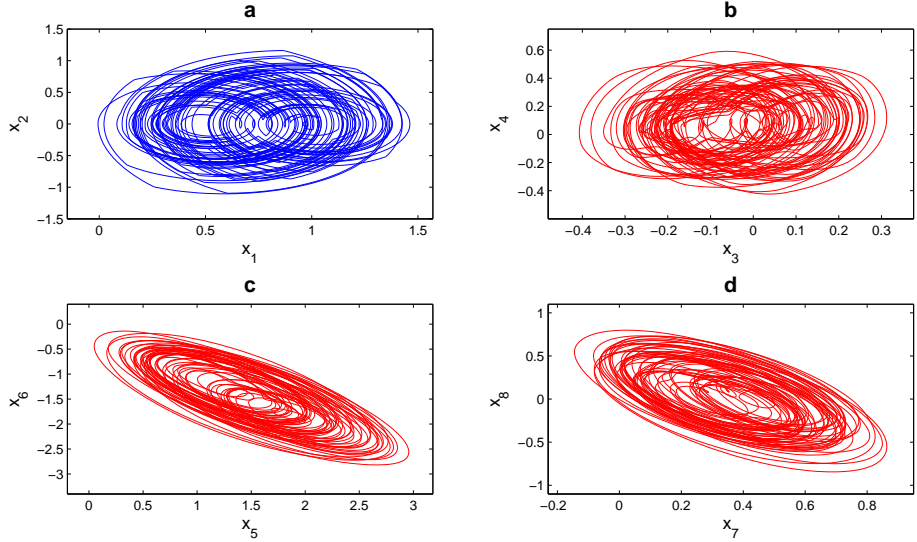}
\caption{\footnotesize{$2-$dimensional projections of the chaotic attractor of the result-system $(\ref{relay_system}).$ (a) Projection on the $x_1-x_2$ plane, (b) Projection on the $x_3-x_4$ plane, (c) Projection on the $x_5-x_6$ plane, (d) Projection on the $x_7-x_8$ plane. The picture in $(a)$ shows the attractor of the prior chaos produced by the generator $(\ref{forced_Duffing_system}),$ which is a relay-system, and in $(b)-(d)$ the chaotic attractors of the replicator systems are observable. The illustrations in $(b)-(d)$ repeated the structure of the attractor shown in $(a),$ and the mimicry between these pictures is an indicator of the morphogenesis of chaos.}}
\label{2d_relay}
\end{figure}

Our theoretical results show that system $(\ref{relay_system}),$ as well as the replicators, admit the chaos through period-doubling cascade and obey the universal behavior of Feigenbaum. Figure $\ref{2d_relay}$ shows the $2-$dimensional projections on the $x_1-x_2, x_3-x_4, x_5-x_6$ and $x_7-x_8$ planes of the trajectory of the result-relay-system $(\ref{relay_system})$ with initial data $x_1(0)=1.37, x_2(0)=-0.05, x_3(0)=0.05, x_4(0)=-0.1,  x_5(0)=1.09, x_6(0)=-0.81,  x_7(0)=0.08, x_8(0)=0.21.$ The picture seen in Figure $\ref{2d_relay}, (a)$ is the attractor of the generator $(\ref{forced_Duffing_system})$ and accordingly Figure $\ref{2d_relay}, (b)-(d)$ represents the attractors of the first, second and the third replicator systems, respectively. It can be easily verified that all replicators used inside the system $(\ref{relay_system})$ satisfy condition $(A7).$ The resemblance of the chaotic attractors of the generator and the replicators is a consequence of morphogenesis of chaos. 

\begin{figure}[ht] 
\centering
\includegraphics[width=14.5cm]{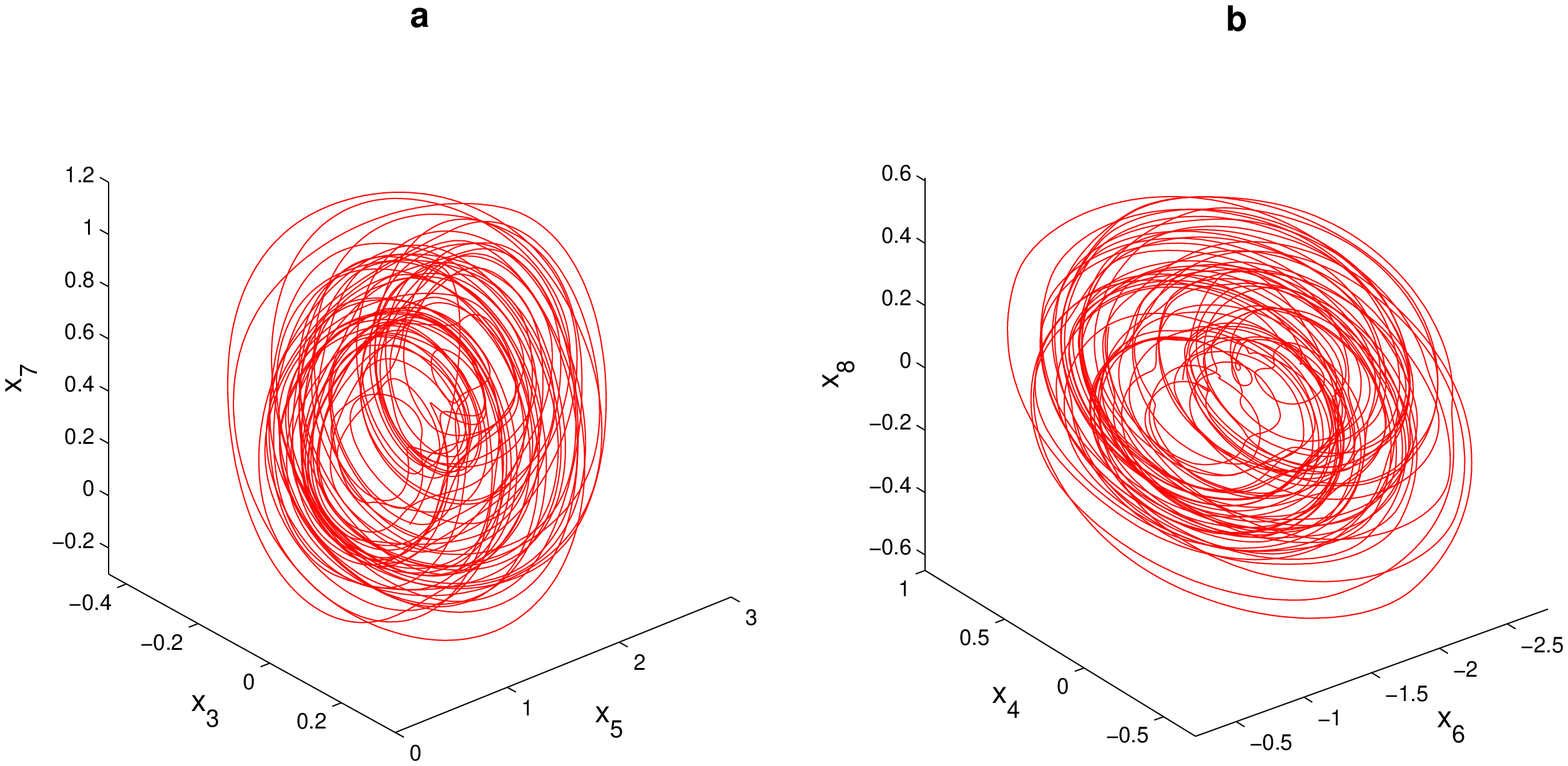}
\caption{\footnotesize{$3-$dimensional projections of the chaotic attractor of the result-relay-system $(\ref{relay_system}).$ (a) Projection on the $x_3-x_5-x_7$ space, (b) Projection on the $x_4-x_6-x_8$ space. The illustrations presented in $(a)$ and $(b)$ inspire us about the shape of the impressive chaotic attractor in the $8-$dimensional space. }}
\label{3d_relay}
\end{figure}

Next, we make use of the same trajectory to illustrate the $3-$dimensional projections on the $x_3-x_5-x_7$ and $x_4-x_6-x_8$ spaces in Figure $\ref{3d_relay}.$ The visualizations shown in Figure $\ref{3d_relay}, (a)$ and $(b),$ inform us about the shape of the result attractor of system $(\ref{relay_system}).$

Now, let us continue with the  control of morphogenesis of chaos. In paper \cite{Akh7}, it is indicated that to control chaos of system $(\ref{forced_Duffing_system})$ generated through the period-doubling cascade, unstable periodic orbits of the logistic map $h(s,\mu_{\infty})=\mu_{\infty} s(1-s)$ must be necessarily controlled, and the OGY control method \cite{Ott90,Sch99} is proposed to stabilize the periodic solutions of system $(\ref{forced_Duffing_system})$ by favour of the conjugate control system 
\begin{eqnarray}
\begin{array}{l}
x_{1}'=x_2  \label{control_system} \\
x_{2}'=-0.18x_2-2x_1-0.00004x_1^3+0.02\displaystyle\cos(2\pi t)+\nu(t,t_{1},\bar{\mu}_i).
\end{array}
\end{eqnarray}
In system $(\ref{control_system}),$ the sequence $\bar{\mu}_i$ is given by the formula
\begin{eqnarray}\label{mu}
\bar \mu_{i}=\mu_\infty\left(1+\frac{[2\kappa^{(j)}-1][\kappa_{i}-\kappa^{(j)}]}{\kappa^{(j)}[1-\kappa^{(j)}]} \right),
\end{eqnarray}
where $\kappa^{(j)}$, $j\geq 0,$ with $\kappa^{(0)}=t_0 \in [0,1]$ is the target periodic orbit of the logistic map $h(s,\mu_{\infty})$ to be controlled, and $\kappa_{i+1}=h(\kappa_{i},\mu_{\infty}),$ $i\geq 0,$ with $\kappa_0=t_1\in [0,1].$

In this method, the sequence $\bar{\mu}_i$ is supposed to vary in a small range around the value $\mu_{\infty}$, that is, $\mu$ can take values in the range $[\mu_{\infty}-\delta, \mu_{\infty}+\delta]$, where $\delta$ is a given small positive number. When the trajectory is outside the neighborhood of the target periodic orbit, no parameter perturbation is applied and the system evolves at its nominal parameter value $\mu_{\infty}$. That is, we set $ \bar \mu_{i} = \mu_{\infty}$ when $\left|\bar \mu_{i}-\mu_{\infty}\right| \geq \delta.$

To simulate the result, let us consider the system 
\begin{eqnarray}
\begin{array}{l}
x_1'=x_2  \label{relay_system_control} \\
x_2'=-0.18x_2-2x_1-0.00004x_1^3+0.02\displaystyle\cos(2\pi t)+\nu(t,t_{1},\bar{\mu}_i) \\
x_3'=x_4-0.1x_1 \\
x_4'=-10x_3-6x_4-0.03x_3^3+4x_2 \\
x_5'=x_6+2x_1 \\
x_6'=-2x_5-2x_6+0.007x_5^3+0.6x_2 \\
x_7'=x_8-0.5x_2 \\
x_8'=-5x_7-4x_8-0.05x_7^3+2.5x_1 \\
\end{array}
\end{eqnarray}
which is the control system conjugate to the result-relay-system $(\ref{relay_system}),$ where again $m_0=2$ and $m_1=1.$

\begin{figure}[ht] 
\centering
\includegraphics[width=14.5cm]{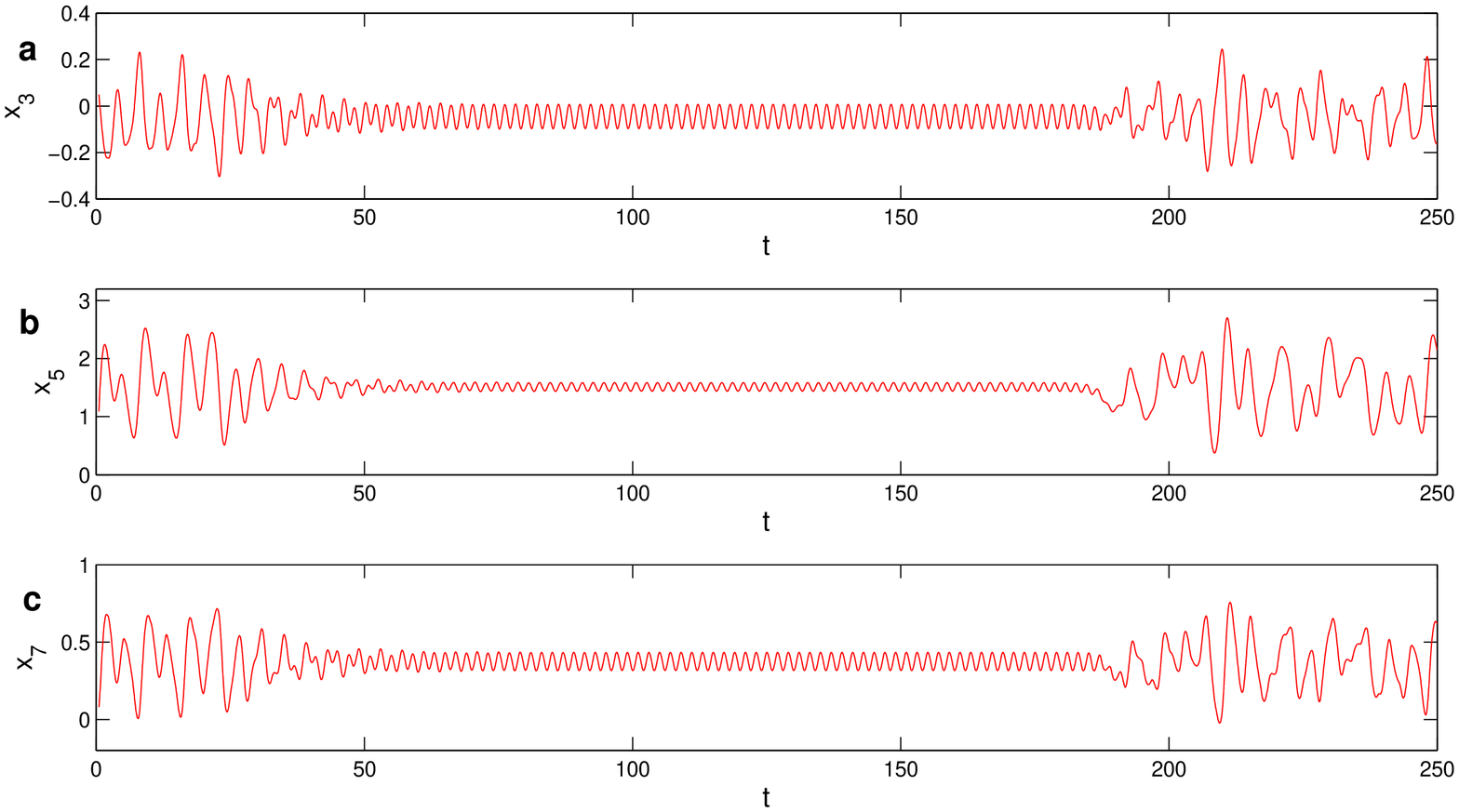}
\caption{\footnotesize{OGY control method applied to the result-relay-system $(\ref{relay_system}).$ (a) The graph of the $x_3-$ coordinate,  (b) The graph of the $x_5-$ coordinate, (c) The graph of the $x_7-$ coordinate. For stabilizing the $2-$periodic solution of the result-relay-system $(\ref{relay_system}),$ the OGY control method is applied to the generator system $(\ref{forced_Duffing_system})$ around the fixed point $2.8/3.8$ of the logistic map. The value $\delta=0.19$ is used, and the control starts at $t=25$ and ends at $t=125$. After approximately $60$ iterations when the control is switched off, the chaos becomes dominant again and irregular motion reappears. The presented coordinates $x_3,x_5$ and $x_7,$ all correspond to replicator systems and it is possible to obtain similar illustrations for the remaining ones, which are not just simulated here.}}
\label{control_relay}
\end{figure}

To simulate the control results, we make use of the values  $\delta=0.19, t_1=0.5, t_0=2.8/3.8$ and the trajectory of system $(\ref{relay_system_control})$
with the initial data $x_1(0)=1.37, x_2(0)=-0.05, x_3(0)=0.05, x_4(0)=-0.1,  x_5(0)=1.09, x_6(0)=-0.81,  x_7(0)=0.08, x_8(0)=0.21.$ Taking the value $t_0=2.8/3.8$ means that the control mechanism is applied around the fixed point of the logistic map, and consequently stabilizes the $2-$periodic solutions of the generator and  the existing replicators. The control is switched on at $t=25$ and switched off at $t=125.$ The graphs of the coordinates $x_3,x_5$ and $x_7$ are shown in Figure $\ref{control_relay}.$ It is observable that the $2-$periodic solutions of the replicators and hence of the result-relay-system $(\ref{relay_system})$ is stabilized. In other words, the morphogenesis of chaos is controlled and the results of Theorem \ref{control_of_chaos} are validated also through simulations. One can see that after approximately $60$ iterations when the control is switched off, the chaos becomes dominant again and irregular motion reappears.

\subsection{\textbf{Quasiperiodicity through morphogenesis}}\label{quasiperiodicity}

Now, let us indicate that if there are more than one generator system, morphogenesis will have some new forms such as periodicity gives birth to quasiperiodicity. 

In paper \cite{Dowell86}, it is shown that the Duffing equation
\begin{eqnarray} 
\begin{array}{l} \label{quasi1}
x''+0.168 x' - \frac{x}{2} \left(1-x^2\right) = \mu \displaystyle \sin(t)
\end{array}
\end{eqnarray}
where $\mu$ is a parameter, admits the chaos through period-doubling cascade at the accumulation point $\mu_{\infty}=0.21$ of the sequence $\left\{\mu_i\right\}$ of parameter values, that is, for $\mu =\mu_{\infty},$ the equation $(\ref{quasi1})$ admits infinitely many periodic solutions with periods $2k\pi,$ for each natural number $k.$ Making the change of variables $t=2\pi s$ and $x(t)=y(s),$ one can obtain the equation
\begin{eqnarray} 
\begin{array}{l} \label{quasi2}
y''+0.336\pi y' - 2\pi^2 y \left(1-y^2\right) = 4 \pi^2\mu \displaystyle \sin(2\pi t).
\end{array}
\end{eqnarray}
Likewise equation $(\ref{quasi1}),$ it is clear that the equation  $(\ref{quasi2})$ also admits the chaos through period-doubling cascade and admits infinitely many periodic solutions with periods $1,2,4,\ldots$ when one considers the equation with the parameter value $\mu =\mu_{\infty}.$ 
Using the new variables $x_1=x,x_2=x',x_3=t$ and $x_4=y,x_5=y',x_6=s,$ respectively, one can convert the equations $(\ref{quasi1})$ and $(\ref{quasi2})$ to the systems
\begin{eqnarray} 
\begin{array}{l}
x'_1=x_2 \\  \label{quasi3}
x'_2=-0.168x_2+\frac{x_1}{2} \left(1-x_1^2\right)+\mu_{\infty}\displaystyle \sin(x_3) \\
x'_3=1
\end{array}
\end{eqnarray}
and
\begin{eqnarray} 
\begin{array}{l} \label{quasi4}
x'_4=x_5 \\
x'_5=-0.336\pi x_5+2\pi^2x_4(1-x_4^2)+4 \pi^2 \mu_{\infty} \sin(2\pi x_6)  \\
x'_6=1
\end{array}
\end{eqnarray}
respectively.
Now, we shall make use both of the systems $(\ref{quasi3})$ and $(\ref{quasi4})$ as generator systems to obtain a chaotic system with infinitely many quasiperiodic solutions.
We mean that the two systems admit incommensurate periods and consequently their influence on the replicator will be quasiperiodic and one can expect that replicator will expose infinitely many quasiperiodic solutions. Consider now the $9-$dimensional result-system 
\begin{eqnarray} 
\begin{array}{l}
x'_1=x_2 \\ \label{quasi5}
x'_2=-0.168x_2+0.5x_1(1-x_1^2)+\mu_{\infty} \sin(x_3)+C_1(x_2(t-2\pi)-x_2(t)) \\
x'_3=1 \\
x'_4=x_5 \\
x'_5=-0.336\pi x_5+2\pi^2x_4(1-x_4^2)+4 \pi^2 \mu_{\infty} \sin(2\pi x_6)+C_2(x_5(t-1)-x_5(t)) \\
x'_6=1 \\
x'_7=x_8+x_1+x_4 \\
x'_8=-3x_7-2x_8-0.008x_7^3+x_2+x_5 \\
x'_9=-x_9+x_3+x_6,
\end{array}
\end{eqnarray}
where the last three equations are of a replicator system.

\begin{figure}[ht] 
\centering
\includegraphics[width=13.5cm]{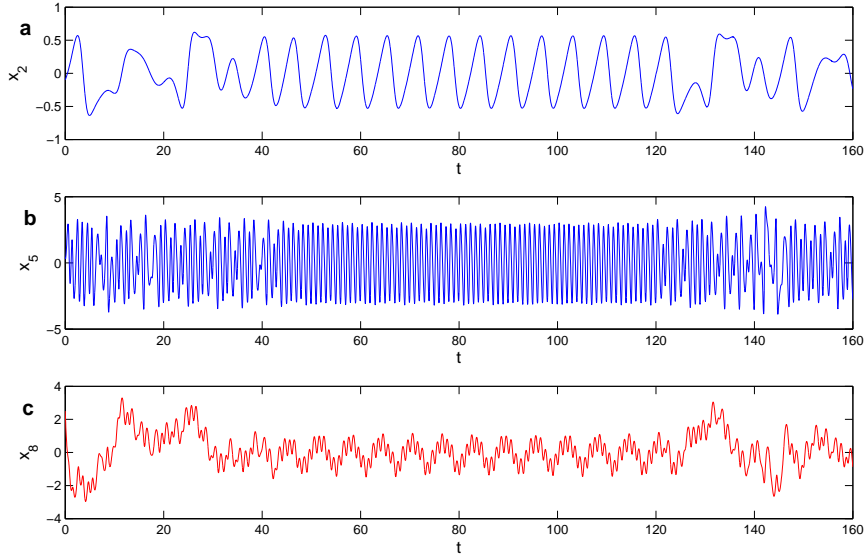}
\caption{\footnotesize{Pyragas control method applied to the result-system $(\ref{quasi5}).$ (a) The graph of the $x_2-$coordinate, (b) The graph of the $x_5-$coordinate, (c) The graph of the $x_8-$coordinate. The simulation result for the result-system $(\ref{quasi5})$ is provided such that in $(a)$ and $(b)$ periodic solutions with incommensurate periods $1$ and $2\pi$ are controlled by Pyragas method and in $(c)$ quasiperiodic solution of the replicator system is pictured. For the coordinates $x_1,x_4$ and $x_7$ we have the similar results which are not illustrated here.}}
\label{quasi1fig}
\end{figure}

We make use of the Pyragas control method together with the values of $C_1=0.1,$ and $C_2=0.62,$ to stabilize the $2\pi$ and $1$ periodic solutions of the generators systems (\ref{quasi3}) and (\ref{quasi4}), respectively. For our purposes, let us simulate a solution of system $(\ref{quasi5})$ with initial data $x_1(0)=0.4,  x_2(0)= -0.1, x_3(0)= 0,  x_4(0)= -0.15,  x_5(0) = 0.4,  x_6(0) =0, x_7(0)=1.1, x_8(0)= 2.5, x_9(0)= 0.12.$ The simulation results are seen in Figure \ref{quasi1fig}. The control mechanism starts at $t=40$ and ends at $t=120.$  Up to $t=40,$ the chaos not only in the generator systems, but also in the replicator is observable. Between $t=40$ and $t=120$ the quasiperiodic solution of the replicator is stablized. After $t=120,$ chaos in the system (\ref{quasi5}) is develops again.

Possibly the obtained simulation result and previous theoretical discussions can give a support to the idea of \textit{quasiperiodical cascade} for the appearance of chaos which can be considered as a development of the popular period-doubling route to chaos. 

In paper \cite{Shilnikov02}, it has been mentioned that, in general, in the place of countable set of periodic solutions to form chaos, one can take an uncountable collection of Poisson stable motions which are dense in a quasi-minimal set. This can be also observed in Horseshoe attractor \cite{Smale67}. These emphasize that our simulation of quasiperiodic solutions can be considered as another evidence for the theoretical results.

\subsection{\textbf{Replicators with nonnegative eigenvalues}}\label{open}

%
%


We recall that in our theoretical discussions all eigenvalues of the real valued constant matrix $A,$ used in system $(\ref{2}),$ are assumed to have negative real parts. Now, as open problems from the theoretical point of view, we shall discuss through simulations the problem of morphogenesis of chaos in the case when the matrix $A$ possesses an eigenvalue with positive or zero real part. 

\begin{figure}[ht] 
\centering
\includegraphics[width=14.5cm]{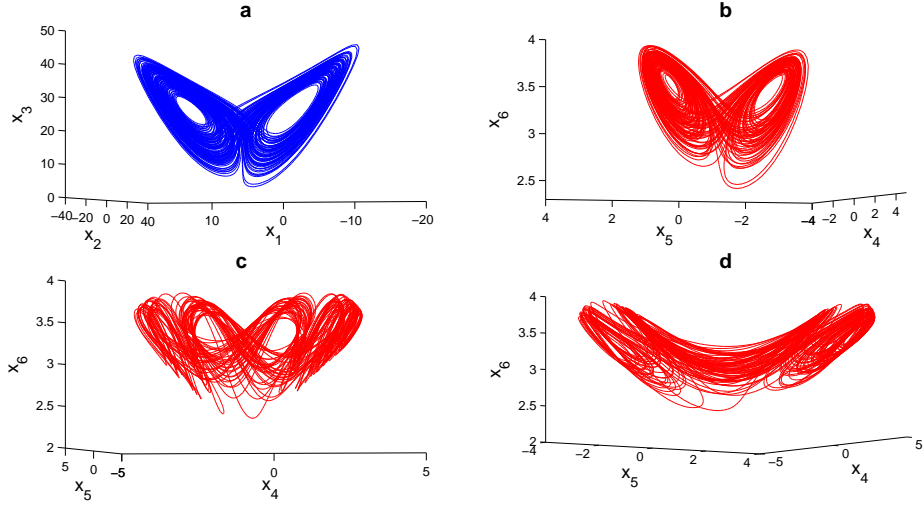}
\caption{\footnotesize{$3-$dimensional projections of the chaotic attractor of the result-system $(\ref{open_problem1}).$  (a) Projection on the $x_1-x_2-x_3$ space, (b) Projection on the $x_4-x_5-x_6$ space. (c)-(d) Projections on the $x_4-x_5-x_6$ space with different standpoints. In $(a),$ the famous Lorenz attractor produced by the generator system $(\ref{Lorenz_system})$ with coefficients $\sigma=10,$ $r=28$ and $b=\frac{8}{3}$ is shown. In $(b)-(d),$ as in usual way, the projection of the chaotic attractor of the result-system $(\ref{open_problem1}),$ which can separately be considered as a chaotic attractor, is presented with three different standpoints. Possibly one can call the attractor of the result-system as $6D$ Lorenz attractor.} }
\label{positive_eigenvalue}
\end{figure}

First, we are going to concentrate on the case of the existence of an eigenvalue with positive real part. Let us make use of the Lorenz system $(\ref{Lorenz_system})$ together with the coefficients $\sigma=10,$ $r=28$ and $b=\frac{8}{3}$ as the generator, which is known to be chaotic \cite{Lorenz63,Sprott10}, and set up the $6-$dimensional result-system 
\begin{eqnarray}
\begin{array}{l}
x_1'=-10x_1+10x_2 \label{open_problem1} \\
x_2'=-x_2+28x_1-x_1x_3 \\
x_3'=x_1x_2-\frac{8}{3}x_3 \\
x_4'=-2x_4+x_1 \\
x_5'=-3x_5+x_2 \\
x_6'=4x_6-x_6^3+x_3.
\end{array}
\end{eqnarray}
It is crucial to note that the system $(\ref{open_problem1})$ is of the form of system $(\ref{1})+(\ref{2})$ where the matrix $A$ admits the eigenvalues $-2,-3$ and $4,$ such that one of them is positive. 
We take into account the solution of system $(\ref{open_problem1})$ with the initial data $x_1(0)=-12.7,x_2(0)=-8.5,x_3(0)=36.5,x_4(0)=-3.4,x_5(0)=-3.2,x_6(0)=3.7$  and visualize in Figure \ref{positive_eigenvalue} the projections of the corresponding trajectory on the $x_1-x_2-x_3$ space and on the $x_4-x_5-x_6$ space with three different standpoints. It is seen that the replicator system admits the chaos and the morphogenesis procedure works for system $(\ref{open_problem1}).$

\begin{figure}[ht] 
\centering
\includegraphics[width=14.5cm]{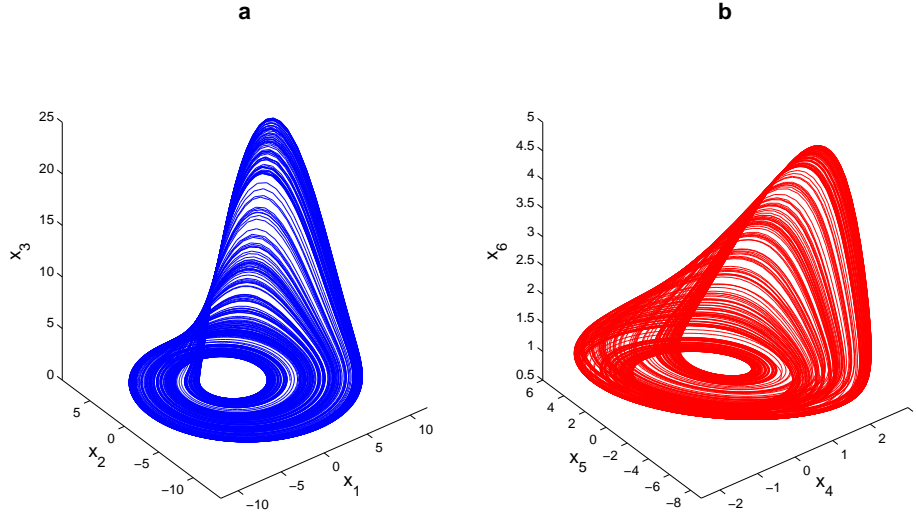}
\caption{\footnotesize{ $3-$dimensional projections of the chaotic attractor of the result-system $(\ref{open_problem3}).$ (a) Projection on the $x_1-x_2-x_3$ space, (b) Projection on the $x_4-x_5-x_6$ space. The picture in $(a)$ indicates the famous R$\ddot{o}$ssler attractor produced by the generator system $(\ref{open_problem2}).$ The similarity between the illustrations presented in $(a)$ and $(b)$ supports the morphogenesis of chaos. The attractor of the result-system $(\ref{open_problem3})$ can be possibly called as $6D$ R$\ddot{o}$ssler attractor.}}
\label{zero_eigenvalue}
\end{figure}

Next, we continue to our discussion with the case of the existence of an eigenvalue with a zero real part. This time we consider the chaotic R$\ddot{o}$ssler system \cite{Sprott10,Rossler76} described by 
\begin{eqnarray}
\begin{array}{l}
x_1'=-(x_2+x_3) \label{open_problem2} \\
x_2'=x_1+0.2x_2 \\
x_3'=0.2+x_3(x_1-5.7)
\end{array}
\end{eqnarray}
as the generator and constitute the result-system 
\begin{eqnarray}
\begin{array}{l}
x_1'=-(x_2+x_3) \label{open_problem3} \\
x_2'=x_1+0.2x_2 \\
x_3'=0.2+x_3(x_1-5.7) \\
x_4'=-4x_4+x_1 \\
x_5'=-x_5+x_2 \\
x_6'=-0.2x_6^3+x_3.
\end{array}
\end{eqnarray}
In this case, one can consider system $(\ref{open_problem3})$ as in the form of $(\ref{1})+(\ref{2})$ where the matrix $A$ is a diagonal matrix with entries $-4,-1,0$ on the diagonal and admits the number $0$ as an eigenvalue. We simulate the solution of system $(\ref{open_problem3})$ with the initial data $x_1(0)=4.6,x_2(0)=-3.3,x_3(0)=0,x_4(0)=1,x_5(0)=-3.7,x_6(0)=0.8.$ The projections of the trajectory on the $x_1-x_2-x_3$ and $x_4-x_5-x_6$ spaces are seen in Figure $\ref{zero_eigenvalue}.$ The simulation results confirm that the replicator mimics the complex behavior of the generator system.

These results of the simulations request more detailed investigation which concern not only the theoretical existence of chaos, but also its resistance and stability. 

\section{Conclusion} \label{conclusion}

Morphogenesis of different types of chaos such as obtained by period-doubling cascade, Devaney's and Li-Yorke chaos is recognized in the paper. The definitions of chaotic sets as well as the hyperbolic sets of functions are introduced and the self-replication of the chaos is proved rigorously. The morphogenesis mechanism is based on a generating chaotic element inserted in a network of systems.  Morphogenesis of intermittency as well as the double-scroll Chua's attractor and quasiperiodicity are discussed. Moreover, control problem of the self-replicated chaos is realized. Some of the results are illustrated through relay system's dynamics. Appropriate simulations are presented using the indicated method successively.

One can construct the morphogenesis mechanism by the formation of consecutive replications of chaos or replication of chaos from a core system. It is also possible to construct a result-system using these two mechanisms in a mixed style.

In the present paper, we use the term ``morphogenesis" in the meaning of ``processes creating forms" where we accept the \textit{form} not only as a type of chaos, but also accompanying concepts as the structure of the chaotic attractor, its fractal dimension, form of the bifurcation diagram, the spectra of Lyapunov exponents, inheritance of intermittency, etc. Replication of a known type of chaos in systems with arbitrary large dimension is a significant consequence of our paper. More precisely, by the method presented, we show that a known type of chaos, such as period-doubling cascade, Devaney's or Li-Yorke chaos, can be extended to systems with arbitrary large dimension. We provide a new mechanism through the insertion of chaos from one system to another where the first is not necessarily a simple one as an ``embryo". To be more precise,  we provide replication of chaos between unidirectionally coupled systems such that finally to obtain a result-system admitting the same type of chaos.


Our method can be useful to explain extension of irregular behavior like crisis in economic models \cite{Hanswalter89} as well as reproduction of chaos in mechanical systems \cite{Moon} and electric circuits \cite{Chua93}.

{\par\noindent}

\noindent {\large \textbf{Acknowledgements}}

{\par\noindent}

This research was supported by a grant (111T320) from TUBITAK, the Scientific and Technological Research Council of Turkey.



\end{document}